\documentclass[aps, prx, twocolumn, footinbib, superscriptaddress]{revtex4-2}

\usepackage{amsmath}
\usepackage{bm}
\usepackage{braket}
\usepackage{float}
\usepackage[acronym]{glossaries}
\usepackage{graphicx}
\usepackage[
  colorlinks=true, 
  allcolors=blue
]{hyperref}
\usepackage[capitalise]{cleveref}
\usepackage[separate-uncertainty]{siunitx}
\usepackage{xcolor}

\newacronym{ase}{ASE}{Anomalous Skin Effect}
\newacronym{cse}{CSE}{Classical Skin Effect}
\newacronym{vse}{VSE}{Viscous Skin Effect}
\newacronym{mr}{MR}{momentum-relaxing}
\newacronym{mc}{MC}{momentum-conserving}

\begin{document}
    
\title{Non-local electrodynamics in ultra-pure PdCoO\texorpdfstring{\textsubscript{2}}{2}}
\author{Graham Baker}
\author{Timothy W. Branch}
\affiliation{
  Stewart Blusson Quantum Matter Institute, 
  University of British Columbia, 
  Vancouver, British Columbia, Canada V6T 1Z4
}
\affiliation{
  Department of Physics and Astronomy, 
  University of British Columbia, 
  Vancouver, British Columbia, Canada V6T 1Z1
}
\author{J. S. Bobowski}
\affiliation{
  Stewart Blusson Quantum Matter Institute, 
  University of British Columbia, 
  Vancouver, British Columbia, Canada V6T 1Z4
}
\affiliation{
  Department of Physics and Astronomy, 
  University of British Columbia, 
  Vancouver, British Columbia, Canada V6T 1Z1
}
\affiliation{
  Department of Physics,
  Kyoto University, 
  Kyoto 606-8502, Japan
}
\author{James Day}
\affiliation{
  Stewart Blusson Quantum Matter Institute, 
  University of British Columbia, 
  Vancouver, British Columbia, Canada V6T 1Z4
}
\affiliation{
  Department of Physics and Astronomy, 
  University of British Columbia, 
  Vancouver, British Columbia, Canada V6T 1Z1
}
\author{Davide Valentinis}
\affiliation{
  Institute for Theory of Condensed Matter, 
  Karlsruhe Institute of Technology, 
  Karlsruhe 76131, Germany
}
\affiliation{
  Institute for Quantum Materials and Technologies,
  Karlsruhe Institute of Technology, 
  Karlsruhe 76131, Germany
}
\author{Mohamed Oudah}
\affiliation{
  Stewart Blusson Quantum Matter Institute, 
  University of British Columbia, 
  Vancouver, British Columbia, Canada V6T 1Z4
}
\author{Philippa McGuinness}
\author{Seunghyun Khim}
\affiliation{
  Max Planck Institute for Chemical Physics of Solids, 
  N{\"o}thnitzer Straße 40, 01187 Dresden, Germany
}
\author{Piotr Surówka}
\affiliation{Department of Theoretical Physics, Wroc\l{}aw  University  of  Science  and  Technology,  50-370  Wroc\l{}aw,  Poland}
\affiliation{Institute for Theoretical Physics, University of Amsterdam, 1090 GL Amsterdam, The Netherlands}
\affiliation{Dutch Institute for Emergent Phenomena (DIEP), University of Amsterdam, 1090 GL Amsterdam, The Netherlands}
\author{Yoshiteru Maeno}
\affiliation{
  Toyota Riken - Kyoto University Research Center (TRiKUC),
  Kyoto 606-8501, Japan
}
\affiliation{
  Department of Physics,
  Kyoto University, 
  Kyoto 606-8502, Japan
}
\author{Roderich Moessner}
\affiliation{
  Max Planck Institute for the Physics of Complex Systems, 
  N{\"o}thnitzer Straße 38, 01187 Dresden, Germany
}
\author{Jörg Schmalian}
\affiliation{
  Institute for Theory of Condensed Matter, 
  Karlsruhe Institute of Technology, 
  Karlsruhe 76131, Germany
}
\affiliation{
  Institute for Quantum Materials and Technologies,
  Karlsruhe Institute of Technology, 
  Karlsruhe 76131, Germany
}
\author{Andrew P. Mackenzie}
\affiliation{
  Max Planck Institute for Chemical Physics of Solids, 
  N{\"o}thnitzer Straße 40, 01187 Dresden, Germany
}
\affiliation{
  Scottish Universities Physics Alliance, School of Physics and Astronomy, 
  University of St Andrews, St Andrews KY16 9SS, United Kingdom
}
\author{D. A. Bonn}
\affiliation{
  Stewart Blusson Quantum Matter Institute, 
  University of British Columbia, 
  Vancouver, British Columbia, Canada V6T 1Z4
}
\affiliation{
  Department of Physics and Astronomy, 
  University of British Columbia, 
  Vancouver, British Columbia, Canada V6T 1Z1
}
\date{\today}

\begin{abstract}
    The motion of electrons in the vast majority of conductors is diffusive, obeying Ohm's law. However, the recent discovery and growth of high-purity materials with extremely long electronic mean free paths has sparked interest in non-ohmic alternatives, including viscous and ballistic flow. Although non-ohmic transport regimes have been discovered across a range of materials, including two-dimensional electron gases, graphene, topological semimetals, and the delafossite metals, determining their nature has proved to be challenging. Here, we report on a new approach to the problem, employing broadband microwave spectroscopy of the delafossite metal PdCoO$_2$ in three distinct sample geometries that would be identical for diffusive transport. The observed differences, which go as far as differing power laws, take advantage of the hexagonal symmetry of PdCoO$_2$. This permits a particularly elegant symmetry-based diagnostic for non-local electrodynamics, with the result favoring ballistic over strictly hydrodynamic flow. Furthermore, it uncovers a new effect for ballistic electron flow, owing to the highly facetted shape of the hexagonal Fermi surface. We combine our extensive dataset with an analysis of the Boltzmann equation to characterize the non-local regime in PdCoO$_2$. More broadly, our results highlight the potential of broadband microwave spectroscopy to play a central role in investigating exotic transport regimes in the new generation of ultra-high conductivity materials.
\end{abstract}

\maketitle

There has been significant recent interest in unconventional electronic transport regimes in conductors in which a local relationship between electric current and electric field, as described by Ohm's law, breaks down. In the absence of frequent \gls{mr} scattering, non-local effects can develop: when \gls{mc} scattering is sufficiently frequent, electrons flow collectively as a viscous fluid; when neither form of scattering is frequent, electrons propagate ballistically. While viscous and ballistic effects are conceptually distinct, they can lead to similar experimental signatures \cite{Nazaryan2021}, making it dangerous to interpret experiments by comparing to theory rooted in one origin or the other. 
Additional impetus for comparing to theory incorporating both effects comes from the expectation that the hierarchy of scattering rates in most ultra-pure materials places them near the ballistic-to-viscous crossover, rather than at an extreme. 
Adding to the complexity, while early experiments and theories were based on materials with isotropic Fermi surfaces \cite{Molenkamp1994,DeJong1995}, a recent theoretical focus has been extending these ideas to anisotropic systems \cite{Cook2019,Varnavides2020,Cook2021,Qi2021}.
Many questions remain unresolved, however, and it is vital to extend the range of experimental tools with which to investigate these new transport regimes.

To date, the vast majority of work has been performed in the DC limit. The reason for this is the difficulty of performing broadband AC experiments in the GHz region of the spectrum; as we will discuss in more detail below, this is an area in which theory has been ahead of experiment, with predictions for the expected frequency dependence in different non-local regimes, but no actual data for comparison. In this paper, we explore this frequency-dependent physics using unique, bespoke experimental apparatus: a broadband bolometric microwave spectrometer \cite{Turner2004}. We report results from a number of ultra-pure metals over 1.5 decades of frequency, from 0.6 to \SI{20}{GHz}. Through measurements on Sr$_2$RuO$_4$ and Sn, we verify the predictions for the frequency dependence of the surface resistance for the Classical and Anomalous Skin Effects. Data from the ultra-pure delafossite PdCoO$_2$, in contrast, deviate from any previous prediction of frequency-dependent surface resistance. Analysis of such data required the construction of a more complete electrodynamic theory of metals than previously existed, and allows us to conclude that the PdCoO$_2$ data are due to the combination of a highly anisotropic Fermi surface and a contribution from momentum-conserving scattering. 

The paper is organized as follows: after introducing our broadband bolometric technique in \cref{sec:spectrometer}, we discuss the AC electrodynamics of high-purity metals in \cref{sec:skin_effects}, and results from Sr$_2$RuO$_4$ and Sn in \cref{sec:sr2ruo4_sn}. We then review what is known about non-local transport in PdCoO$_2$ in \cref{sec:pdcoo2_background}, before presenting and analyzing its broadband AC response in \cref{sec:pdcoo2_measurements}. We close the paper with \cref{sec:discussion_outlook}, discussing the implications of these results and the future role that AC measurements can play in the study of unconventional transport regimes in metals.

\section{Microwave spectroscopy}\label{sec:spectrometer}

Differentiating between non-local effects requires transport measurements on length scales comparable to the \gls{mr} and \gls{mc} mean free paths, $\lambda_{\text{mr}}$ and $\lambda_{\text{mc}}$. To date, this has been approached by studying how DC transport properties vary with the dimensions of micro-structured samples. Here we take a new approach by measuring the AC properties of bulk samples, using the skin effect to impose a \textit{tunable} length scale, the skin depth. In any metal, AC electromagnetic fields decay as they propagate. The result is that the electromagnetic fields---and resulting current density---are confined to a ``skin layer'' at the surface. 
A key advantage to AC measurements, in principle, is that this skin depth is frequency dependent: the DC approach necessitates additional fabrication each time sample dimensions are varied, but AC measurements offer the possibility of continuously varying the skin depth. Although conceptually simple, this requires broadband measurement over the microwave range of the spectrum.

Microwave frequencies fall between the range of conventional electrical and optical techniques, presenting a unique challenge for broadband spectroscopy---particularly for high-conductivity samples \cite{Turner2004,Huttema2006,Bonn2007}. While reflectivity measurements are commonly used at higher frequencies, at microwave frequencies two difficulties arise: (1) the free-space wavelength becomes comparable to or greater than typical sample dimensions, leading to diffraction, and (2) the reflectivity of high-conductivity metals becomes indistinguishable from unity. At lower frequencies, electrical leads are attached directly to the sample. However, the impedance of high-conductivity samples at microwave frequencies is on the order of \si{\milli\ohm}---much lower than typical values of either contact resistance or transmission line impedance. 

The most widely-adopted technique at microwave frequencies is cavity perturbation, in which a sample's electromagnetic properties are measured via its effect when introduced into a high-$Q$ resonator. Cavity perturbation achieves high sensitivity by amplifying the sample-radiation interaction through repeated reflections. However, the technique suffers the drawback of being at a fixed frequency. As will be discussed in the following section, this eliminates a key means of distinguishing between electrodynamic regimes and  necessitates the use of restrictive assumptions when interpreting data.

Here, we report the use of a bespoke technique giving us the capability of covering the range from 0.6 to \SI{20}{\giga\hertz} continuously \cite{Turner2004}. The basis of the technique is a bolometric measurement of the power absorbed by the sample from microwave electromagnetic fields, which is proportional to its surface resistance. Compared to other approaches to spectroscopic measurements at microwave frequencies \cite{Matsuda1994,Booth1994}, our technique is uniquely capable of resolving non-local effects in high-purity metals as a result of three distinct technical advantages: (1) We achieve high-sensitivity bolometric detection via a miniaturized thermal stage, careful thermometer selection, and home-built signal conditioning electronics. This provides us the sensitivity and dynamic range necessary to cover different skin effect regimes. (2) We use a custom-made transmission line to achieve a contactless, uniform, and well-defined microwave field configuration. This permits the separation of anisotropic components and gives us flexibility of measurement geometry---as we shall see, the geometric requirements are even more stringent for non-local versus local measurements. (3) We use an in-situ reference sample to measure the microwave field strength at the sample, eliminating frequency-dependent standing-wave effects and yielding an absolute calibration. Taken together, these technical advantages give us the unprecedented and unmatched ability to measure non-local electrodynamics in anisotropic, ultra-pure metals.

\section{Non-local electrodynamics of metals}\label{sec:skin_effects}

\begin{figure*}[!htbp]
  \centering
  \includegraphics{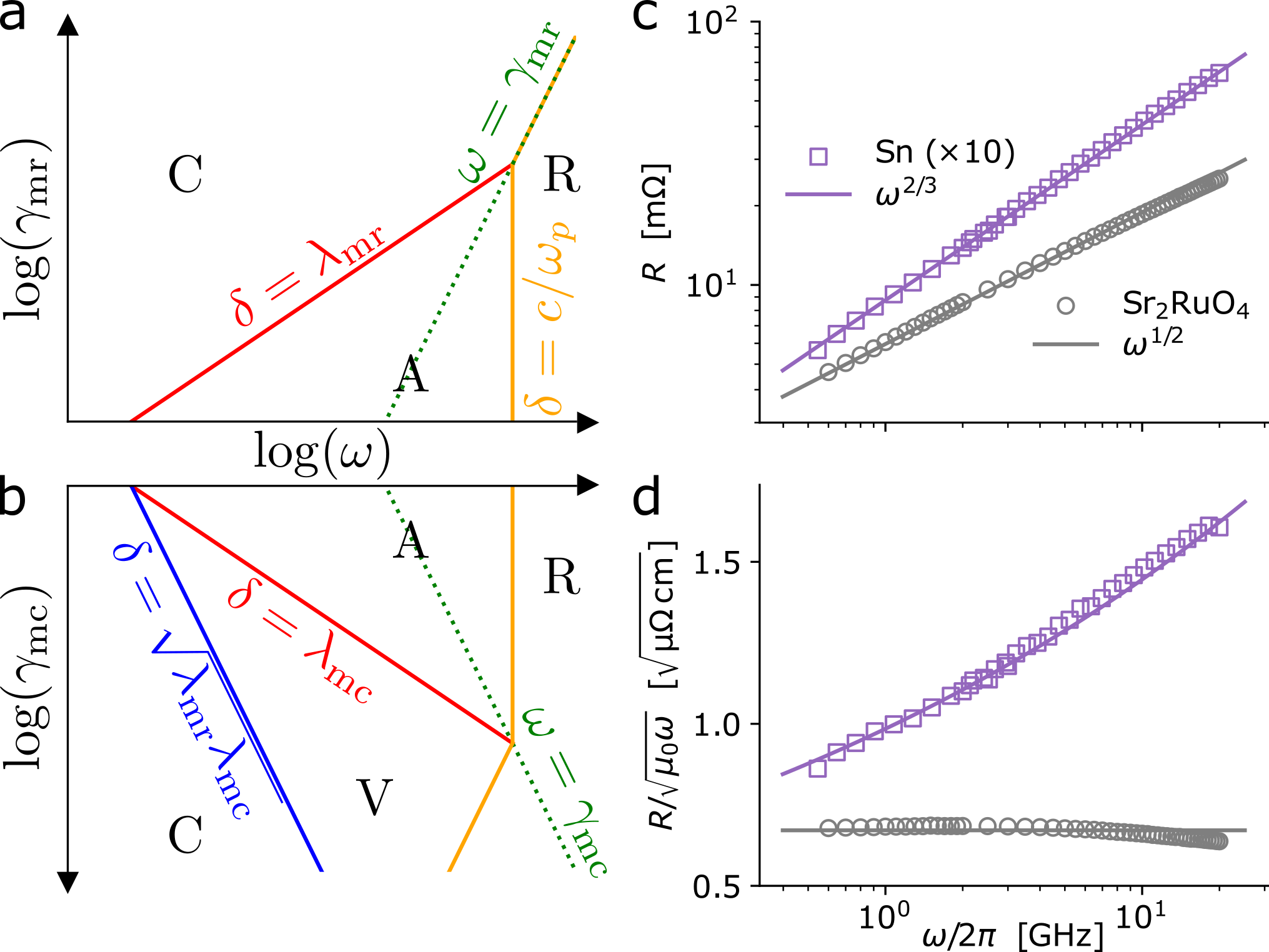}
  \caption[Test]{(a),(b) Predicted skin effect regimes according to conventional theory as a function of frequency $\omega$, momentum-relaxing scattering rate $\gamma_{\text{mr}}$, and momentum-conserving scattering rate $\gamma_{\text{mc}}$. (a) The transition from \underline{\textbf{C}}lassical to \underline{\textbf{A}}nomalous Skin Effect occurs when the momentum-relaxing mean free path $\lambda_{\text{mr}}$ is longer than the skin depth $\delta$. In this case, electrons propagate ballistically within the skin layer. (b) For sufficiently strong momentum-conserving scattering within the skin layer, a \underline{\textbf{V}}iscous Skin Effect is predicted to emerge. To our knowledge, this has not yet been experimentally observed. In both (a) and (b), the \underline{\textbf{R}}elaxation regime occurs at the highest frequencies, reflecting finite-frequency effects in the conductivity. (c),(d) Spectroscopic surface resistance measurements of Sr$_2$RuO$_4$ and Sn. (c) The power law behavior of $R(\omega)$ indicates that Sr$_2$RuO$_4$ exhibits the Classical Skin Effect while Sn exhibits the Anomalous Skin Effect. (d) Dividing $R$ by $\sqrt{\omega}$ provides a sensitive visual test for deviations from classical $R\sim\sqrt{\omega}$ behavior.}
  \label{fig:skin_effects}
\end{figure*}

The skin effect takes several forms depending on the nature of the electron dynamics within the skin layer, each characterized by unique frequency dependence and symmetry constraints of the AC surface resistance.
Mathematically, electromagnetic propagation can be conveniently described via a propagator $\mathcal{A}$ which depends on wavevector $q$ and frequency $\omega$:
\begin{equation}
    \mathcal{A}_{ij}(\bm{q},\omega)
    =\frac{\mu_{0}}{i\mu_{0}\omega\sigma_{ij}+\omega^{2}/c^{2}-q^{2}}.
\end{equation}
The poles of $\mathcal{A}$ are solutions to Maxwell's equations and give the dispersion relations $q(\omega)$ governing the propagation of electromagnetic modes within the metal. The effect of a metal's conductivity $\sigma$ is to increase the wavevector and to induce an imaginary component; the skin depth is given by $\delta=1/\text{Im}(q)$.

In most metals, the relationship between the AC electric field and the induced current density is local---the electric current at a given point in space depends only on the electric field at that same point. This is expressed by Ohm's law:
\begin{equation}
  J_{i}(\bm{r},\omega)=\sigma_{ij}(\omega)E_{j}(\bm{r},\omega) . 
\end{equation}
The reason Ohm's law is valid is that frequent scattering randomizes an electron's momentum on a scale much shorter than the variation of the decaying electric field.
This gives rise to the \gls{cse}, for which surface resistance is directly related to the local conductivity:
\begin{equation}
    R_{i}(\omega)
    =\text{Re}\sqrt{\frac{i\mu_{0}\omega}{\sigma_{ii}(\omega)}} .
\end{equation}
At low frequency, $R\sim\omega^{1/2}$. The symmetry of $R$ is that of the local conductivity tensor $\sigma_{ij}$, which is set by the crystal system.

However, Ohm's law cannot always be valid: in the absence of \gls{mr} collisions, an electron's momentum will depend on the entire history of the varying electric field along its trajectory. This can be resolved via a generalized, non-local version of Ohm's law using a wavevector-dependent conductivity:
\begin{equation}
  J_{i}(\bm{q},\omega)=\sigma_{ij}(\bm{q},\omega)E_{j}(\bm{q},\omega).
\end{equation}
In this case, $R$ is a wavevector-integrated function of the non-local conductivity, gaining an additional source of anisotropy via the direction of the wavevector $\bm{q}$:
\begin{equation}
    R_{i,\bm{\hat{q}}}(\omega)
    =\text{Re}{\int}dq\,\mathcal{A}_{ii}(\bm{q},\omega) .
\end{equation}

An established instance of non-local electrodynamics is the \gls{ase}, in which electrons propagate ballistically within the skin layer. The \gls{ase} is expected to occur when $\delta\ll\lambda_{\text{mr}}$ (as marked by the red line in \cref{fig:skin_effects}(a)), and is predicted to exhibit a surface resistance that becomes independent of the bulk conductivity and whose magnitude exceeds the \gls{cse} expectation \cite{Reuter1948,Sondheimer1954,Pippard1954}. The \gls{ase} has been observed experimentally at low temperature in a select number of high-purity elemental metals \cite{Pippard1947,Chambers1950,Chambers1952,Smith1959}, though always at one fixed frequency. It has also long been predicted that surface resistance follows $\omega^{2/3}$ behavior in the \gls{ase} \cite{Reuter1948,Sondheimer1954,Pippard1954}. To our knowledge, while this frequency dependence has been assumed in the interpretation of fixed-frequency measurements, it has never been measured directly.

Another instance of non-local electrodynamics---the \gls{vse}---is predicted to occur in an intermediate regime $\sqrt{\lambda_{\text{mr}}\lambda_{\text{mc}}}\ll\delta\ll \lambda_{mc}$ (as shown by the blue and red lines in \cref{fig:skin_effects}(b)) and is characterized by $\omega^{3/4}$ behavior \cite{Gurzhi1968}. To our knowledge, it has not yet been observed experimentally.

The upper frequency boundary for all of the above skin effects is determined by the onset of the relaxation regime, in which the conductivity leaves the zero-frequency limit. This boundary is given by $\delta=c/\omega_{p}$ where $c$ is the vacuum speed of light and $\omega_{p}$ is the plasma frequency, as marked by the orange line in \cref{fig:skin_effects}(a) and (b). For local electrodynamics, this aligns with the conventional Drude criterion $\omega=\gamma_{\text{mr}}$, shown by the dashed green line in \cref{fig:skin_effects}(a).

\section{Measurements on S\lowercase{r}$_2$R\lowercase{u}O$_4$ and S\lowercase{n}}\label{sec:sr2ruo4_sn}

To set the stage for discussing more complex behavior, we begin by showing spectroscopic measurements of single-crystal Sr$_2$RuO$_4$ and  polycrystalline Sn in \cref{fig:skin_effects}(c) and (d). Both measurements were taken using our unique, home-built spectrometer and slightly above the respective superconducting transition temperatures. 
The Sr$_2$RuO$_4$ data, taken at \SI{2.6}{\kelvin}, exhibits the $\omega^{1/2}$ behavior expected for the \gls{cse}. The magnitude of a \gls{cse} fit gives an in-plane resistivity of approximately \SI{50}{\nano\ohm\centi\metre}, consistent with that of the highest-quality samples as measured by conventional DC four-point electrical measurements \cite{Mackenzie1998,Barber2018}. As a consistency check, the measured resistivity value implies that the conditions $\delta<\lambda_{\text{mr}}$ and $\omega<\gamma_{\text{mr}}$ are met over our measurement frequencies, meaning that we should indeed expect to observe the \gls{cse} \footnote[1]{See Supplemental Material at [URL will be inserted by publisher]}. 
The Sn data, taken at \SI{5.0}{\kelvin}, exhibits the $\omega^{2/3}$ behavior expected for the \gls{ase}. The magnitude of an \gls{ase} fit gives a Fermi velocity of of \SI[per-mode=symbol]{2.2e6}{\metre\per\second} \footnotemark[1], within 20\% of the published value of \SI[per-mode=symbol]{1.9e6}{\metre\per\second} \cite{Ashcroft1976}.
Because the \gls{ase} surface resistance is insensitive to the \gls{mr} scattering rate, in this case we cannot directly use the information from our fit to compute the expected skin effect boundaries. However, for $\delta=\lambda_{\text{mr}}$ to be satisfied over our measurement range implies an upper bound on $\gamma_{\text{mr}}$, values below which are known to be achievable in high-purity Sn \cite{Pippard1947}\footnotemark[1].
To our knowledge, this is the first ever spectroscopic measurement of the \gls{ase}, confirming the frequency dependence predicted by theory \cite{Reuter1948,Sondheimer1954,Pippard1954}.
Since we will be concerned mainly with departures from the \gls{cse}, we introduce a plot that emphasizes deviation from this simple $\sqrt{\omega}$ behavior. In \cref{fig:skin_effects}(d), we plot the same data sets, but rescaled by $\sqrt{\omega}$. For Sr$_2$RuO$_4$, the result is quite flat, as expected for the CSE; for Sn, there is a clear upward trend.
These measurements highlight our ability to make precise measurements over a broad frequency range, confirming existing theory both in frequency dependence and in absolute magnitude.

\section{DC evidence for non-local transport in \texorpdfstring{P\lowercase{d}C\lowercase{o}O$_2$}{PdCoO2}}\label{sec:pdcoo2_background}

Having established the capability of our spectroscopic technique to differentiate between transport regimes, we now turn our attention to measurements of PdCoO$_2$, whose crystal structure is shown in \cref{fig:measurement}(a). This material is of considerable interest for AC conductivity measurements because DC studies indicate that it has extremely high purity \cite{Hicks2012,Sunko2020} and shows signatures of non-local transport. Initial analysis of the dependence of its resistivity on channel width suggested viscous corrections to ballistic transport \cite{Moll2016} but the data were analyzed using models previously employed to study non-ohmic transport in two-dimensional electron gases, in which a circular Fermi surface was assumed \cite{Molenkamp1994,DeJong1995}. The real Fermi surface of PdCoO$_2$ (\cref{fig:measurement}(b)) is nearly cylindrical, as expected for a nearly two-dimensional metal, but has a faceted, nearly hexagonal cross-section \cite{Noh2009,Hicks2012,Sunko2017}. Although this makes no difference to the local DC transport properties at high temperatures, when a non-local DC transport regime is entered, strong directional effects are seen. Specifically, devices of a given geometry yield different responses depending on their orientation relative to the Fermi surface facets \cite{Bachmann2019,McGuinness2021,Bachmann2022}. To date, the combined role of directional and viscous effects in PdCoO$_2$ has not been analyzed. This background motivated us to investigate whether the AC response of PdCoO$_2$ might be even richer than that predicted for the usual \gls{ase}, and  provide new insight into the physics of PdCoO$_2$.  

\section{Measurements on \texorpdfstring{P\lowercase{d}C\lowercase{o}O$_2$}{PdCoO2}}\label{sec:pdcoo2_measurements}

\begin{figure*}[!htbp]
  \includegraphics{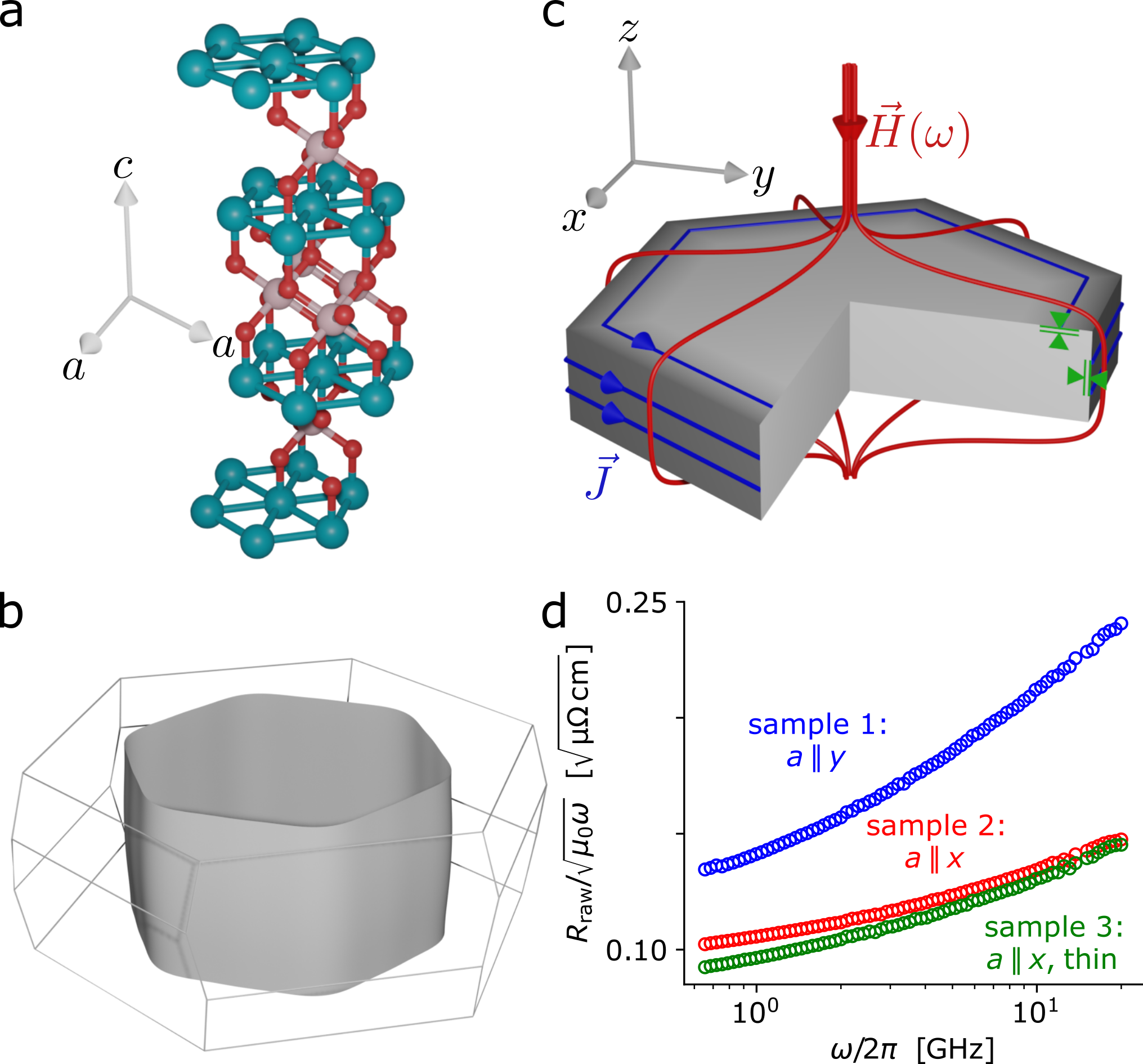}
  \caption{(a) PdCoO\textsubscript{2} crystal structure, belonging to the trigonal crystal system. (b) PdCoO\textsubscript{2}'s hexagonally faceted Fermi surface, as determined via ARPES and quantum oscillations \cite{Hicks2012}. (c) Measurement geometry. Samples, which grow naturally as thin platelets, were cut into hexagons to reflect the 6-fold rotational symmetry of the crystal structure. A microwave-frequency magnetic field is applied along $z$. This induces eddy currents (blue) acting to screen the magnetic field (red) from the interior of the sample. The resulting field strength at the sample's surface (grey shading) is highest on the side faces but also becomes appreciable toward the edges of the top and bottom faces. Because the skin depth is much smaller than the sample dimensions, this gives rise to two separate ``skin regions'' (indicated by green arrows). In both, current flows in the (001) plane and the wavevector is perpendicular to the surface. The measured signal contains a mixture of the two skin regions that depends on the sample's aspect ratio. (d) Raw data. We measured three samples, with two different cut orientations and two different aspect ratios. When the sample is cut with $a\parallel x$, $\bm{q}\parallel\braket{110}$ (``30") on the side faces; when the sample is cut with $a\parallel y$, $\bm{q}\parallel\braket{100}$ (``0") on the side faces. Varying the aspect ratio enables isolating the contribution from the top and bottom faces, where $\bm{q}\parallel c$. All three measurements differ---by symmetry, this can only occur for non-local electrodynamics.}
  \label{fig:measurement}
\end{figure*}

Our measurements were performed at \SI{2}{\kelvin} so as to match previous non-local DC transport studies \cite{Moll2016,Bachmann2019,McGuinness2021,Bachmann2022}. At this temperature, published values imply that the skin depth in our frequency range satisfies $c/\omega_{p}\ll\delta\ll\lambda_{\text{mr}}$. This is an ideal regime for our investigation; for a single scattering rate, we would expect to observe the \gls{ase}; for sufficient \gls{mc} scattering, we would be in a position to observe the \gls{vse}.
The faceted Fermi surface of \cref{fig:measurement}(b) immediately suggests three extremal wavevector directions for which to perform measurements: $\bm{q}\parallel\braket{100}$ (``0''), $\bm{q}\parallel\braket{110}$ (``30''), and $\bm{q}\parallel\braket{001}$ (``c''). Samples of PdCoO\textsubscript{2} grow as platelets with in-plane dimensions around \SI{1}{\milli\metre} and typical thicknesses of tens of microns. To reflect the underlying symmetry of the crystal structure, we cut samples to have hexagonal cross section, with lateral dimensions of about \SI{0.5}{\milli\metre} (\cref{fig:measurement}(c)). We applied a spatially-uniform, microwave-frequency magnetic field parallel to the $c$ axis, inducing eddy currents which flow in loops perpendicular to the magnetic field. 
This establishes two distinct skin regions: one for currents on the two large hexagonal faces, with wavevector along the $c$ axis, and the other for currents on the six small rectangular faces, with wavevector perpendicular to the $c$ axis. The hexagonal cross section of the sample ensures that the wavevectors for each of the six rectangular faces are along symmetry equivalent directions. A measurement of a given sample thus contains a mixture of two distinct surface resistance components---with in- and out-of-plane wavevector directions---with weights depending on the sample's aspect ratio. 
Our raw measurements are shown in (\cref{fig:measurement}(d)). Sample 1 was cut with $a\parallel y$ so that the in-plane wavevector had $\bm{q}\parallel0$; sample 2 was cut with $a\parallel x$ so that the in-plane wavevector had $\bm{q}\parallel30$. Sample 3 was cut with the same orientation as sample 2, but was thinner, increasing the relative weight of the contribution from $\bm{q}\parallel c$. With measurements of these three samples, we obtained sufficient information to disentangle the surface resistance components for the three wavevector directions of interest. However, even without disentangling the components, the fact that the three raw measurements differ provides immediate evidence of non-local electrodynamics.  
For all three components, current flows in the plane perpendicular to the $c$ axis; for local electrodynamics, the triangular in-plane lattice dictates that the conductivity tensor has no in-plane anisotropy and so all three components---and therefore the three raw measurements which mix them---would be identical. To proceed further, we used electromagnetic simulations to disentangle the surface resistance components for the three wavevector directions from the raw measurements of the three samples \footnotemark[1].

\begin{figure*}[!htbp]
  \includegraphics{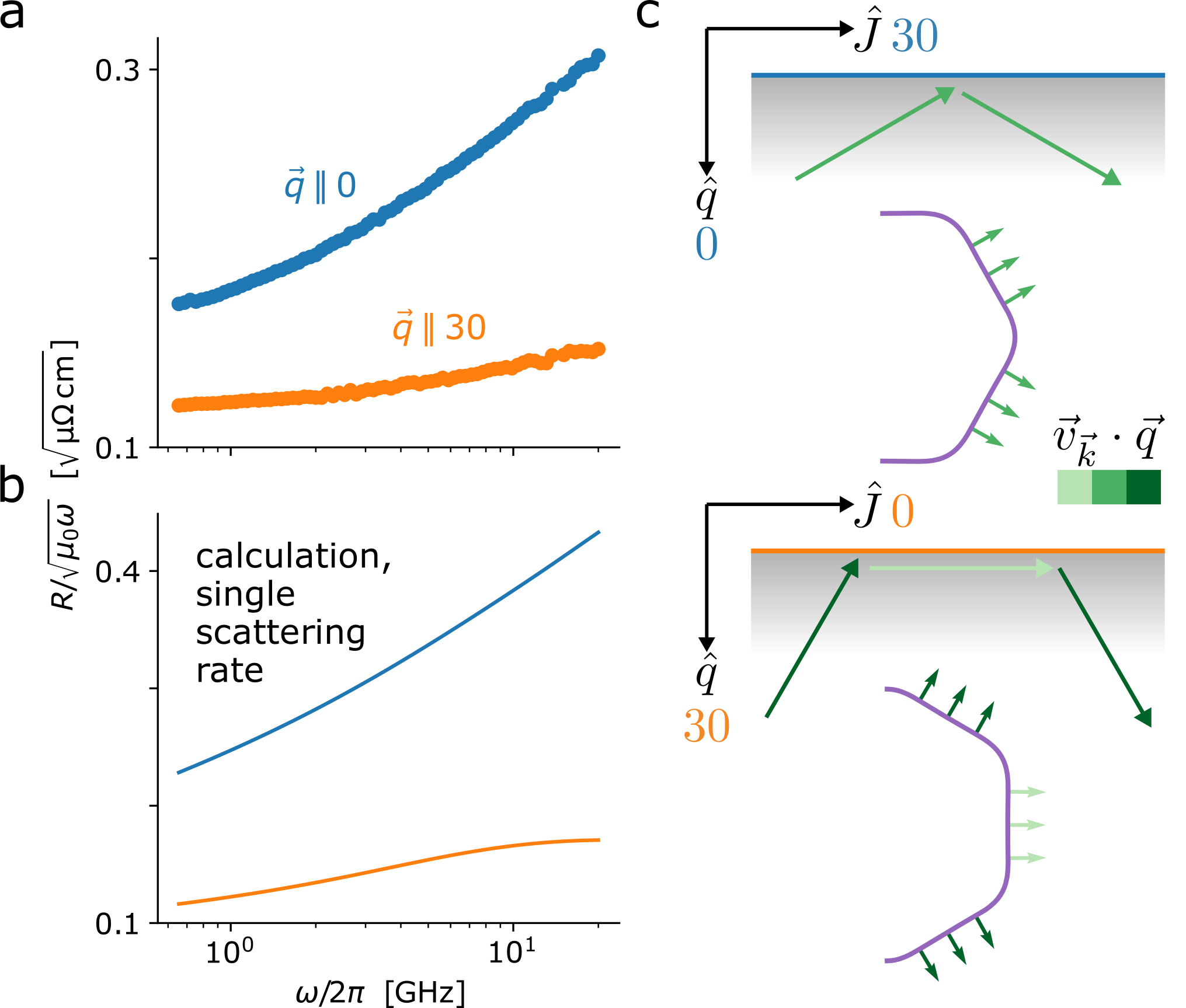}
  \caption{(a) Surface resistance data with in-plane wavevector, obtained by subtracting the $\bm{q}\parallel c$ component from the raw measurements. The different power-law behavior originates from predominantly ballistic propagation within the skin layer coupled with a strongly-facetted Fermi surface, as illustrated in (b) and (c). (b) Calculated surface resistance based on the experimentally-determined Fermi surface, capturing the different power-law behavior of the two orientations. The calculation is for a single relaxation rate, i.e. $\gamma_{\text{mc}}=\gamma_{\text{mr}}$, and uses published values with no free parameters \cite{Note1}.  (c) Illustration of ballistic propagation within the skin layer. Top: There are two main directions of electron propagation, both at an angle to the surface. As frequency increases, the skin depth becomes shallower. The electrons spend an increasingly smaller fraction of a mean free path inside the skin layer, leading to an increasing surface resistance—the Anomalous Skin Effect. Bottom: There are three main directions of electron propagation. Electrons propagating parallel to the sample's surface spend the entirety of a mean free path within the skin layer, regardless of how shallow the skin depth becomes. Often, this is a negligible fraction of the Fermi surface; in PdCoO\textsubscript{2}, approximately a third of the Fermi surface propagates parallel to the sample's surface. The Anomalous Skin E
  Effect is largely suppressed even when the mean free path is much larger than the skin depth.}
  \label{fig:in_plane}
\end{figure*}

The measured surface resistance for the two in-plane wavevectors is shown in \cref{fig:in_plane}(a). Surprisingly, the two orientations exhibit distinct power-law behaviors. 
A useful property of the viscosity tensor in a plane with six-fold rotational symmetry provides an elegant avenue for differentiating ballistic and viscous effects: in this setting, as is the case in PdCoO\textsubscript{2}, the in-plane viscosity tensor is isotropic \cite{Frisch1986}. This implies the qualitative insight that the anisotropy in the surface resistance at \SI{2}{\kelvin} for the two orientations cannot be due to purely viscous effects. 

With this in mind, we turn to the possible ballistic origin of this effect. The standard theory of the \gls{ase} (i.e., ballistic propagation within the skin layer)---Pippard theory \cite{Pippard1954}---predicts that any orientation should exhibit $R\sim\omega^{2/3}$, with only the pre-factor being orientation-dependent. Our data is at odds with Pippard theory: while one orientation exhibits behavior close to $\omega^{2/3}$, the other exhibits only a weak deviation from classical behavior. This breakdown of Pippard theory is all the more surprising because---aside from its ubiquity---Pippard theory has previously demonstrated success in describing the behavior of anisotropic Fermi surfaces. Famously, Pippard performed the first ever experimental determination of a Fermi surface by applying his eponymous theory to measurements of the ASE in Cu, revealing deviation from a spherical Fermi surface \cite{Pippard1957}. Nonetheless, Pippard theory treats Fermi surface geometry phenomenologically, and was originally justified by agreement with more rigorous treatments based on solving the Boltzmann equation for spherical \cite{Reuter1948} and spheroidal \cite{Sondheimer1954} Fermi surfaces.

Clearly, we need to go further. To model our results, we solved the Boltzmann equation using a realistic, three-dimensional parameterization of the Fermi surface of PdCoO\textsubscript{2} based on ARPES and quantum oscillation measurements \cite{Hicks2012}. As seen in \cref{fig:in_plane}(a) and (b), our calculations qualitatively reproduce the difference in power-law behavior between the two orientations. 
An intuitive explanation for the difference in power laws comes from applying Pippard's ``ineffectiveness concept'' \cite{Pippard1947,Pippard1954} to the Fermi surface of PdCoO\textsubscript{2} (\cref{fig:in_plane}(c)):
only those electrons that spend an entirety of a mean free path within the skin layer are effective at screening electromagnetic fields. As the ratio of mean free path to skin depth increases, electrons spend an increasingly small fraction of a mean free path within the skin layer, so the surface resistance becomes increasingly larger than the classical value. Mathematically, this can be described as an effective mean free path for each state $\bm{k}$, which represents that state's contribution to the overall conductivity:
\begin{equation}
    \lambda_{\bm{k}}^{\text{eff}}
    =\frac
        {(\bm{\hat{v}}_{\bm{k}}\cdot\bm{\hat{E}})^{2}\,\lambda_{0}}
        {1+i(\bm{\hat{v}}_{\bm{k}}\cdot\bm{\hat{q}})\,q\lambda_{0}}
\end{equation}
where $\lambda_{0}$ is the bare mean free path and $\bm{\hat{v}}_{\bm{k}}$ is the unit velocity vector. (Because the present discussion is focused on purely ballistic effects, here we have taken $\lambda_{mr}=\lambda_{mc}=\lambda_{0}$).
In PdCoO\textsubscript{2} in the lower orientation in \cref{fig:in_plane}(c), a third of electrons propagate nearly parallel to the sample's surface, such that $\bm{\hat{v}}_{\bm{k}}\cdot\bm{\hat{q}}=0$. These electrons remain effective at screening regardless of the ratio of mean free path to skin depth, largely suppressing the increase in surface resistance. Indeed, there have been several theoretical works predicting extreme Fermi surface geometries for which Pippard theory would break down \cite{Glasser1968,Kaganov1994,Zimbovskaya2006}. To our knowledge, the present results represent the first experimental confirmation of these ideas. 

\begin{figure}[!htbp]
  \includegraphics{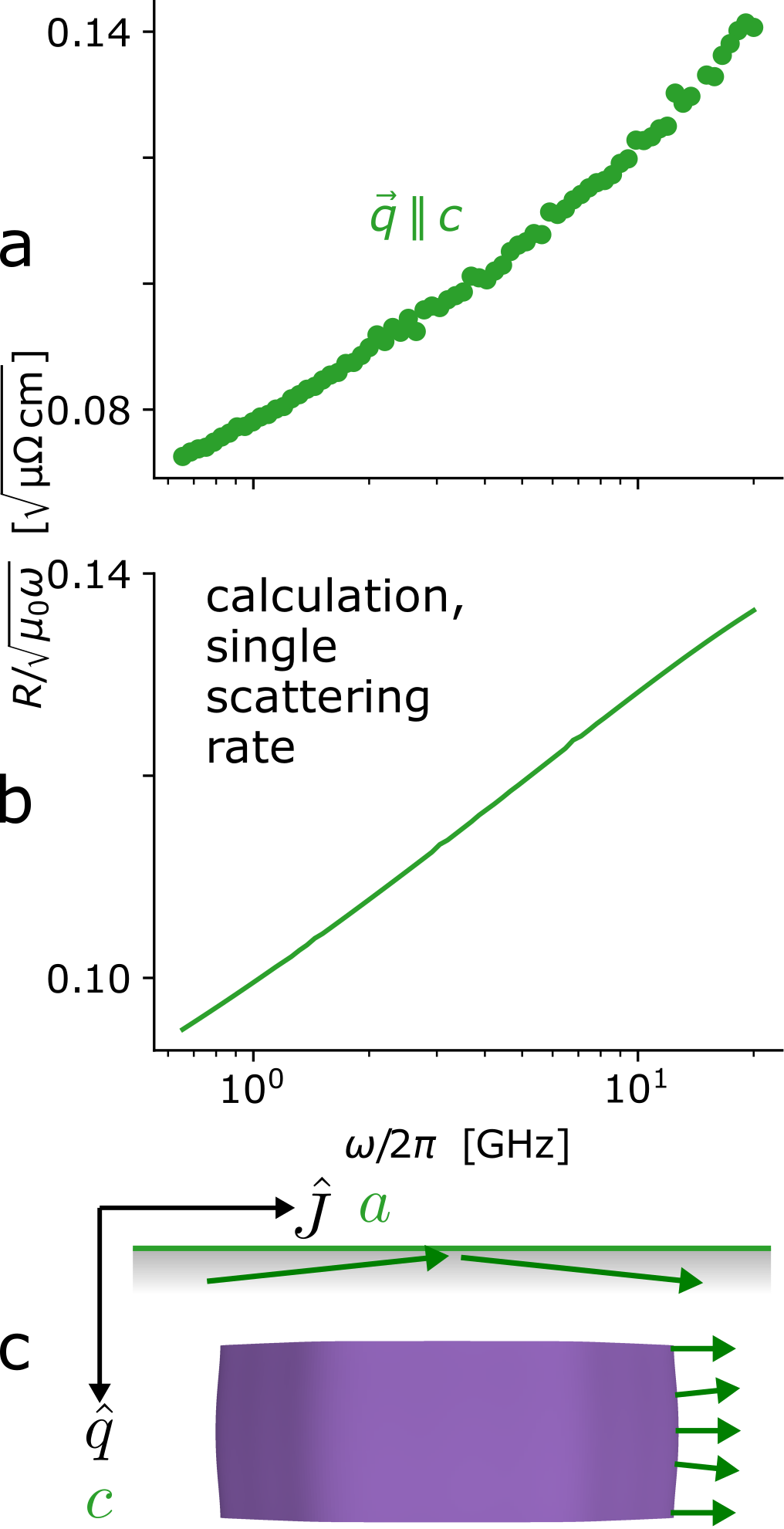}
  \caption{ (a) Surface resistance data with out-of-plane wavevector, obtained by comparing raw measurements from samples of different thicknesses. The data shows upward deviation from classical $R\sim\sqrt{\omega}$ behavior. Because of the high ratio of mean free path to skin depth reached toward the upper end of our frequency range, even the small amount of Fermi surface warping along $k_{z}$ is sufficient to produce non-local effects. (b) Calculated surface resistance based on the experimentally-determined Fermi surface. The calculation is for a single relaxation rate, i.e. $\gamma_{\text{mc}}=\gamma_{\text{mr}}$, and uses published values with no free parameters \cite{Note1}. (c) Illustration of ballistic propagation within the skin layer.}
  \label{fig:out_of_plane}
\end{figure}

The measured surface resistance for wavevector along the $c$ axis, shown in \cref{fig:out_of_plane}(a), exhibits a clear deviation from classical $R\sim\sqrt{\omega}$ behavior. The observation of non-local electrodynamics in this orientation is surprising: as per the ineffectiveness concept, this means that electrons must be able to propagate in and out of the skin layer within a single mean free path. However, the nearly cylindrical geometry of the Fermi surface means that in this orientation, electrons propagate at a shallow angle relative to the skin layer (\cref{fig:out_of_plane}(c)). PdCoO\textsubscript{2} is often described as electronically two-dimensional or quasi-two-dimensional, as supported by its low-temperature resistivity anisotropy of $\rho_{c}/\rho_{a}\approx1000$ \cite{Hicks2012}. However, in a perfectly two-dimensional material, the ASE would be completely suppressed. Its presence here is a result of the subtle warping of the Fermi surface along $k_{z}$, as was resolved by quantum oscillations, and highlights the limitations of a purely two-dimensional description of transport properties in PdCoO\textsubscript{2}. This observation has implications for DC transport measurements. To date, studies have focused on how resistivity varies when restricting in-plane dimensions; these results imply that size effects will also be present while varying thickness along the $c$ axis. We estimate that the maximum skin depth over our measured frequency range is on the order of \SI{100}{\nano\metre}. This implies that size effects are likely to be especially important to thin films, which have been the subject of recent growth efforts \cite{Harada2018,Brahlek2019,Sun2019}.

\begin{figure*}[!htbp]
  \centering
  \includegraphics{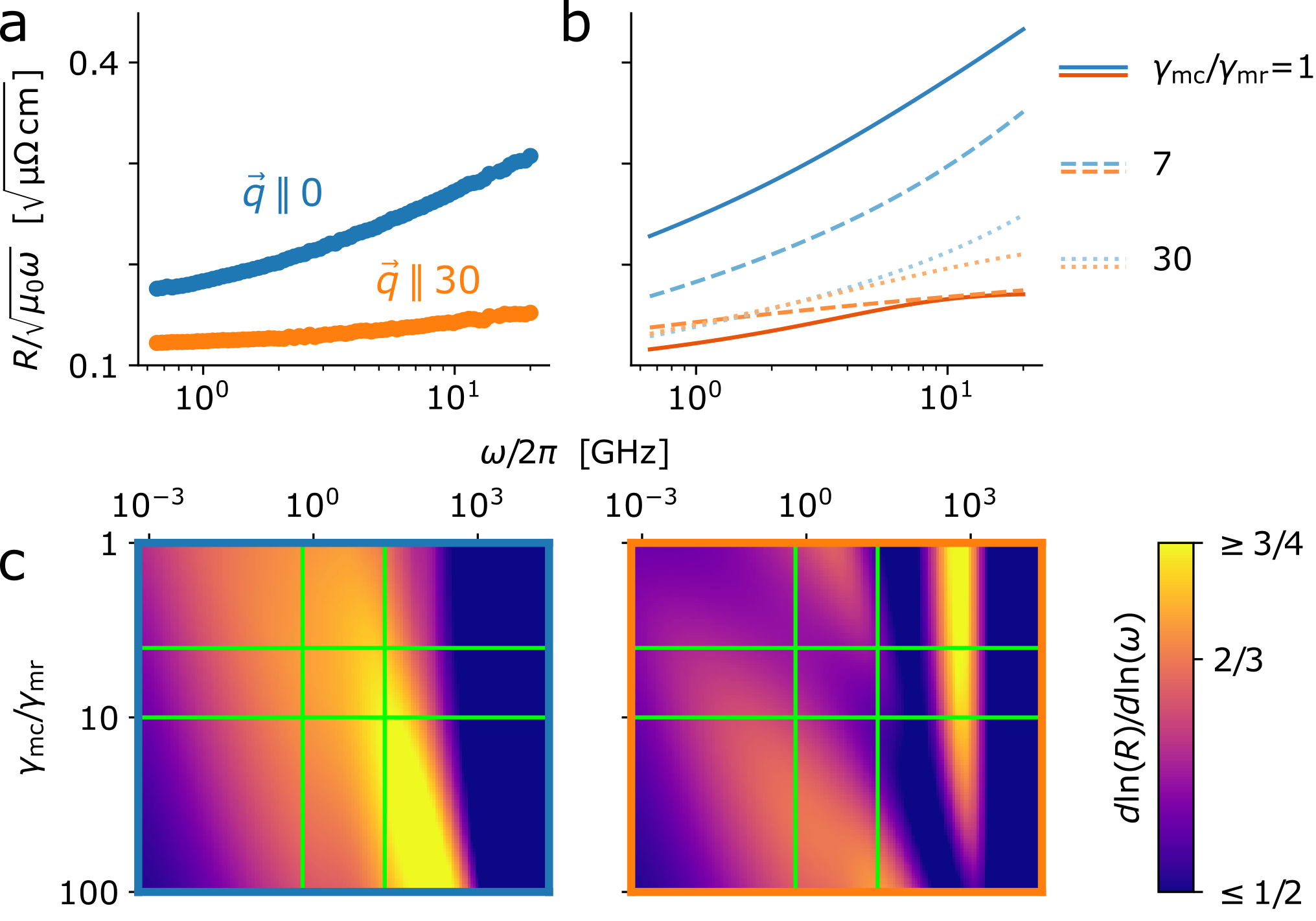}
  \caption{Determination of momentum conserving scattering rate. (a) Experimental data for in-plane wavevector, on the same scale as (b). (b) Calculated surface resistance for various values of $\gamma_{\text{mc}}$, with $\gamma_{\text{mr}}$ and all other parameters fixed by their literature values \cite{Mackenzie2017}. For $\gamma_{\text{mc}}/\gamma_{\text{mr}}=1$, corresponding to the conventional relaxation time approximation, the calculated anisotropy between the two surface resistances is too large. For $\gamma_{\text{mc}}/\gamma_{\text{mr}}=30$, the anisotropy is too small; the two surface resistances are almost equal, as expected deep in the viscous regime. For our best fit value of $\gamma_{\text{mc}}/\gamma_{\text{mr}}=7$, a quantitative match is achieved. (c),(d) Calculated surface resistance for (c) $R_{a,\bm{\hat{q}}\parallel0}$and (d)$R_{a,\bm{\hat{q}}\parallel30}$ over a broader range of frequency and momentum-conserving scattering rate. Vertical green lines show our measurement range while horizontal green lines show our finding that $\gamma_{\text{mc}}/\gamma_{\text{mr}}$ is between 4 and 10. The color plot shows the power-law behavior of the surface resistance, showing its orientation-dependent modification as well as where our measurements fit relative to other skin effect regimes.}
  \label{fig:mc}
\end{figure*} 

Having identified a ballistic origin for the main features of the three measured surface resistances, we now turn to placing a quantitative bound on the rate of \gls{mc} scattering. To do so, we developed a model that allows for arbitrary rates of \gls{mr} and \gls{mc} scattering. Combined with using a realistic, three-dimensional Fermi surface parameterization, these ingredients allow our model to encompass both directional and viscous effects. In \cref{fig:mc}, we compare the data for in-plane wavevector to calculations in which we no longer make our earlier assumption of a single relaxation rate. The only free parameter of these calculations is the \gls{mc} scattering rate; the \gls{mr} scattering rate and other material parameters are taken from the literature. As seen in \cref{fig:mc}(b), the higher the value of $\gamma_{\text{mc}}$, the lower the anisotropy between the two surface resistances. This is in accord with our earlier insight that deep in the viscous regime, symmetry dictates that there should be no anisotropy. The calculations for $\gamma_{\text{mc}}/\gamma_{\text{mr}}=1$, shown previously, produce too much anisotropy. For $\gamma_{\text{mc}}/\gamma_{\text{mr}}=30$, the isotropic, viscous limit is nearly reached. For intermediate values, a quantitative match to the data is achieved. From this comparison, we determined that $\gamma_{\text{mc}}=(7\pm3)\gamma_{\text{mr}}$ at \SI{2}{\kelvin}. In \cref{fig:mc}(c) and (d), we show the calculated power-law behavior of the surface resistance over a wider range of frequency, showing the orientation-dependent modification of power-law behavior as well as the relation of our measurements to other skin effect regimes. 

\section{Discussion \& outlook}\label{sec:discussion_outlook}

The fact that our experiments are best matched by $\gamma_{\text{mc}}\approx7\gamma_{\text{mr}}$ at \SI{2}{\kelvin} is striking for several reasons. Firstly, it agrees within experimental error with the deduction made from the width dependence of DC transport in  Ref. \cite{Moll2016}. That DC measurement, however, was performed on a device of unknown orientation relative to the Fermi surface facets, and analyzed using the assumption of a circular Fermi surface cross-section, so the agreement may be fortuitous. We believe that the deduction we report here is on a firm footing because orientation and Fermi surface faceting have been taken carefully into account. 

Secondly, the observation of a  contribution from \gls{mc} scattering raises the question of its source at this low temperature of \SI{2}{\kelvin}. Given that $T{\ll}T_F$, it is expected that direct electron-electron scattering is negligible, even taking into account an enhanced cross-section due to Fermi surface faceting \cite{Cook2019}. A recent work has proposed a phonon-mediated electron-electron interaction as a source of sufficient \gls{mc} scattering \cite{Wang2021}, but such a mechanism---indeed any mechanism invoking electron-electron scattering---would be predominantly \gls{mr} in PdCoO\textsubscript{2} due to Umklapp processes. 
While electron-phonon processes can in principle be a source of MC scattering \cite{Moll2016,Levchenko2020,Huang2021}, we estimate that it is insufficient to give rise to our experimentally-determined value of $\gamma_{\text{mc}}$ \footnotemark[1], and there is no sign in DC measurements at temperatures below \SI{10}{\kelvin} of the strong temperature dependence predicted in such scenarios. Instead, the low temperature of our measurements is suggestive of impurity scattering as the source of the observed $\gamma_{\text{mc}}$ \cite{Cook2019}. 
While recent theoretical work has begun to explore the connection between electron-impurity scattering and \gls{mc} scattering \cite{Hui2020}, these experimental observations motivate further theoretical work to establish whether this explanation is realistic in the present setting.

The results presented in this paper have future implications both for understanding the physics of PdCoO$_2$ and for the broader field of non-local effects in ultra-clean metals. 
For the first time, we have combined experiment and analysis to investigate the combined effects of momentum-conserving scattering and Fermi surface anisotropy---including both the effect of in-plane faceting and out-of-plane warping.
In PdCoO$_2$ and other delafossites, the existence of Boltzmann transport code capable of working with realistic Fermi surface parameterizations will enable further in-depth investigation of the balance between MC and MR scattering in both the AC and DC regimes. By extending the range of frequencies over which there is information to \SI{20}{\giga\hertz} and showing the unconventional frequency dependencies revealed in our measurements, we have provided a stringent test for this analysis model. Its ability to reproduce those frequency dependences gives more confidence in its correctness (and in the assumptions on which it is based) than could ever be achieved from DC data alone.
While Refs. \cite{Bachmann2019,McGuinness2021,Bachmann2022} showed that Fermi surface geometry and orientation can have an effect on non-local transport, here we have shown that they can go as far as modifying the power-law relationship between transport properties and an imposed, extrinsic lengthscale. Both in the skin effect and the size effect in narrow channels, these power laws are often taken as an unambiguous means to differentiate between diffusive, viscous, and ballistic transport. Our finding highlights that Fermi surface geometry and orientation can act as confounding factors when differentiating between transport regimes.
Here we observed that predominantly ballistic electron dynamics lead to behaviour close to the expected $\omega^{2/3}$ dependence in one orientation, but in another orientation lead to behaviour much closer to the $\omega^{1/2}$ dependence typically associated with diffusive dynamics.
This indicates that in DC experiments, orientation-dependent differences from the canonical quadratic relationship between conductance and width in the ballistic regime are likely to be seen in width-restricted channel experiments performed in carefully oriented channels.

The success of our AC Boltzmann transport theory in turn enables extension of this kind of study to other ultra-high-conductivity materials with known Fermi surfaces, and prediction of the non-local AC transport effects likely to be found in them. Experimentally, we have demonstrated the utility of broadband microwave spectroscopy in the investigation of non-local electrodynamics. While the foundational measurements of the ASE were performed at fixed frequencies \cite{Pippard1947,Pippard1957,Chambers1950,Chambers1952}, in the present work, continuous-frequency measurements were critical to the interpretation of our results: in particular, in identifying a ballistic- rather than viscous-dominated regime, and in revealing the predicted breakdown of Pippard theory as a result of a strongly-faceted Fermi surface. These effects are also technologically relevant, as future applications of ultra-high-conductivity materials are likely to operate at GHz frequencies. The ASE is known to limit the conductance of interconnects in integrated circuits when operated at these high frequencies \cite{Sarvari2008}. The present results demonstrate that conductance can be improved by aligning interconnects along a direction for which the ASE is suppressed. Finally, to our knowledge, our findings represent the first experimental observation of the ASE outside of elemental metals, suggesting experimental opportunities among new-generation ultra-high-conductivity materials. The interplay of frequency, scattering rates, carrier density, and Fermi surface geometry gives rise to a rich phenomenological landscape for non-local electrodynamics, particularly in the microwave and terahertz range---which, to date, remains largely unexplored \cite{Casimir1967,Forcella2014,Valentinis2021,Valentinis2021-2,Matus2022,Valentinis}.

\begin{acknowledgments}
We are grateful to Alex Levchenko for helpful discussions. 
G.B., T.W.B., J.D., M.H., and D.A.B. acknowledge support from the Max Planck-UBC-UTokyo Center for Quantum Materials and the Canada First Research Excellence Fund, Quantum Materials and Future Technologies Program, as well as the Natural Sciences and Engineering Research Council of Canada (RGPIN-2018-04280).
D. V. acknowledges partial support by the Swiss National Science Foundation (SNSF) through the SNSF Early Postdoc. Mobility Grant P2GEP2\_18145.
D.V. and J.S. acknowledge support by the European Commission's Horizon 2020 RISE program Hydrotronics (Grant No. 873028).
P.S. acknowledges the support of the Narodowe Centrum Nauki (NCN) Sonata Bis grant 2019/34/E/ST3/00405 and the Nederlandse Organisatie voor Wetenschappelijk Onderzoek (NWO) Klein grant via NWA route 2.
Y.M. acknowledges the support by JSPS KAKENHI JP22H01168.
The work in Dresden was in part supported by the Deutsche Forschungsgemeinschaft (DFG) through the W\"{u}rzburg-Dresden Cluster of Excellence on Complexity and Topology in Quantum Matter -- ct.qmat (EXC 2147, project ID 39085490) and the Leibniz Prize programme.
\end{acknowledgments}

\newpage
\bibliography{main.bib}

\begin{thebibliography}{52}%
\makeatletter
\providecommand \@ifxundefined [1]{%
 \@ifx{#1\undefined}
}%
\providecommand \@ifnum [1]{%
 \ifnum #1\expandafter \@firstoftwo
 \else \expandafter \@secondoftwo
 \fi
}%
\providecommand \@ifx [1]{%
 \ifx #1\expandafter \@firstoftwo
 \else \expandafter \@secondoftwo
 \fi
}%
\providecommand \natexlab [1]{#1}%
\providecommand \enquote  [1]{``#1''}%
\providecommand \bibnamefont  [1]{#1}%
\providecommand \bibfnamefont [1]{#1}%
\providecommand \citenamefont [1]{#1}%
\providecommand \href@noop [0]{\@secondoftwo}%
\providecommand \href [0]{\begingroup \@sanitize@url \@href}%
\providecommand \@href[1]{\@@startlink{#1}\@@href}%
\providecommand \@@href[1]{\endgroup#1\@@endlink}%
\providecommand \@sanitize@url [0]{\catcode `\\12\catcode `\$12\catcode
  `\&12\catcode `\#12\catcode `\^12\catcode `\_12\catcode `\%12\relax}%
\providecommand \@@startlink[1]{}%
\providecommand \@@endlink[0]{}%
\providecommand \url  [0]{\begingroup\@sanitize@url \@url }%
\providecommand \@url [1]{\endgroup\@href {#1}{\urlprefix }}%
\providecommand \urlprefix  [0]{URL }%
\providecommand \Eprint [0]{\href }%
\providecommand \doibase [0]{https://doi.org/}%
\providecommand \selectlanguage [0]{\@gobble}%
\providecommand \bibinfo  [0]{\@secondoftwo}%
\providecommand \bibfield  [0]{\@secondoftwo}%
\providecommand \translation [1]{[#1]}%
\providecommand \BibitemOpen [0]{}%
\providecommand \bibitemStop [0]{}%
\providecommand \bibitemNoStop [0]{.\EOS\space}%
\providecommand \EOS [0]{\spacefactor3000\relax}%
\providecommand \BibitemShut  [1]{\csname bibitem#1\endcsname}%
\let\auto@bib@innerbib\@empty
%</preamble>
\bibitem [{\citenamefont {Nazaryan}\ and\ \citenamefont
  {Levitov}(2021)}]{Nazaryan2021}%
  \BibitemOpen
  \bibfield  {author} {\bibinfo {author} {\bibfnamefont {K.~G.}\ \bibnamefont
  {Nazaryan}}\ and\ \bibinfo {author} {\bibfnamefont {L.}~\bibnamefont
  {Levitov}},\ }\bibfield  {title} {\bibinfo {title} {{Robustness of vorticity
  in electron fluids}},\ }\Eprint {https://arxiv.org/abs/2111.09878}
  {arXiv:2111.09878}  (\bibinfo {year} {2021})\BibitemShut {NoStop}%
\bibitem [{\citenamefont {Molenkamp}\ and\ \citenamefont
  {de~Jong}(1994)}]{Molenkamp1994}%
  \BibitemOpen
  \bibfield  {author} {\bibinfo {author} {\bibfnamefont {L.~W.}\ \bibnamefont
  {Molenkamp}}\ and\ \bibinfo {author} {\bibfnamefont {M.~J.~M.}\ \bibnamefont
  {de~Jong}},\ }\bibfield  {title} {\bibinfo {title}
  {{Electron-electron-scattering-induced size effects in a two-dimensional
  wire}},\ }\href {https://doi.org/10.1103/PhysRevB.49.5038} {\bibfield
  {journal} {\bibinfo  {journal} {Physical Review B}\ }\textbf {\bibinfo
  {volume} {49}},\ \bibinfo {pages} {5038} (\bibinfo {year}
  {1994})}\BibitemShut {NoStop}%
\bibitem [{\citenamefont {{De Jong}}\ and\ \citenamefont
  {Molenkamp}(1995)}]{DeJong1995}%
  \BibitemOpen
  \bibfield  {author} {\bibinfo {author} {\bibfnamefont {M.~J.~M.}\
  \bibnamefont {{De Jong}}}\ and\ \bibinfo {author} {\bibfnamefont {L.~W.}\
  \bibnamefont {Molenkamp}},\ }\bibfield  {title} {\bibinfo {title}
  {{Hydrodynamic electron flow in high-mobility wires}},\ }\href
  {https://doi.org/10.1103/PhysRevB.51.13389} {\bibfield  {journal} {\bibinfo
  {journal} {Physical Review B}\ }\textbf {\bibinfo {volume} {51}},\ \bibinfo
  {pages} {13389} (\bibinfo {year} {1995})}\BibitemShut {NoStop}%
\bibitem [{\citenamefont {Cook}\ and\ \citenamefont {Lucas}(2019)}]{Cook2019}%
  \BibitemOpen
  \bibfield  {author} {\bibinfo {author} {\bibfnamefont {C.~Q.}\ \bibnamefont
  {Cook}}\ and\ \bibinfo {author} {\bibfnamefont {A.}~\bibnamefont {Lucas}},\
  }\bibfield  {title} {\bibinfo {title} {{Electron hydrodynamics with a
  polygonal Fermi surface}},\ }\href
  {https://doi.org/10.1103/PhysRevB.99.235148} {\bibfield  {journal} {\bibinfo
  {journal} {Physical Review B}\ }\textbf {\bibinfo {volume} {99}},\ \bibinfo
  {pages} {235148} (\bibinfo {year} {2019})}\BibitemShut {NoStop}%
\bibitem [{\citenamefont {Varnavides}\ \emph {et~al.}(2020)\citenamefont
  {Varnavides}, \citenamefont {Jermyn}, \citenamefont {Anikeeva}, \citenamefont
  {Felser},\ and\ \citenamefont {Narang}}]{Varnavides2020}%
  \BibitemOpen
  \bibfield  {author} {\bibinfo {author} {\bibfnamefont {G.}~\bibnamefont
  {Varnavides}}, \bibinfo {author} {\bibfnamefont {A.~S.}\ \bibnamefont
  {Jermyn}}, \bibinfo {author} {\bibfnamefont {P.}~\bibnamefont {Anikeeva}},
  \bibinfo {author} {\bibfnamefont {C.}~\bibnamefont {Felser}},\ and\ \bibinfo
  {author} {\bibfnamefont {P.}~\bibnamefont {Narang}},\ }\bibfield  {title}
  {\bibinfo {title} {{Electron hydrodynamics in anisotropic materials}},\
  }\href {https://doi.org/10.1038/s41467-020-18553-y} {\bibfield  {journal}
  {\bibinfo  {journal} {Nature Communications}\ }\textbf {\bibinfo {volume}
  {11}},\ \bibinfo {pages} {4710} (\bibinfo {year} {2020})}\BibitemShut
  {NoStop}%
\bibitem [{\citenamefont {Cook}\ and\ \citenamefont {Lucas}(2021)}]{Cook2021}%
  \BibitemOpen
  \bibfield  {author} {\bibinfo {author} {\bibfnamefont {C.~Q.}\ \bibnamefont
  {Cook}}\ and\ \bibinfo {author} {\bibfnamefont {A.}~\bibnamefont {Lucas}},\
  }\bibfield  {title} {\bibinfo {title} {{Viscometry of Electron Fluids from
  Symmetry}},\ }\href {https://doi.org/10.1103/PhysRevLett.127.176603}
  {\bibfield  {journal} {\bibinfo  {journal} {Physical Review Letters}\
  }\textbf {\bibinfo {volume} {127}},\ \bibinfo {pages} {176603} (\bibinfo
  {year} {2021})}\BibitemShut {NoStop}%
\bibitem [{\citenamefont {Qi}\ and\ \citenamefont {Lucas}(2021)}]{Qi2021}%
  \BibitemOpen
  \bibfield  {author} {\bibinfo {author} {\bibfnamefont {M.}~\bibnamefont
  {Qi}}\ and\ \bibinfo {author} {\bibfnamefont {A.}~\bibnamefont {Lucas}},\
  }\bibfield  {title} {\bibinfo {title} {{Distinguishing viscous, ballistic,
  and diffusive current flows in anisotropic metals}},\ }\href
  {https://doi.org/10.1103/PhysRevB.104.195106} {\bibfield  {journal} {\bibinfo
   {journal} {Physical Review B}\ }\textbf {\bibinfo {volume} {104}},\ \bibinfo
  {pages} {195106} (\bibinfo {year} {2021})}\BibitemShut {NoStop}%
\bibitem [{\citenamefont {Turner}\ \emph {et~al.}(2004)\citenamefont {Turner},
  \citenamefont {Broun}, \citenamefont {Kamal}, \citenamefont {Hayden},
  \citenamefont {Bobowski}, \citenamefont {Harris}, \citenamefont {Morgan},
  \citenamefont {Preston}, \citenamefont {Bonn},\ and\ \citenamefont
  {Hardy}}]{Turner2004}%
  \BibitemOpen
  \bibfield  {author} {\bibinfo {author} {\bibfnamefont {P.~J.}\ \bibnamefont
  {Turner}}, \bibinfo {author} {\bibfnamefont {D.~M.}\ \bibnamefont {Broun}},
  \bibinfo {author} {\bibfnamefont {S.}~\bibnamefont {Kamal}}, \bibinfo
  {author} {\bibfnamefont {M.~E.}\ \bibnamefont {Hayden}}, \bibinfo {author}
  {\bibfnamefont {J.~S.}\ \bibnamefont {Bobowski}}, \bibinfo {author}
  {\bibfnamefont {R.}~\bibnamefont {Harris}}, \bibinfo {author} {\bibfnamefont
  {D.~C.}\ \bibnamefont {Morgan}}, \bibinfo {author} {\bibfnamefont {J.~S.}\
  \bibnamefont {Preston}}, \bibinfo {author} {\bibfnamefont {D.~A.}\
  \bibnamefont {Bonn}},\ and\ \bibinfo {author} {\bibfnamefont {W.~N.}\
  \bibnamefont {Hardy}},\ }\bibfield  {title} {\bibinfo {title} {{Bolometric
  technique for high-resolution broadband microwave spectroscopy of
  ultra-low-loss samples}},\ }\href {https://doi.org/10.1063/1.1633001}
  {\bibfield  {journal} {\bibinfo  {journal} {Review of Scientific
  Instruments}\ }\textbf {\bibinfo {volume} {75}},\ \bibinfo {pages} {124}
  (\bibinfo {year} {2004})}\BibitemShut {NoStop}%
\bibitem [{\citenamefont {Huttema}\ \emph {et~al.}(2006)\citenamefont
  {Huttema}, \citenamefont {Morgan}, \citenamefont {Turner}, \citenamefont
  {Hardy}, \citenamefont {Zhou}, \citenamefont {Bonn}, \citenamefont {Liang},\
  and\ \citenamefont {Broun}}]{Huttema2006}%
  \BibitemOpen
  \bibfield  {author} {\bibinfo {author} {\bibfnamefont {W.~A.}\ \bibnamefont
  {Huttema}}, \bibinfo {author} {\bibfnamefont {B.}~\bibnamefont {Morgan}},
  \bibinfo {author} {\bibfnamefont {P.~J.}\ \bibnamefont {Turner}}, \bibinfo
  {author} {\bibfnamefont {W.~N.}\ \bibnamefont {Hardy}}, \bibinfo {author}
  {\bibfnamefont {X.}~\bibnamefont {Zhou}}, \bibinfo {author} {\bibfnamefont
  {D.~A.}\ \bibnamefont {Bonn}}, \bibinfo {author} {\bibfnamefont
  {R.}~\bibnamefont {Liang}},\ and\ \bibinfo {author} {\bibfnamefont {D.~M.}\
  \bibnamefont {Broun}},\ }\bibfield  {title} {\bibinfo {title} {{Apparatus for
  high-resolution microwave spectroscopy in strong magnetic fields}},\ }\href
  {https://doi.org/10.1063/1.2167127} {\bibfield  {journal} {\bibinfo
  {journal} {Review of Scientific Instruments}\ }\textbf {\bibinfo {volume}
  {77}},\ \bibinfo {pages} {023901} (\bibinfo {year} {2006})}\BibitemShut
  {NoStop}%
\bibitem [{\citenamefont {Bonn}\ and\ \citenamefont {Hardy}(2007)}]{Bonn2007}%
  \BibitemOpen
  \bibfield  {author} {\bibinfo {author} {\bibfnamefont {D.}~\bibnamefont
  {Bonn}}\ and\ \bibinfo {author} {\bibfnamefont {W.}~\bibnamefont {Hardy}},\
  }\bibfield  {title} {\bibinfo {title} {{Microwave Electrodynamics of High
  Temperature Superconductors}},\ }in\ \href
  {https://doi.org/10.1007/978-0-387-68734-6_4} {\emph {\bibinfo {booktitle}
  {Handbook of High-Temperature Superconductivity: Theory and Experiment}}}\
  (\bibinfo {year} {2007})\ pp.\ \bibinfo {pages} {145--214}\BibitemShut
  {NoStop}%
\bibitem [{\citenamefont {Matsuda}\ \emph {et~al.}(1994)\citenamefont
  {Matsuda}, \citenamefont {Ong}, \citenamefont {Yan}, \citenamefont {Harris},\
  and\ \citenamefont {Peterson}}]{Matsuda1994}%
  \BibitemOpen
  \bibfield  {author} {\bibinfo {author} {\bibfnamefont {Y.}~\bibnamefont
  {Matsuda}}, \bibinfo {author} {\bibfnamefont {N.~P.}\ \bibnamefont {Ong}},
  \bibinfo {author} {\bibfnamefont {Y.~F.}\ \bibnamefont {Yan}}, \bibinfo
  {author} {\bibfnamefont {J.~M.}\ \bibnamefont {Harris}},\ and\ \bibinfo
  {author} {\bibfnamefont {J.~B.}\ \bibnamefont {Peterson}},\ }\bibfield
  {title} {\bibinfo {title} {{Vortex viscosity in YBa$_2$Cu$_3$O$_{7-\delta}$
  at low temperatures}},\ }\href {https://doi.org/10.1103/PhysRevB.49.4380}
  {\bibfield  {journal} {\bibinfo  {journal} {Physical Review B}\ }\textbf
  {\bibinfo {volume} {49}},\ \bibinfo {pages} {4380} (\bibinfo {year}
  {1994})}\BibitemShut {NoStop}%
\bibitem [{\citenamefont {Booth}\ \emph {et~al.}(1994)\citenamefont {Booth},
  \citenamefont {Wu},\ and\ \citenamefont {Anlage}}]{Booth1994}%
  \BibitemOpen
  \bibfield  {author} {\bibinfo {author} {\bibfnamefont {J.~C.}\ \bibnamefont
  {Booth}}, \bibinfo {author} {\bibfnamefont {D.~H.}\ \bibnamefont {Wu}},\ and\
  \bibinfo {author} {\bibfnamefont {S.~M.}\ \bibnamefont {Anlage}},\ }\bibfield
   {title} {\bibinfo {title} {{A broadband method for the measurement of the
  surface impedance of thin films at microwave frequencies}},\ }\href
  {https://doi.org/10.1063/1.1144816} {\bibfield  {journal} {\bibinfo
  {journal} {Review of Scientific Instruments}\ }\textbf {\bibinfo {volume}
  {65}},\ \bibinfo {pages} {2082} (\bibinfo {year} {1994})}\BibitemShut
  {NoStop}%
\bibitem [{\citenamefont {Reuter}\ and\ \citenamefont
  {Sondheimer}(1948)}]{Reuter1948}%
  \BibitemOpen
  \bibfield  {author} {\bibinfo {author} {\bibfnamefont {G.~E.~H.}\
  \bibnamefont {Reuter}}\ and\ \bibinfo {author} {\bibfnamefont {E.~H.}\
  \bibnamefont {Sondheimer}},\ }\bibfield  {title} {\bibinfo {title} {{The
  theory of the anomalous skin effect in metals}},\ }\href
  {https://doi.org/10.1098/rspa.1948.0123} {\bibfield  {journal} {\bibinfo
  {journal} {Proceedings of the Royal Society A}\ }\textbf {\bibinfo {volume}
  {195}},\ \bibinfo {pages} {336} (\bibinfo {year} {1948})}\BibitemShut
  {NoStop}%
\bibitem [{\citenamefont {Sondheimer}(1954)}]{Sondheimer1954}%
  \BibitemOpen
  \bibfield  {author} {\bibinfo {author} {\bibfnamefont {E.~H.}\ \bibnamefont
  {Sondheimer}},\ }\bibfield  {title} {\bibinfo {title} {{The Theory of the
  Anomalous Skin Effect in Anisotropic Metals}},\ }\href
  {https://doi.org/https://doi.org/10.1098/rspa.1954.0156} {\bibfield
  {journal} {\bibinfo  {journal} {Proceedings of the Royal Society A}\ }\textbf
  {\bibinfo {volume} {224}},\ \bibinfo {pages} {260} (\bibinfo {year}
  {1954})}\BibitemShut {NoStop}%
\bibitem [{\citenamefont {Pippard}(1954)}]{Pippard1954}%
  \BibitemOpen
  \bibfield  {author} {\bibinfo {author} {\bibfnamefont {A.}~\bibnamefont
  {Pippard}},\ }\bibfield  {title} {\bibinfo {title} {{The anomalous skin
  effect in anisotropic metals}},\ }\href
  {https://doi.org/10.1098/rspa.1954.0157} {\bibfield  {journal} {\bibinfo
  {journal} {Proceedings of the Royal Society A}\ }\textbf {\bibinfo {volume}
  {224}},\ \bibinfo {pages} {273} (\bibinfo {year} {1954})}\BibitemShut
  {NoStop}%
\bibitem [{\citenamefont {Pippard}(1947)}]{Pippard1947}%
  \BibitemOpen
  \bibfield  {author} {\bibinfo {author} {\bibfnamefont {A.}~\bibnamefont
  {Pippard}},\ }\bibfield  {title} {\bibinfo {title} {{The surface impedance of
  superconductors and normal metals at high frequencies II. The anomalous skin
  effect in normal metals}},\ }\href {https://doi.org/10.1098/rspa.1947.0122}
  {\bibfield  {journal} {\bibinfo  {journal} {Proceedings of the Royal Society
  A}\ }\textbf {\bibinfo {volume} {191}},\ \bibinfo {pages} {385} (\bibinfo
  {year} {1947})}\BibitemShut {NoStop}%
\bibitem [{\citenamefont {Chambers}(1950)}]{Chambers1950}%
  \BibitemOpen
  \bibfield  {author} {\bibinfo {author} {\bibfnamefont {R.~G.}\ \bibnamefont
  {Chambers}},\ }\bibfield  {title} {\bibinfo {title} {{Anomalous Skin Effect
  in Metals}},\ }\href {https://doi.org/https://doi.org/10.1038/165239b0}
  {\bibfield  {journal} {\bibinfo  {journal} {Nature}\ }\textbf {\bibinfo
  {volume} {165}},\ \bibinfo {pages} {289} (\bibinfo {year}
  {1950})}\BibitemShut {NoStop}%
\bibitem [{\citenamefont {Chambers}(1952)}]{Chambers1952}%
  \BibitemOpen
  \bibfield  {author} {\bibinfo {author} {\bibfnamefont {R.~G.}\ \bibnamefont
  {Chambers}},\ }\bibfield  {title} {\bibinfo {title} {{The anomalous skin
  effect}},\ }\href {https://doi.org/https://doi.org/10.1098/rspa.1952.0226}
  {\bibfield  {journal} {\bibinfo  {journal} {Proceedings of the Royal Society
  A}\ }\textbf {\bibinfo {volume} {215}},\ \bibinfo {pages} {481} (\bibinfo
  {year} {1952})}\BibitemShut {NoStop}%
\bibitem [{\citenamefont {Smith}(1959)}]{Smith1959}%
  \BibitemOpen
  \bibfield  {author} {\bibinfo {author} {\bibfnamefont {G.}~\bibnamefont
  {Smith}},\ }\bibfield  {title} {\bibinfo {title} {{Anomalous Skin Effect in
  Bismuth}},\ }\href {https://doi.org/https://doi.org/10.1103/PhysRev.115.1561}
  {\bibfield  {journal} {\bibinfo  {journal} {Physical Review}\ }\textbf
  {\bibinfo {volume} {115}},\ \bibinfo {pages} {1561} (\bibinfo {year}
  {1959})}\BibitemShut {NoStop}%
\bibitem [{\citenamefont {Gurzhi}(1968)}]{Gurzhi1968}%
  \BibitemOpen
  \bibfield  {author} {\bibinfo {author} {\bibfnamefont {R.~N.}\ \bibnamefont
  {Gurzhi}},\ }\bibfield  {title} {\bibinfo {title} {{Hydrodynamic effects in
  solids at low temperature}},\ }\href
  {https://doi.org/10.1070/PU1968v011n02ABEH003815} {\bibfield  {journal}
  {\bibinfo  {journal} {Soviet Physics Uspekhi}\ }\textbf {\bibinfo {volume}
  {11}},\ \bibinfo {pages} {255} (\bibinfo {year} {1968})}\BibitemShut
  {NoStop}%
\bibitem [{\citenamefont {Mackenzie}\ \emph {et~al.}(1998)\citenamefont
  {Mackenzie}, \citenamefont {Haselwimmer}, \citenamefont {Tyler},
  \citenamefont {Lonzarich}, \citenamefont {Mori}, \citenamefont {Nishizaki},\
  and\ \citenamefont {Maeno}}]{Mackenzie1998}%
  \BibitemOpen
  \bibfield  {author} {\bibinfo {author} {\bibfnamefont {A.~P.}\ \bibnamefont
  {Mackenzie}}, \bibinfo {author} {\bibfnamefont {R.~K.}\ \bibnamefont
  {Haselwimmer}}, \bibinfo {author} {\bibfnamefont {A.~W.}\ \bibnamefont
  {Tyler}}, \bibinfo {author} {\bibfnamefont {G.~G.}\ \bibnamefont
  {Lonzarich}}, \bibinfo {author} {\bibfnamefont {Y.}~\bibnamefont {Mori}},
  \bibinfo {author} {\bibfnamefont {S.}~\bibnamefont {Nishizaki}},\ and\
  \bibinfo {author} {\bibfnamefont {Y.}~\bibnamefont {Maeno}},\ }\bibfield
  {title} {\bibinfo {title} {{Extremely Strong Dependence of Superconductivity
  on Disorder in Sr$_2$RuO$_4$}},\ }\href
  {https://doi.org/10.1103/PhysRevLett.80.161} {\bibfield  {journal} {\bibinfo
  {journal} {Physical Review Letters}\ }\textbf {\bibinfo {volume} {80}},\
  \bibinfo {pages} {161} (\bibinfo {year} {1998})}\BibitemShut {NoStop}%
\bibitem [{\citenamefont {Barber}\ \emph {et~al.}(2018)\citenamefont {Barber},
  \citenamefont {Gibbs}, \citenamefont {Maeno}, \citenamefont {Mackenzie},\
  and\ \citenamefont {Hicks}}]{Barber2018}%
  \BibitemOpen
  \bibfield  {author} {\bibinfo {author} {\bibfnamefont {M.~E.}\ \bibnamefont
  {Barber}}, \bibinfo {author} {\bibfnamefont {A.~S.}\ \bibnamefont {Gibbs}},
  \bibinfo {author} {\bibfnamefont {Y.}~\bibnamefont {Maeno}}, \bibinfo
  {author} {\bibfnamefont {A.~P.}\ \bibnamefont {Mackenzie}},\ and\ \bibinfo
  {author} {\bibfnamefont {C.~W.}\ \bibnamefont {Hicks}},\ }\bibfield  {title}
  {\bibinfo {title} {{Resistivity in the Vicinity of a van Hove Singularity:
  Sr$_2$RuO$_4$ under Uniaxial Pressure}},\ }\href
  {https://doi.org/10.1103/PhysRevLett.120.076602} {\bibfield  {journal}
  {\bibinfo  {journal} {Physical Review Letters}\ }\textbf {\bibinfo {volume}
  {120}},\ \bibinfo {pages} {76602} (\bibinfo {year} {2018})}\BibitemShut
  {NoStop}%
\bibitem [{Note1()}]{Note1}%
  \BibitemOpen
  \bibinfo {note} {See Supplemental Material at [URL will be inserted by
  publisher]}\BibitemShut {NoStop}%
\bibitem [{\citenamefont {Ashcroft}\ and\ \citenamefont
  {Mermin}(1976)}]{Ashcroft1976}%
  \BibitemOpen
  \bibfield  {author} {\bibinfo {author} {\bibfnamefont {N.}~\bibnamefont
  {Ashcroft}}\ and\ \bibinfo {author} {\bibfnamefont {N.}~\bibnamefont
  {Mermin}},\ }\href@noop {} {\emph {\bibinfo {title} {{Solid State
  Physics}}}}\ (\bibinfo  {publisher} {Holt-Saunders},\ \bibinfo {year}
  {1976})\BibitemShut {NoStop}%
\bibitem [{\citenamefont {Hicks}\ \emph {et~al.}(2012)\citenamefont {Hicks},
  \citenamefont {Gibbs}, \citenamefont {Mackenzie}, \citenamefont {Takatsu},
  \citenamefont {Maeno},\ and\ \citenamefont {Yelland}}]{Hicks2012}%
  \BibitemOpen
  \bibfield  {author} {\bibinfo {author} {\bibfnamefont {C.~W.}\ \bibnamefont
  {Hicks}}, \bibinfo {author} {\bibfnamefont {A.~S.}\ \bibnamefont {Gibbs}},
  \bibinfo {author} {\bibfnamefont {A.~P.}\ \bibnamefont {Mackenzie}}, \bibinfo
  {author} {\bibfnamefont {H.}~\bibnamefont {Takatsu}}, \bibinfo {author}
  {\bibfnamefont {Y.}~\bibnamefont {Maeno}},\ and\ \bibinfo {author}
  {\bibfnamefont {E.~A.}\ \bibnamefont {Yelland}},\ }\bibfield  {title}
  {\bibinfo {title} {{Quantum Oscillations and High Carrier Mobility in the
  Delafossite PdCoO$_2$}},\ }\href
  {https://doi.org/10.1103/PhysRevLett.109.116401} {\bibfield  {journal}
  {\bibinfo  {journal} {Physical Review Letters}\ }\textbf {\bibinfo {volume}
  {109}},\ \bibinfo {pages} {116401} (\bibinfo {year} {2012})}\BibitemShut
  {NoStop}%
\bibitem [{\citenamefont {Sunko}\ \emph {et~al.}(2020)\citenamefont {Sunko},
  \citenamefont {Mcguinness}, \citenamefont {Chang}, \citenamefont {Zhakina},
  \citenamefont {Khim}, \citenamefont {Dreyer}, \citenamefont {Konczykowski},
  \citenamefont {Borrmann}, \citenamefont {Moll}, \citenamefont {K{\"{o}}nig},
  \citenamefont {Muller},\ and\ \citenamefont {Mackenzie}}]{Sunko2020}%
  \BibitemOpen
  \bibfield  {author} {\bibinfo {author} {\bibfnamefont {V.}~\bibnamefont
  {Sunko}}, \bibinfo {author} {\bibfnamefont {P.~H.}\ \bibnamefont
  {Mcguinness}}, \bibinfo {author} {\bibfnamefont {C.~S.}\ \bibnamefont
  {Chang}}, \bibinfo {author} {\bibfnamefont {E.}~\bibnamefont {Zhakina}},
  \bibinfo {author} {\bibfnamefont {S.}~\bibnamefont {Khim}}, \bibinfo {author}
  {\bibfnamefont {C.~E.}\ \bibnamefont {Dreyer}}, \bibinfo {author}
  {\bibfnamefont {M.}~\bibnamefont {Konczykowski}}, \bibinfo {author}
  {\bibfnamefont {H.}~\bibnamefont {Borrmann}}, \bibinfo {author}
  {\bibfnamefont {P.~J.~W.}\ \bibnamefont {Moll}}, \bibinfo {author}
  {\bibfnamefont {M.}~\bibnamefont {K{\"{o}}nig}}, \bibinfo {author}
  {\bibfnamefont {D.~A.}\ \bibnamefont {Muller}},\ and\ \bibinfo {author}
  {\bibfnamefont {A.~P.}\ \bibnamefont {Mackenzie}},\ }\bibfield  {title}
  {\bibinfo {title} {{Controlled Introduction of Defects to Delafossite Metals
  by Electron Irradiation}},\ }\href
  {https://doi.org/10.1103/PhysRevX.10.021018} {\bibfield  {journal} {\bibinfo
  {journal} {Physical Review X}\ }\textbf {\bibinfo {volume} {10}},\ \bibinfo
  {pages} {021018} (\bibinfo {year} {2020})}\BibitemShut {NoStop}%
\bibitem [{\citenamefont {Moll}\ \emph {et~al.}(2016)\citenamefont {Moll},
  \citenamefont {Kushwaha}, \citenamefont {Nandi}, \citenamefont {Schmidt},\
  and\ \citenamefont {Mackenzie}}]{Moll2016}%
  \BibitemOpen
  \bibfield  {author} {\bibinfo {author} {\bibfnamefont {P.~J.~W.}\
  \bibnamefont {Moll}}, \bibinfo {author} {\bibfnamefont {P.}~\bibnamefont
  {Kushwaha}}, \bibinfo {author} {\bibfnamefont {N.}~\bibnamefont {Nandi}},
  \bibinfo {author} {\bibfnamefont {B.}~\bibnamefont {Schmidt}},\ and\ \bibinfo
  {author} {\bibfnamefont {A.~P.}\ \bibnamefont {Mackenzie}},\ }\bibfield
  {title} {\bibinfo {title} {{Evidence for hydrodynamic electron flow in
  PdCoO$_2$}},\ }\href {https://doi.org/10.1126/science.aac8385} {\bibfield
  {journal} {\bibinfo  {journal} {Science}\ }\textbf {\bibinfo {volume}
  {351}},\ \bibinfo {pages} {1061} (\bibinfo {year} {2016})}\BibitemShut
  {NoStop}%
\bibitem [{\citenamefont {Noh}\ \emph {et~al.}(2009)\citenamefont {Noh},
  \citenamefont {Jeong}, \citenamefont {Jeong}, \citenamefont {Cho},
  \citenamefont {Kim}, \citenamefont {Kim}, \citenamefont {Min},\ and\
  \citenamefont {Kim}}]{Noh2009}%
  \BibitemOpen
  \bibfield  {author} {\bibinfo {author} {\bibfnamefont {H.-J.}\ \bibnamefont
  {Noh}}, \bibinfo {author} {\bibfnamefont {J.}~\bibnamefont {Jeong}}, \bibinfo
  {author} {\bibfnamefont {J.}~\bibnamefont {Jeong}}, \bibinfo {author}
  {\bibfnamefont {E.-J.}\ \bibnamefont {Cho}}, \bibinfo {author} {\bibfnamefont
  {S.~B.}\ \bibnamefont {Kim}}, \bibinfo {author} {\bibfnamefont
  {K.}~\bibnamefont {Kim}}, \bibinfo {author} {\bibfnamefont {B.~I.}\
  \bibnamefont {Min}},\ and\ \bibinfo {author} {\bibfnamefont {H.-D.}\
  \bibnamefont {Kim}},\ }\bibfield  {title} {\bibinfo {title} {{Anisotropic
  Electric Conductivity of Delafossite PdCoO$_2$ Studied by Angle-Resolved
  Photoemission Spectroscopy}},\ }\href
  {https://doi.org/10.1103/PhysRevLett.102.256404} {\bibfield  {journal}
  {\bibinfo  {journal} {Physical Review Letters}\ }\textbf {\bibinfo {volume}
  {102}},\ \bibinfo {pages} {256404} (\bibinfo {year} {2009})}\BibitemShut
  {NoStop}%
\bibitem [{\citenamefont {Sunko}\ \emph {et~al.}(2017)\citenamefont {Sunko},
  \citenamefont {Rosner}, \citenamefont {Kushwaha}, \citenamefont {Khim},
  \citenamefont {Mazzola}, \citenamefont {Bawden}, \citenamefont {Clark},
  \citenamefont {Riley}, \citenamefont {Kasinathan}, \citenamefont {Haverkort},
  \citenamefont {Kim}, \citenamefont {Hoesch}, \citenamefont {Fujii},
  \citenamefont {Vobornik}, \citenamefont {Mackenzie},\ and\ \citenamefont
  {King}}]{Sunko2017}%
  \BibitemOpen
  \bibfield  {author} {\bibinfo {author} {\bibfnamefont {V.}~\bibnamefont
  {Sunko}}, \bibinfo {author} {\bibfnamefont {H.}~\bibnamefont {Rosner}},
  \bibinfo {author} {\bibfnamefont {P.}~\bibnamefont {Kushwaha}}, \bibinfo
  {author} {\bibfnamefont {S.}~\bibnamefont {Khim}}, \bibinfo {author}
  {\bibfnamefont {F.}~\bibnamefont {Mazzola}}, \bibinfo {author} {\bibfnamefont
  {L.}~\bibnamefont {Bawden}}, \bibinfo {author} {\bibfnamefont {O.~J.}\
  \bibnamefont {Clark}}, \bibinfo {author} {\bibfnamefont {J.~M.}\ \bibnamefont
  {Riley}}, \bibinfo {author} {\bibfnamefont {D.}~\bibnamefont {Kasinathan}},
  \bibinfo {author} {\bibfnamefont {M.~W.}\ \bibnamefont {Haverkort}}, \bibinfo
  {author} {\bibfnamefont {T.~K.}\ \bibnamefont {Kim}}, \bibinfo {author}
  {\bibfnamefont {M.}~\bibnamefont {Hoesch}}, \bibinfo {author} {\bibfnamefont
  {J.}~\bibnamefont {Fujii}}, \bibinfo {author} {\bibfnamefont
  {I.}~\bibnamefont {Vobornik}}, \bibinfo {author} {\bibfnamefont {A.~P.}\
  \bibnamefont {Mackenzie}},\ and\ \bibinfo {author} {\bibfnamefont {P.~D.}\
  \bibnamefont {King}},\ }\bibfield  {title} {\bibinfo {title} {{Maximal
  Rashba-like spin splitting via kinetic-energy-coupled inversion-symmetry
  breaking}},\ }\href {https://doi.org/10.1038/nature23898} {\bibfield
  {journal} {\bibinfo  {journal} {Nature}\ }\textbf {\bibinfo {volume} {549}},\
  \bibinfo {pages} {492} (\bibinfo {year} {2017})}\BibitemShut {NoStop}%
\bibitem [{\citenamefont {Bachmann}\ \emph {et~al.}(2019)\citenamefont
  {Bachmann}, \citenamefont {Sharpe}, \citenamefont {Barnard}, \citenamefont
  {Putzke}, \citenamefont {K{\"{o}}nig}, \citenamefont {Khim}, \citenamefont
  {Goldhaber-Gordon}, \citenamefont {Mackenzie},\ and\ \citenamefont
  {Moll}}]{Bachmann2019}%
  \BibitemOpen
  \bibfield  {author} {\bibinfo {author} {\bibfnamefont {M.~D.}\ \bibnamefont
  {Bachmann}}, \bibinfo {author} {\bibfnamefont {A.~L.}\ \bibnamefont
  {Sharpe}}, \bibinfo {author} {\bibfnamefont {A.~W.}\ \bibnamefont {Barnard}},
  \bibinfo {author} {\bibfnamefont {C.}~\bibnamefont {Putzke}}, \bibinfo
  {author} {\bibfnamefont {M.}~\bibnamefont {K{\"{o}}nig}}, \bibinfo {author}
  {\bibfnamefont {S.}~\bibnamefont {Khim}}, \bibinfo {author} {\bibfnamefont
  {D.}~\bibnamefont {Goldhaber-Gordon}}, \bibinfo {author} {\bibfnamefont
  {A.~P.}\ \bibnamefont {Mackenzie}},\ and\ \bibinfo {author} {\bibfnamefont
  {P.~J.~W.}\ \bibnamefont {Moll}},\ }\bibfield  {title} {\bibinfo {title}
  {{Super-geometric electron focusing on the hexagonal Fermi surface of
  PdCoO$_2$}},\ }\href {https://doi.org/10.1038/s41467-019-13020-9} {\bibfield
  {journal} {\bibinfo  {journal} {Nature Communications}\ }\textbf {\bibinfo
  {volume} {10}},\ \bibinfo {pages} {5081} (\bibinfo {year}
  {2019})}\BibitemShut {NoStop}%
\bibitem [{\citenamefont {McGuinness}\ \emph {et~al.}(2021)\citenamefont
  {McGuinness}, \citenamefont {Zhakina}, \citenamefont {K{\"{o}}nig},
  \citenamefont {Bachmann}, \citenamefont {Putzke}, \citenamefont {Moll},
  \citenamefont {Khim},\ and\ \citenamefont {Mackenzie}}]{McGuinness2021}%
  \BibitemOpen
  \bibfield  {author} {\bibinfo {author} {\bibfnamefont {P.~H.}\ \bibnamefont
  {McGuinness}}, \bibinfo {author} {\bibfnamefont {E.}~\bibnamefont {Zhakina}},
  \bibinfo {author} {\bibfnamefont {M.}~\bibnamefont {K{\"{o}}nig}}, \bibinfo
  {author} {\bibfnamefont {M.~D.}\ \bibnamefont {Bachmann}}, \bibinfo {author}
  {\bibfnamefont {C.}~\bibnamefont {Putzke}}, \bibinfo {author} {\bibfnamefont
  {P.~J.~W.}\ \bibnamefont {Moll}}, \bibinfo {author} {\bibfnamefont
  {S.}~\bibnamefont {Khim}},\ and\ \bibinfo {author} {\bibfnamefont {A.~P.}\
  \bibnamefont {Mackenzie}},\ }\bibfield  {title} {\bibinfo {title}
  {{Low-symmetry nonlocal transport in microstructured squares of delafossite
  metals}},\ }\href {https://doi.org/10.1073/PNAS.2113185118} {\bibfield
  {journal} {\bibinfo  {journal} {Proceedings of the National Academy of
  Sciences}\ }\textbf {\bibinfo {volume} {118}},\ \bibinfo {pages} {2113185118}
  (\bibinfo {year} {2021})}\BibitemShut {NoStop}%
\bibitem [{\citenamefont {Bachmann}\ \emph {et~al.}(2022)\citenamefont
  {Bachmann}, \citenamefont {Sharpe}, \citenamefont {Baker}, \citenamefont
  {Barnard}, \citenamefont {Putzke}, \citenamefont {Scaffidi}, \citenamefont
  {Nandi}, \citenamefont {Mcguinness}, \citenamefont {Zhakina}, \citenamefont
  {Moravec}, \citenamefont {Khim}, \citenamefont {K{\"{o}}nig}, \citenamefont
  {Goldhaber-Gordon}, \citenamefont {Mackenzie},\ and\ \citenamefont
  {Moll}}]{Bachmann2022}%
  \BibitemOpen
  \bibfield  {author} {\bibinfo {author} {\bibfnamefont {M.~D.}\ \bibnamefont
  {Bachmann}}, \bibinfo {author} {\bibfnamefont {A.~L.}\ \bibnamefont
  {Sharpe}}, \bibinfo {author} {\bibfnamefont {G.}~\bibnamefont {Baker}},
  \bibinfo {author} {\bibfnamefont {A.~W.}\ \bibnamefont {Barnard}}, \bibinfo
  {author} {\bibfnamefont {C.}~\bibnamefont {Putzke}}, \bibinfo {author}
  {\bibfnamefont {T.}~\bibnamefont {Scaffidi}}, \bibinfo {author}
  {\bibfnamefont {N.}~\bibnamefont {Nandi}}, \bibinfo {author} {\bibfnamefont
  {P.~H.}\ \bibnamefont {Mcguinness}}, \bibinfo {author} {\bibfnamefont
  {E.}~\bibnamefont {Zhakina}}, \bibinfo {author} {\bibfnamefont
  {M.}~\bibnamefont {Moravec}}, \bibinfo {author} {\bibfnamefont
  {S.}~\bibnamefont {Khim}}, \bibinfo {author} {\bibfnamefont {M.}~\bibnamefont
  {K{\"{o}}nig}}, \bibinfo {author} {\bibfnamefont {D.}~\bibnamefont
  {Goldhaber-Gordon}}, \bibinfo {author} {\bibfnamefont {A.~P.}\ \bibnamefont
  {Mackenzie}},\ and\ \bibinfo {author} {\bibfnamefont {P.~J.~W.}\ \bibnamefont
  {Moll}},\ }\bibfield  {title} {\bibinfo {title} {{Directional ballistic
  transport in the two-dimensional metal PdCoO$_2$}},\ }\bibfield  {journal}
  {\bibinfo  {journal} {Nature Physics}\ }\href
  {https://doi.org/10.1038/s41567-022-01570-7} {10.1038/s41567-022-01570-7}
  (\bibinfo {year} {2022})\BibitemShut {NoStop}%
\bibitem [{\citenamefont {Frisch}\ \emph {et~al.}(1986)\citenamefont {Frisch},
  \citenamefont {Hasslacher},\ and\ \citenamefont {Pomeau}}]{Frisch1986}%
  \BibitemOpen
  \bibfield  {author} {\bibinfo {author} {\bibfnamefont {U.}~\bibnamefont
  {Frisch}}, \bibinfo {author} {\bibfnamefont {B.}~\bibnamefont {Hasslacher}},\
  and\ \bibinfo {author} {\bibfnamefont {Y.}~\bibnamefont {Pomeau}},\
  }\bibfield  {title} {\bibinfo {title} {{Lattice-gas Automata for the
  Navier-Stokes Equation}},\ }\href
  {https://doi.org/10.1103/PhysRevLett.56.1505} {\bibfield  {journal} {\bibinfo
   {journal} {Physical Review Letters}\ }\textbf {\bibinfo {volume} {56}},\
  \bibinfo {pages} {1505} (\bibinfo {year} {1986})}\BibitemShut {NoStop}%
\bibitem [{\citenamefont {Pippard}(1957)}]{Pippard1957}%
  \BibitemOpen
  \bibfield  {author} {\bibinfo {author} {\bibfnamefont {A.}~\bibnamefont
  {Pippard}},\ }\bibfield  {title} {\bibinfo {title} {{An Experimental
  Determination of the Fermi Surface in Copper}},\ }\href
  {https://doi.org/10.1098/rsta.1957.0023} {\bibfield  {journal} {\bibinfo
  {journal} {Philosophical Transactions of the Royal Society A}\ }\textbf
  {\bibinfo {volume} {250}},\ \bibinfo {pages} {325} (\bibinfo {year}
  {1957})}\BibitemShut {NoStop}%
\bibitem [{\citenamefont {Glasser}(1968)}]{Glasser1968}%
  \BibitemOpen
  \bibfield  {author} {\bibinfo {author} {\bibfnamefont {M.~L.}\ \bibnamefont
  {Glasser}},\ }\bibfield  {title} {\bibinfo {title} {{Influence of Band
  Structure on the Nonlocal Conductivity of Metals and the Anomalous Skin
  Effect}},\ }\href {https://doi.org/https://doi.org/10.1103/PhysRev.176.1110}
  {\bibfield  {journal} {\bibinfo  {journal} {Physical Review}\ }\textbf
  {\bibinfo {volume} {176}},\ \bibinfo {pages} {1110} (\bibinfo {year}
  {1968})}\BibitemShut {NoStop}%
\bibitem [{\citenamefont {Kaganov}\ and\ \citenamefont
  {Contreras}(1994)}]{Kaganov1994}%
  \BibitemOpen
  \bibfield  {author} {\bibinfo {author} {\bibfnamefont {M.~I.}\ \bibnamefont
  {Kaganov}}\ and\ \bibinfo {author} {\bibfnamefont {P.}~\bibnamefont
  {Contreras}},\ }\bibfield  {title} {\bibinfo {title} {{Theory of the
  anomalous skin effect in metals with complicated Fermi surfaces}},\ }\href
  {http://www.jetp.ac.ru/cgi-bin/e/index/e/79/6/p985?a=list} {\bibfield
  {journal} {\bibinfo  {journal} {Journal of Experimental and Theoretical
  Physics}\ }\textbf {\bibinfo {volume} {79}},\ \bibinfo {pages} {985}
  (\bibinfo {year} {1994})}\BibitemShut {NoStop}%
\bibitem [{\citenamefont {Zimbovskaya}(2006)}]{Zimbovskaya2006}%
  \BibitemOpen
  \bibfield  {author} {\bibinfo {author} {\bibfnamefont {N.~A.}\ \bibnamefont
  {Zimbovskaya}},\ }\bibfield  {title} {\bibinfo {title} {{Local features of
  the Fermi surface curvature and the anomalous skin effect in metals}},\
  }\href {https://doi.org/10.1088/0953-8984/18/35/003} {\bibfield  {journal}
  {\bibinfo  {journal} {Journal of Physics: Condensed Matter}\ }\textbf
  {\bibinfo {volume} {18}},\ \bibinfo {pages} {8149} (\bibinfo {year}
  {2006})}\BibitemShut {NoStop}%
\bibitem [{\citenamefont {Harada}\ \emph {et~al.}(2018)\citenamefont {Harada},
  \citenamefont {Fujiwara},\ and\ \citenamefont {Tsukazaki}}]{Harada2018}%
  \BibitemOpen
  \bibfield  {author} {\bibinfo {author} {\bibfnamefont {T.}~\bibnamefont
  {Harada}}, \bibinfo {author} {\bibfnamefont {K.}~\bibnamefont {Fujiwara}},\
  and\ \bibinfo {author} {\bibfnamefont {A.}~\bibnamefont {Tsukazaki}},\
  }\bibfield  {title} {\bibinfo {title} {{Highly conductive PdCoO$_2$ ultrathin
  films for transparent electrodes}},\ }\href
  {https://doi.org/10.1063/1.5027579} {\bibfield  {journal} {\bibinfo
  {journal} {APL Materials}\ }\textbf {\bibinfo {volume} {6}},\ \bibinfo
  {pages} {046107} (\bibinfo {year} {2018})}\BibitemShut {NoStop}%
\bibitem [{\citenamefont {Brahlek}\ \emph {et~al.}(2019)\citenamefont
  {Brahlek}, \citenamefont {Rimal}, \citenamefont {Ok}, \citenamefont
  {Mukherjee}, \citenamefont {Mazza}, \citenamefont {Lu}, \citenamefont {Lee},
  \citenamefont {{Zac Ward}}, \citenamefont {Unocic}, \citenamefont {Eres},\
  and\ \citenamefont {Oh}}]{Brahlek2019}%
  \BibitemOpen
  \bibfield  {author} {\bibinfo {author} {\bibfnamefont {M.}~\bibnamefont
  {Brahlek}}, \bibinfo {author} {\bibfnamefont {G.}~\bibnamefont {Rimal}},
  \bibinfo {author} {\bibfnamefont {J.~M.}\ \bibnamefont {Ok}}, \bibinfo
  {author} {\bibfnamefont {D.}~\bibnamefont {Mukherjee}}, \bibinfo {author}
  {\bibfnamefont {A.~R.}\ \bibnamefont {Mazza}}, \bibinfo {author}
  {\bibfnamefont {Q.}~\bibnamefont {Lu}}, \bibinfo {author} {\bibfnamefont
  {H.~N.}\ \bibnamefont {Lee}}, \bibinfo {author} {\bibfnamefont
  {T.}~\bibnamefont {{Zac Ward}}}, \bibinfo {author} {\bibfnamefont {R.~R.}\
  \bibnamefont {Unocic}}, \bibinfo {author} {\bibfnamefont {G.}~\bibnamefont
  {Eres}},\ and\ \bibinfo {author} {\bibfnamefont {S.}~\bibnamefont {Oh}},\
  }\bibfield  {title} {\bibinfo {title} {{Growth of metallic delafossite
  PdCoO$_2$ by molecular beam epitaxy}},\ }\href
  {https://doi.org/10.1103/PhysRevMaterials.3.093401} {\bibfield  {journal}
  {\bibinfo  {journal} {Physical Review Materials}\ }\textbf {\bibinfo {volume}
  {3}},\ \bibinfo {pages} {093401} (\bibinfo {year} {2019})}\BibitemShut
  {NoStop}%
\bibitem [{\citenamefont {Sun}\ \emph {et~al.}(2019)\citenamefont {Sun},
  \citenamefont {Barone}, \citenamefont {Chang}, \citenamefont {Holtz},
  \citenamefont {Paik}, \citenamefont {Schubert}, \citenamefont {Muller},\ and\
  \citenamefont {Schlom}}]{Sun2019}%
  \BibitemOpen
  \bibfield  {author} {\bibinfo {author} {\bibfnamefont {J.}~\bibnamefont
  {Sun}}, \bibinfo {author} {\bibfnamefont {M.~R.}\ \bibnamefont {Barone}},
  \bibinfo {author} {\bibfnamefont {C.~S.}\ \bibnamefont {Chang}}, \bibinfo
  {author} {\bibfnamefont {M.~E.}\ \bibnamefont {Holtz}}, \bibinfo {author}
  {\bibfnamefont {H.}~\bibnamefont {Paik}}, \bibinfo {author} {\bibfnamefont
  {J.}~\bibnamefont {Schubert}}, \bibinfo {author} {\bibfnamefont {D.~A.}\
  \bibnamefont {Muller}},\ and\ \bibinfo {author} {\bibfnamefont {D.~G.}\
  \bibnamefont {Schlom}},\ }\bibfield  {title} {\bibinfo {title} {{Growth of
  PdCoO$_2$ by ozone-assisted molecular-beam epitaxy}},\ }\href
  {https://doi.org/10.1063/1.5130627} {\bibfield  {journal} {\bibinfo
  {journal} {APL Materials}\ }\textbf {\bibinfo {volume} {7}},\ \bibinfo
  {pages} {121112} (\bibinfo {year} {2019})}\BibitemShut {NoStop}%
\bibitem [{\citenamefont {Mackenzie}(2017)}]{Mackenzie2017}%
  \BibitemOpen
  \bibfield  {author} {\bibinfo {author} {\bibfnamefont {A.~P.}\ \bibnamefont
  {Mackenzie}},\ }\bibfield  {title} {\bibinfo {title} {{The properties of
  ultrapure delafossite metals}},\ }\href
  {https://doi.org/10.1088/1361-6633/aa50e5} {\bibfield  {journal} {\bibinfo
  {journal} {Reports on Progress in Physics}\ }\textbf {\bibinfo {volume}
  {80}},\ \bibinfo {pages} {32501} (\bibinfo {year} {2017})}\BibitemShut
  {NoStop}%
\bibitem [{\citenamefont {Wang}\ \emph {et~al.}(2021)\citenamefont {Wang},
  \citenamefont {Varnavides}, \citenamefont {Anikeeva}, \citenamefont {Gooth},
  \citenamefont {Felser},\ and\ \citenamefont {Narang}}]{Wang2021}%
  \BibitemOpen
  \bibfield  {author} {\bibinfo {author} {\bibfnamefont {Y.}~\bibnamefont
  {Wang}}, \bibinfo {author} {\bibfnamefont {G.}~\bibnamefont {Varnavides}},
  \bibinfo {author} {\bibfnamefont {P.}~\bibnamefont {Anikeeva}}, \bibinfo
  {author} {\bibfnamefont {J.}~\bibnamefont {Gooth}}, \bibinfo {author}
  {\bibfnamefont {C.}~\bibnamefont {Felser}},\ and\ \bibinfo {author}
  {\bibfnamefont {P.}~\bibnamefont {Narang}},\ }\bibfield  {title} {\bibinfo
  {title} {{Generalized Design Principles for Hydrodynamic Electron Transport
  in Anisotropic Metals}},\ }\Eprint {https://arxiv.org/abs/2109.00550}
  {arXiv:2109.00550}  (\bibinfo {year} {2021})\BibitemShut {NoStop}%
\bibitem [{\citenamefont {Levchenko}\ and\ \citenamefont
  {Schmalian}(2020)}]{Levchenko2020}%
  \BibitemOpen
  \bibfield  {author} {\bibinfo {author} {\bibfnamefont {A.}~\bibnamefont
  {Levchenko}}\ and\ \bibinfo {author} {\bibfnamefont {J.}~\bibnamefont
  {Schmalian}},\ }\bibfield  {title} {\bibinfo {title} {{Transport properties
  of strongly coupled electron–phonon liquids}},\ }\href
  {https://doi.org/10.1016/j.aop.2020.168218} {\bibfield  {journal} {\bibinfo
  {journal} {Annals of Physics}\ }\textbf {\bibinfo {volume} {419}},\ \bibinfo
  {pages} {168218} (\bibinfo {year} {2020})}\BibitemShut {NoStop}%
\bibitem [{\citenamefont {Huang}\ and\ \citenamefont
  {Lucas}(2021)}]{Huang2021}%
  \BibitemOpen
  \bibfield  {author} {\bibinfo {author} {\bibfnamefont {X.}~\bibnamefont
  {Huang}}\ and\ \bibinfo {author} {\bibfnamefont {A.}~\bibnamefont {Lucas}},\
  }\bibfield  {title} {\bibinfo {title} {{Electron-phonon hydrodynamics}},\
  }\href {https://doi.org/10.1103/PhysRevB.103.155128} {\bibfield  {journal}
  {\bibinfo  {journal} {Physical Review B}\ }\textbf {\bibinfo {volume}
  {103}},\ \bibinfo {pages} {155128} (\bibinfo {year} {2021})}\BibitemShut
  {NoStop}%
\bibitem [{\citenamefont {Hui}\ \emph {et~al.}(2020)\citenamefont {Hui},
  \citenamefont {Lederer}, \citenamefont {Oganesyan},\ and\ \citenamefont
  {Kim}}]{Hui2020}%
  \BibitemOpen
  \bibfield  {author} {\bibinfo {author} {\bibfnamefont {A.}~\bibnamefont
  {Hui}}, \bibinfo {author} {\bibfnamefont {S.}~\bibnamefont {Lederer}},
  \bibinfo {author} {\bibfnamefont {V.}~\bibnamefont {Oganesyan}},\ and\
  \bibinfo {author} {\bibfnamefont {E.-A.}\ \bibnamefont {Kim}},\ }\bibfield
  {title} {\bibinfo {title} {{Quantum aspects of hydrodynamic transport from
  weak electron-impurity scattering}},\ }\href
  {https://doi.org/10.1103/PhysRevB.101.121107} {\bibfield  {journal} {\bibinfo
   {journal} {Physical Review B}\ }\textbf {\bibinfo {volume} {101}},\ \bibinfo
  {pages} {121107} (\bibinfo {year} {2020})}\BibitemShut {NoStop}%
\bibitem [{\citenamefont {Sarvari}(2008)}]{Sarvari2008}%
  \BibitemOpen
  \bibfield  {author} {\bibinfo {author} {\bibfnamefont {R.}~\bibnamefont
  {Sarvari}},\ }\emph {\bibinfo {title} {{Impact of Size Effects and Anomalous
  Skin Effect on Metallic Wires as GSI Interconnects}}},\ \href@noop {} {Ph.D.
  thesis},\ \bibinfo  {school} {Georgia Institute of Technology} (\bibinfo
  {year} {2008})\BibitemShut {NoStop}%
\bibitem [{\citenamefont {Casimir}\ and\ \citenamefont
  {Ubbink}(1967)}]{Casimir1967}%
  \BibitemOpen
  \bibfield  {author} {\bibinfo {author} {\bibfnamefont {H.~B.~G.}\
  \bibnamefont {Casimir}}\ and\ \bibinfo {author} {\bibfnamefont
  {J.}~\bibnamefont {Ubbink}},\ }\bibfield  {title} {\bibinfo {title} {{The
  skin effect II. The skin effect at high frequencies}},\ }\href
  {https://www.pearl-hifi.com/06_Lit_Archive/02_PEARL_Arch/Vol_16/Sec_53/Philips_Tech_Review/PTechReview-28-1967-300.pdf}
  {\bibfield  {journal} {\bibinfo  {journal} {Philips Technical Review}\
  }\textbf {\bibinfo {volume} {28}},\ \bibinfo {pages} {300} (\bibinfo {year}
  {1967})}\BibitemShut {NoStop}%
\bibitem [{\citenamefont {Forcella}\ \emph {et~al.}(2014)\citenamefont
  {Forcella}, \citenamefont {Zaanen}, \citenamefont {Valentinis},\ and\
  \citenamefont {van~der Marel}}]{Forcella2014}%
  \BibitemOpen
  \bibfield  {author} {\bibinfo {author} {\bibfnamefont {D.}~\bibnamefont
  {Forcella}}, \bibinfo {author} {\bibfnamefont {J.}~\bibnamefont {Zaanen}},
  \bibinfo {author} {\bibfnamefont {D.}~\bibnamefont {Valentinis}},\ and\
  \bibinfo {author} {\bibfnamefont {D.}~\bibnamefont {van~der Marel}},\
  }\bibfield  {title} {\bibinfo {title} {{Electromagnetic properties of viscous
  charged fluids}},\ }\href {https://doi.org/10.1103/PhysRevB.90.035143}
  {\bibfield  {journal} {\bibinfo  {journal} {Physical Review B}\ }\textbf
  {\bibinfo {volume} {90}},\ \bibinfo {pages} {035143} (\bibinfo {year}
  {2014})}\BibitemShut {NoStop}%
\bibitem [{\citenamefont {Valentinis}\ \emph {et~al.}(2021)\citenamefont
  {Valentinis}, \citenamefont {Zaanen},\ and\ \citenamefont {{Van Der
  Marel}}}]{Valentinis2021}%
  \BibitemOpen
  \bibfield  {author} {\bibinfo {author} {\bibfnamefont {D.}~\bibnamefont
  {Valentinis}}, \bibinfo {author} {\bibfnamefont {J.}~\bibnamefont {Zaanen}},\
  and\ \bibinfo {author} {\bibfnamefont {D.}~\bibnamefont {{Van Der Marel}}},\
  }\bibfield  {title} {\bibinfo {title} {{Propagation of shear stress in
  strongly interacting metallic Fermi liquids enhances transmission of
  terahertz radiation}},\ }\href {https://doi.org/10.1038/s41598-021-86356-2}
  {\bibfield  {journal} {\bibinfo  {journal} {Scientific Reports}\ }\textbf
  {\bibinfo {volume} {11}},\ \bibinfo {pages} {7105} (\bibinfo {year}
  {2021})}\BibitemShut {NoStop}%
\bibitem [{\citenamefont {Valentinis}(2021)}]{Valentinis2021-2}%
  \BibitemOpen
  \bibfield  {author} {\bibinfo {author} {\bibfnamefont {D.}~\bibnamefont
  {Valentinis}},\ }\bibfield  {title} {\bibinfo {title} {{Optical signatures of
  shear collective modes in strongly interacting Fermi liquids}},\ }\href
  {https://doi.org/10.1103/PhysRevResearch.3.023076} {\bibfield  {journal}
  {\bibinfo  {journal} {Physical Review Research}\ }\textbf {\bibinfo {volume}
  {3}},\ \bibinfo {pages} {023076} (\bibinfo {year} {2021})}\BibitemShut
  {NoStop}%
\bibitem [{\citenamefont {Matus}\ \emph {et~al.}(2022)\citenamefont {Matus},
  \citenamefont {Dantas}, \citenamefont {Moessner},\ and\ \citenamefont
  {Sur{\'{o}}wka}}]{Matus2022}%
  \BibitemOpen
  \bibfield  {author} {\bibinfo {author} {\bibfnamefont {P.}~\bibnamefont
  {Matus}}, \bibinfo {author} {\bibfnamefont {R.~M.~A.}\ \bibnamefont
  {Dantas}}, \bibinfo {author} {\bibfnamefont {R.}~\bibnamefont {Moessner}},\
  and\ \bibinfo {author} {\bibfnamefont {P.}~\bibnamefont {Sur{\'{o}}wka}},\
  }\bibfield  {title} {\bibinfo {title} {{Skin effect as a probe of transport
  regimes in Weyl semimetals}},\ }\href
  {https://doi.org/10.1073/pnas.2200367119} {\bibfield  {journal} {\bibinfo
  {journal} {Proceedings of the National Academy of Sciences}\ }\textbf
  {\bibinfo {volume} {119}},\ \bibinfo {pages} {e2200367119} (\bibinfo {year}
  {2022})}\BibitemShut {NoStop}%
\bibitem [{\citenamefont {Valentinis}\ \emph {et~al.}(2022)\citenamefont
  {Valentinis}, \citenamefont {Baker}, \citenamefont {Bonn},\ and\
  \citenamefont {Schmalian}}]{Valentinis}%
  \BibitemOpen
  \bibfield  {author} {\bibinfo {author} {\bibfnamefont {D.}~\bibnamefont
  {Valentinis}}, \bibinfo {author} {\bibfnamefont {G.}~\bibnamefont {Baker}},
  \bibinfo {author} {\bibfnamefont {D.~A.}\ \bibnamefont {Bonn}},\ and\
  \bibinfo {author} {\bibfnamefont {J.}~\bibnamefont {Schmalian}},\ }\bibfield
  {title} {\bibinfo {title} {{Kinetic theory of the non-local electrodynamic
  response in anisotropic metals: skin effect in 2D systems}},\ }\Eprint
  {https://arxiv.org/abs/2204.13344} {arXiv:2204.13344}  (\bibinfo {year}
  {2022})\BibitemShut {NoStop}%
\end{thebibliography}%


\begin{thebibliography}{27}%
\makeatletter
\providecommand \@ifxundefined [1]{%
 \@ifx{#1\undefined}
}%
\providecommand \@ifnum [1]{%
 \ifnum #1\expandafter \@firstoftwo
 \else \expandafter \@secondoftwo
 \fi
}%
\providecommand \@ifx [1]{%
 \ifx #1\expandafter \@firstoftwo
 \else \expandafter \@secondoftwo
 \fi
}%
\providecommand \natexlab [1]{#1}%
\providecommand \enquote  [1]{``#1''}%
\providecommand \bibnamefont  [1]{#1}%
\providecommand \bibfnamefont [1]{#1}%
\providecommand \citenamefont [1]{#1}%
\providecommand \href@noop [0]{\@secondoftwo}%
\providecommand \href [0]{\begingroup \@sanitize@url \@href}%
\providecommand \@href[1]{\@@startlink{#1}\@@href}%
\providecommand \@@href[1]{\endgroup#1\@@endlink}%
\providecommand \@sanitize@url [0]{\catcode `\\12\catcode `\$12\catcode
  `\&12\catcode `\#12\catcode `\^12\catcode `\_12\catcode `\%12\relax}%
\providecommand \@@startlink[1]{}%
\providecommand \@@endlink[0]{}%
\providecommand \url  [0]{\begingroup\@sanitize@url \@url }%
\providecommand \@url [1]{\endgroup\@href {#1}{\urlprefix }}%
\providecommand \urlprefix  [0]{URL }%
\providecommand \Eprint [0]{\href }%
\providecommand \doibase [0]{https://doi.org/}%
\providecommand \selectlanguage [0]{\@gobble}%
\providecommand \bibinfo  [0]{\@secondoftwo}%
\providecommand \bibfield  [0]{\@secondoftwo}%
\providecommand \translation [1]{[#1]}%
\providecommand \BibitemOpen [0]{}%
\providecommand \bibitemStop [0]{}%
\providecommand \bibitemNoStop [0]{.\EOS\space}%
\providecommand \EOS [0]{\spacefactor3000\relax}%
\providecommand \BibitemShut  [1]{\csname bibitem#1\endcsname}%
\let\auto@bib@innerbib\@empty
%</preamble>
\bibitem [{\citenamefont {Mao}\ \emph {et~al.}(2000)\citenamefont {Mao},
  \citenamefont {Maenoab},\ and\ \citenamefont {Fukazawa}}]{Mao2000}%
  \BibitemOpen
  \bibfield  {author} {\bibinfo {author} {\bibfnamefont {Z.~Q.}\ \bibnamefont
  {Mao}}, \bibinfo {author} {\bibfnamefont {Y.}~\bibnamefont {Maenoab}},\ and\
  \bibinfo {author} {\bibfnamefont {H.}~\bibnamefont {Fukazawa}},\ }\bibfield
  {title} {\bibinfo {title} {{Crystal growth of Sr$_2$RuO$_4$}},\ }\href
  {https://doi.org/10.1016/S0025-5408(00)00378-0} {\bibfield  {journal}
  {\bibinfo  {journal} {Materials Research Bulletin}\ }\textbf {\bibinfo
  {volume} {35}},\ \bibinfo {pages} {1813} (\bibinfo {year}
  {2000})}\BibitemShut {NoStop}%
\bibitem [{\citenamefont {Bobowski}\ \emph {et~al.}(2019)\citenamefont
  {Bobowski}, \citenamefont {Kikugawa}, \citenamefont {Miyoshi}, \citenamefont
  {Suwa}, \citenamefont {Xu}, \citenamefont {Yonezawa}, \citenamefont
  {Sokolov}, \citenamefont {Mackenzie},\ and\ \citenamefont
  {Maeno}}]{Bobowski2019}%
  \BibitemOpen
  \bibfield  {author} {\bibinfo {author} {\bibfnamefont {J.~S.}\ \bibnamefont
  {Bobowski}}, \bibinfo {author} {\bibfnamefont {N.}~\bibnamefont {Kikugawa}},
  \bibinfo {author} {\bibfnamefont {T.}~\bibnamefont {Miyoshi}}, \bibinfo
  {author} {\bibfnamefont {H.}~\bibnamefont {Suwa}}, \bibinfo {author}
  {\bibfnamefont {H.~S.}\ \bibnamefont {Xu}}, \bibinfo {author} {\bibfnamefont
  {S.}~\bibnamefont {Yonezawa}}, \bibinfo {author} {\bibfnamefont {D.~A.}\
  \bibnamefont {Sokolov}}, \bibinfo {author} {\bibfnamefont {A.~P.}\
  \bibnamefont {Mackenzie}},\ and\ \bibinfo {author} {\bibfnamefont
  {Y.}~\bibnamefont {Maeno}},\ }\bibfield  {title} {\bibinfo {title} {{Improved
  single-crystal growth of Sr$_2$RuO$_4$}},\ }\href
  {https://doi.org/10.3390/condmat4010006} {\bibfield  {journal} {\bibinfo
  {journal} {Condensed Matter}\ }\textbf {\bibinfo {volume} {4}},\ \bibinfo
  {pages} {6} (\bibinfo {year} {2019})}\BibitemShut {NoStop}%
\bibitem [{\citenamefont {Yonezawa}\ \emph {et~al.}(2015)\citenamefont
  {Yonezawa}, \citenamefont {Higuchi}, \citenamefont {Sugimoto}, \citenamefont
  {Sow},\ and\ \citenamefont {Maeno}}]{Yonezawa2015}%
  \BibitemOpen
  \bibfield  {author} {\bibinfo {author} {\bibfnamefont {S.}~\bibnamefont
  {Yonezawa}}, \bibinfo {author} {\bibfnamefont {T.}~\bibnamefont {Higuchi}},
  \bibinfo {author} {\bibfnamefont {Y.}~\bibnamefont {Sugimoto}}, \bibinfo
  {author} {\bibfnamefont {C.}~\bibnamefont {Sow}},\ and\ \bibinfo {author}
  {\bibfnamefont {Y.}~\bibnamefont {Maeno}},\ }\bibfield  {title} {\bibinfo
  {title} {{Compact AC susceptometer for fast sample characterization down to
  0.1 K}},\ }\href {https://doi.org/10.1063/1.4929871} {\bibfield  {journal}
  {\bibinfo  {journal} {Review of Scientific Instruments}\ }\textbf {\bibinfo
  {volume} {86}},\ \bibinfo {pages} {093903} (\bibinfo {year}
  {2015})}\BibitemShut {NoStop}%
\bibitem [{\citenamefont {Bidinosti}\ and\ \citenamefont
  {Hardy}(2000)}]{Bidinosti2000}%
  \BibitemOpen
  \bibfield  {author} {\bibinfo {author} {\bibfnamefont {C.~P.}\ \bibnamefont
  {Bidinosti}}\ and\ \bibinfo {author} {\bibfnamefont {W.~N.}\ \bibnamefont
  {Hardy}},\ }\bibfield  {title} {\bibinfo {title} {{High precision ac
  susceptometer for measuring the temperature and magnetic field dependence of
  the penetration depth in superconductor single crystals}},\ }\href
  {https://doi.org/10.1063/1.1311945} {\bibfield  {journal} {\bibinfo
  {journal} {Review of Scientific Instruments}\ }\textbf {\bibinfo {volume}
  {71}},\ \bibinfo {pages} {3816} (\bibinfo {year} {2000})}\BibitemShut
  {NoStop}%
\bibitem [{\citenamefont {Maeno}\ \emph {et~al.}(1998)\citenamefont {Maeno},
  \citenamefont {Ando}, \citenamefont {Mori}, \citenamefont {Ohmichi},
  \citenamefont {Ikeda}, \citenamefont {Nishizaki},\ and\ \citenamefont
  {Nakatsuji}}]{Maeno1998}%
  \BibitemOpen
  \bibfield  {author} {\bibinfo {author} {\bibfnamefont {Y.}~\bibnamefont
  {Maeno}}, \bibinfo {author} {\bibfnamefont {T.}~\bibnamefont {Ando}},
  \bibinfo {author} {\bibfnamefont {Y.}~\bibnamefont {Mori}}, \bibinfo {author}
  {\bibfnamefont {E.}~\bibnamefont {Ohmichi}}, \bibinfo {author} {\bibfnamefont
  {S.}~\bibnamefont {Ikeda}}, \bibinfo {author} {\bibfnamefont
  {S.}~\bibnamefont {Nishizaki}},\ and\ \bibinfo {author} {\bibfnamefont
  {S.}~\bibnamefont {Nakatsuji}},\ }\bibfield  {title} {\bibinfo {title}
  {{Enhancement of superconductivity of Sr$_2$RuO$_4$ to 3 K by embedded
  metallic microdomains}},\ }\href
  {https://doi.org/10.1103/PhysRevLett.81.3765} {\bibfield  {journal} {\bibinfo
   {journal} {Physical Review Letters}\ }\textbf {\bibinfo {volume} {81}},\
  \bibinfo {pages} {3765} (\bibinfo {year} {1998})}\BibitemShut {NoStop}%
\bibitem [{\citenamefont {Hicks}\ \emph {et~al.}(2014)\citenamefont {Hicks},
  \citenamefont {Brodsky}, \citenamefont {Yelland}, \citenamefont {Gibbs},
  \citenamefont {Bruin}, \citenamefont {Barber}, \citenamefont {Edkins},
  \citenamefont {Nishimura}, \citenamefont {Yonezawa}, \citenamefont {Maeno},\
  and\ \citenamefont {Mackenzie}}]{Hicks2014}%
  \BibitemOpen
  \bibfield  {author} {\bibinfo {author} {\bibfnamefont {C.~W.}\ \bibnamefont
  {Hicks}}, \bibinfo {author} {\bibfnamefont {D.~O.}\ \bibnamefont {Brodsky}},
  \bibinfo {author} {\bibfnamefont {E.~a.}\ \bibnamefont {Yelland}}, \bibinfo
  {author} {\bibfnamefont {A.~S.}\ \bibnamefont {Gibbs}}, \bibinfo {author}
  {\bibfnamefont {J.~a.~N.}\ \bibnamefont {Bruin}}, \bibinfo {author}
  {\bibfnamefont {M.~E.}\ \bibnamefont {Barber}}, \bibinfo {author}
  {\bibfnamefont {S.~D.}\ \bibnamefont {Edkins}}, \bibinfo {author}
  {\bibfnamefont {K.}~\bibnamefont {Nishimura}}, \bibinfo {author}
  {\bibfnamefont {S.}~\bibnamefont {Yonezawa}}, \bibinfo {author}
  {\bibfnamefont {Y.}~\bibnamefont {Maeno}},\ and\ \bibinfo {author}
  {\bibfnamefont {A.~P.}\ \bibnamefont {Mackenzie}},\ }\bibfield  {title}
  {\bibinfo {title} {{Strong Increase of $T_c$ of Sr$_2$RuO$_4$ Under Both
  Tensile and Compressive Strain}},\ }\href
  {https://doi.org/10.1126/science.1248292} {\bibfield  {journal} {\bibinfo
  {journal} {Science}\ }\textbf {\bibinfo {volume} {344}},\ \bibinfo {pages}
  {283} (\bibinfo {year} {2014})}\BibitemShut {NoStop}%
\bibitem [{Note1()}]{Note1}%
  \BibitemOpen
  \bibinfo {note} {We took $A_{ab}=2(a_{b}b_{r}-(b_{r}-b_{l})(a_{b}-a_{t})/2)$
  and $A_{c}=(b_{l}+\protect \sqrt
  {(b_r-b_l)^2+(a_b-a_t)^2}+b_{r})c$.}\BibitemShut {Stop}%
\bibitem [{\citenamefont {Bergemann}\ \emph {et~al.}(2003)\citenamefont
  {Bergemann}, \citenamefont {Mackenzie}, \citenamefont {Julian}, \citenamefont
  {Forsythe},\ and\ \citenamefont {Ohmichi}}]{Bergemann2003}%
  \BibitemOpen
  \bibfield  {author} {\bibinfo {author} {\bibfnamefont {C.}~\bibnamefont
  {Bergemann}}, \bibinfo {author} {\bibfnamefont {A.~P.}\ \bibnamefont
  {Mackenzie}}, \bibinfo {author} {\bibfnamefont {S.~R.}\ \bibnamefont
  {Julian}}, \bibinfo {author} {\bibfnamefont {D.}~\bibnamefont {Forsythe}},\
  and\ \bibinfo {author} {\bibfnamefont {E.}~\bibnamefont {Ohmichi}},\
  }\bibfield  {title} {\bibinfo {title} {{Quasi-two-dimensional Fermi liquid
  properties of the unconventional superconductor Sr$_2$RuO$_4$}},\ }\href
  {https://doi.org/10.1080/00018730310001621737} {\bibfield  {journal}
  {\bibinfo  {journal} {Advances in Physics}\ }\textbf {\bibinfo {volume}
  {52}},\ \bibinfo {pages} {639} (\bibinfo {year} {2003})}\BibitemShut
  {NoStop}%
\bibitem [{\citenamefont {Barber}\ \emph {et~al.}(2018)\citenamefont {Barber},
  \citenamefont {Gibbs}, \citenamefont {Maeno}, \citenamefont {Mackenzie},\
  and\ \citenamefont {Hicks}}]{Barber2018}%
  \BibitemOpen
  \bibfield  {author} {\bibinfo {author} {\bibfnamefont {M.~E.}\ \bibnamefont
  {Barber}}, \bibinfo {author} {\bibfnamefont {A.~S.}\ \bibnamefont {Gibbs}},
  \bibinfo {author} {\bibfnamefont {Y.}~\bibnamefont {Maeno}}, \bibinfo
  {author} {\bibfnamefont {A.~P.}\ \bibnamefont {Mackenzie}},\ and\ \bibinfo
  {author} {\bibfnamefont {C.~W.}\ \bibnamefont {Hicks}},\ }\bibfield  {title}
  {\bibinfo {title} {{Resistivity in the Vicinity of a van Hove Singularity:
  Sr$_2$RuO$_4$ under Uniaxial Pressure}},\ }\href
  {https://doi.org/10.1103/PhysRevLett.120.076602} {\bibfield  {journal}
  {\bibinfo  {journal} {Physical Review Letters}\ }\textbf {\bibinfo {volume}
  {120}},\ \bibinfo {pages} {76602} (\bibinfo {year} {2018})}\BibitemShut
  {NoStop}%
\bibitem [{\citenamefont {Jerzembeck}\ \emph {et~al.}(2022)\citenamefont
  {Jerzembeck}, \citenamefont {R{\o}ising}, \citenamefont {Steppke},
  \citenamefont {Rosner}, \citenamefont {Sokolov}, \citenamefont {Kikugawa},
  \citenamefont {Scaffidi}, \citenamefont {Simon}, \citenamefont {Mackenzie},\
  and\ \citenamefont {Hicks}}]{Jerzembeck2022}%
  \BibitemOpen
  \bibfield  {author} {\bibinfo {author} {\bibfnamefont {F.}~\bibnamefont
  {Jerzembeck}}, \bibinfo {author} {\bibfnamefont {H.~S.}\ \bibnamefont
  {R{\o}ising}}, \bibinfo {author} {\bibfnamefont {A.}~\bibnamefont {Steppke}},
  \bibinfo {author} {\bibfnamefont {H.}~\bibnamefont {Rosner}}, \bibinfo
  {author} {\bibfnamefont {D.~A.}\ \bibnamefont {Sokolov}}, \bibinfo {author}
  {\bibfnamefont {N.}~\bibnamefont {Kikugawa}}, \bibinfo {author}
  {\bibfnamefont {T.}~\bibnamefont {Scaffidi}}, \bibinfo {author}
  {\bibfnamefont {S.~H.}\ \bibnamefont {Simon}}, \bibinfo {author}
  {\bibfnamefont {A.~P.}\ \bibnamefont {Mackenzie}},\ and\ \bibinfo {author}
  {\bibfnamefont {C.~W.}\ \bibnamefont {Hicks}},\ }\bibfield  {title} {\bibinfo
  {title} {{The superconductivity of Sr$_2$RuO$_4$ under $c$-axis uniaxial
  stress}},\ }\href {https://doi.org/10.1038/s41467-022-32177-4} {\bibfield
  {journal} {\bibinfo  {journal} {Nature Communications}\ }\textbf {\bibinfo
  {volume} {13}},\ \bibinfo {pages} {4596} (\bibinfo {year}
  {2022})}\BibitemShut {NoStop}%
\bibitem [{\citenamefont {Broun}(2000)}]{Broun2000}%
  \BibitemOpen
  \bibfield  {author} {\bibinfo {author} {\bibfnamefont {D.~M.}\ \bibnamefont
  {Broun}},\ }\emph {\bibinfo {title} {{The Microwave Electrodynamics of
  Unconventional Superconductors}}},\ \href@noop {} {Ph.D. thesis} (\bibinfo
  {year} {2000})\BibitemShut {NoStop}%
\bibitem [{Note2()}]{Note2}%
  \BibitemOpen
  \bibinfo {note} {In \protect \cref {sec:corrections} only, the formulae
  implicitly reflect a sign choice of $\exp (+i\omega t)$ time dependence so as
  to match Ref. \cite {Broun2000}. Everywhere else, the formulae implicitly
  reflect a sign choice of $\exp (-i\omega t)$ time dependence.}\BibitemShut
  {Stop}%
\bibitem [{\citenamefont {Haug}(2019)}]{Haug2019}%
  \BibitemOpen
  \bibfield  {author} {\bibinfo {author} {\bibfnamefont {A.}~\bibnamefont
  {Haug}},\ }\emph {\bibinfo {title} {{Microwave response of the heavy-fermion
  superconductor CeCu$_2$Si$_2$}}},\ \href@noop {} {Ph.D. thesis} (\bibinfo
  {year} {2019})\BibitemShut {NoStop}%
\bibitem [{\citenamefont {Reuter}\ and\ \citenamefont
  {Sondheimer}(1948)}]{Reuter1948}%
  \BibitemOpen
  \bibfield  {author} {\bibinfo {author} {\bibfnamefont {G.~E.~H.}\
  \bibnamefont {Reuter}}\ and\ \bibinfo {author} {\bibfnamefont {E.~H.}\
  \bibnamefont {Sondheimer}},\ }\bibfield  {title} {\bibinfo {title} {{The
  theory of the anomalous skin effect in metals}},\ }\href
  {https://doi.org/10.1098/rspa.1948.0123} {\bibfield  {journal} {\bibinfo
  {journal} {Proceedings of the Royal Society A}\ }\textbf {\bibinfo {volume}
  {195}},\ \bibinfo {pages} {336} (\bibinfo {year} {1948})}\BibitemShut
  {NoStop}%
\bibitem [{\citenamefont {Sondheimer}(1954)}]{Sondheimer1954}%
  \BibitemOpen
  \bibfield  {author} {\bibinfo {author} {\bibfnamefont {E.~H.}\ \bibnamefont
  {Sondheimer}},\ }\bibfield  {title} {\bibinfo {title} {{The Theory of the
  Anomalous Skin Effect in Anisotropic Metals}},\ }\href
  {https://doi.org/https://doi.org/10.1098/rspa.1954.0156} {\bibfield
  {journal} {\bibinfo  {journal} {Proceedings of the Royal Society A}\ }\textbf
  {\bibinfo {volume} {224}},\ \bibinfo {pages} {260} (\bibinfo {year}
  {1954})}\BibitemShut {NoStop}%
\bibitem [{\citenamefont {Mackenzie}\ \emph {et~al.}(1996)\citenamefont
  {Mackenzie}, \citenamefont {Julian}, \citenamefont {Diver}, \citenamefont
  {Lonzarich}, \citenamefont {Hussey}, \citenamefont {Maeno}, \citenamefont
  {Nishizaki},\ and\ \citenamefont {Fujita}}]{Mackenzie1996}%
  \BibitemOpen
  \bibfield  {author} {\bibinfo {author} {\bibfnamefont {A.~P.}\ \bibnamefont
  {Mackenzie}}, \bibinfo {author} {\bibfnamefont {S.~R.}\ \bibnamefont
  {Julian}}, \bibinfo {author} {\bibfnamefont {A.~J.}\ \bibnamefont {Diver}},
  \bibinfo {author} {\bibfnamefont {G.~G.}\ \bibnamefont {Lonzarich}}, \bibinfo
  {author} {\bibfnamefont {N.~E.}\ \bibnamefont {Hussey}}, \bibinfo {author}
  {\bibfnamefont {Y.}~\bibnamefont {Maeno}}, \bibinfo {author} {\bibfnamefont
  {S.}~\bibnamefont {Nishizaki}},\ and\ \bibinfo {author} {\bibfnamefont
  {T.}~\bibnamefont {Fujita}},\ }\bibfield  {title} {\bibinfo {title}
  {{Calculation of thermodynamic and transport properties of Sr2RuO4 at low
  temperatures using known Fermi surface parameters}},\ }\href
  {https://doi.org/10.1016/0921-4534(95)00770-9} {\bibfield  {journal}
  {\bibinfo  {journal} {Physica C: Superconductivity and its Applications}\
  }\textbf {\bibinfo {volume} {263}},\ \bibinfo {pages} {510} (\bibinfo {year}
  {1996})}\BibitemShut {NoStop}%
\bibitem [{\citenamefont {Pippard}(1947)}]{Pippard1947}%
  \BibitemOpen
  \bibfield  {author} {\bibinfo {author} {\bibfnamefont {A.}~\bibnamefont
  {Pippard}},\ }\bibfield  {title} {\bibinfo {title} {{The surface impedance of
  superconductors and normal metals at high frequencies II. The anomalous skin
  effect in normal metals}},\ }\href {https://doi.org/10.1098/rspa.1947.0122}
  {\bibfield  {journal} {\bibinfo  {journal} {Proceedings of the Royal Society
  A}\ }\textbf {\bibinfo {volume} {191}},\ \bibinfo {pages} {385} (\bibinfo
  {year} {1947})}\BibitemShut {NoStop}%
\bibitem [{\citenamefont {Prozorov}\ and\ \citenamefont
  {Kogan}(2018)}]{Prozorov2018}%
  \BibitemOpen
  \bibfield  {author} {\bibinfo {author} {\bibfnamefont {R.}~\bibnamefont
  {Prozorov}}\ and\ \bibinfo {author} {\bibfnamefont {V.~G.}\ \bibnamefont
  {Kogan}},\ }\bibfield  {title} {\bibinfo {title} {{Effective Demagnetizing
  Factors of Diamagnetic Samples of Various Shapes}},\ }\href
  {https://doi.org/10.1103/PhysRevApplied.10.014030} {\bibfield  {journal}
  {\bibinfo  {journal} {Physical Review Applied}\ }\textbf {\bibinfo {volume}
  {10}},\ \bibinfo {pages} {14030} (\bibinfo {year} {2018})}\BibitemShut
  {NoStop}%
\bibitem [{\citenamefont {Fawzi}\ \emph {et~al.}(1983)\citenamefont {Fawzi},
  \citenamefont {Ali},\ and\ \citenamefont {Burke}}]{Fawzi1983}%
  \BibitemOpen
  \bibfield  {author} {\bibinfo {author} {\bibfnamefont {T.~H.}\ \bibnamefont
  {Fawzi}}, \bibinfo {author} {\bibfnamefont {K.~F.}\ \bibnamefont {Ali}},\
  and\ \bibinfo {author} {\bibfnamefont {P.~E.}\ \bibnamefont {Burke}},\
  }\bibfield  {title} {\bibinfo {title} {{Eddy current losses in finite length
  conducting cylinders}},\ }\href {https://doi.org/10.1109/TMAG.1983.1062759}
  {\bibfield  {journal} {\bibinfo  {journal} {IEEE Transactions on Magnetics}\
  }\textbf {\bibinfo {volume} {19}},\ \bibinfo {pages} {2216} (\bibinfo {year}
  {1983})}\BibitemShut {NoStop}%
\bibitem [{\citenamefont {Landau}\ and\ \citenamefont
  {Lifshitz}(1984)}]{Landau1984}%
  \BibitemOpen
  \bibfield  {author} {\bibinfo {author} {\bibfnamefont {L.~D.}\ \bibnamefont
  {Landau}}\ and\ \bibinfo {author} {\bibfnamefont {E.~M.}\ \bibnamefont
  {Lifshitz}},\ }\bibfield  {title} {\bibinfo {title} {{Chapter VII}},\ }in\
  \href {https://doi.org/10.1016/B978-0-08-030275-1.50013-8} {\emph {\bibinfo
  {booktitle} {Electrodynamics of Continuous Media}}}\ (\bibinfo  {publisher}
  {Pergamon},\ \bibinfo {year} {1984})\ \bibinfo {edition} {2nd}\ ed.,\ Chap.\
  \bibinfo {chapter} {VII}, pp.\ \bibinfo {pages} {199--224}\BibitemShut
  {NoStop}%
\bibitem [{\citenamefont {Hicks}\ \emph {et~al.}(2012)\citenamefont {Hicks},
  \citenamefont {Gibbs}, \citenamefont {Mackenzie}, \citenamefont {Takatsu},
  \citenamefont {Maeno},\ and\ \citenamefont {Yelland}}]{Hicks2012}%
  \BibitemOpen
  \bibfield  {author} {\bibinfo {author} {\bibfnamefont {C.~W.}\ \bibnamefont
  {Hicks}}, \bibinfo {author} {\bibfnamefont {A.~S.}\ \bibnamefont {Gibbs}},
  \bibinfo {author} {\bibfnamefont {A.~P.}\ \bibnamefont {Mackenzie}}, \bibinfo
  {author} {\bibfnamefont {H.}~\bibnamefont {Takatsu}}, \bibinfo {author}
  {\bibfnamefont {Y.}~\bibnamefont {Maeno}},\ and\ \bibinfo {author}
  {\bibfnamefont {E.~A.}\ \bibnamefont {Yelland}},\ }\bibfield  {title}
  {\bibinfo {title} {{Quantum Oscillations and High Carrier Mobility in the
  Delafossite PdCoO$_2$}},\ }\href
  {https://doi.org/10.1103/PhysRevLett.109.116401} {\bibfield  {journal}
  {\bibinfo  {journal} {Physical Review Letters}\ }\textbf {\bibinfo {volume}
  {109}},\ \bibinfo {pages} {116401} (\bibinfo {year} {2012})}\BibitemShut
  {NoStop}%
\bibitem [{\citenamefont {Baker}(2022)}]{Baker2022}%
  \BibitemOpen
  \bibfield  {author} {\bibinfo {author} {\bibfnamefont {G.}~\bibnamefont
  {Baker}},\ }\emph {\bibinfo {title} {{Non-local electrical conductivity in
  PdCoO$_2$}}},\ \href@noop {} {Ph.D. thesis},\ \bibinfo  {school} {University
  of British Columbia} (\bibinfo {year} {2022})\BibitemShut {NoStop}%
\bibitem [{\citenamefont {Branch}(2021)}]{Branch2021}%
  \BibitemOpen
  \bibfield  {author} {\bibinfo {author} {\bibfnamefont {T.~W.}\ \bibnamefont
  {Branch}},\ }\emph {\bibinfo {title} {{Microwave Responses of Strongly
  Demagnetized Metallic Samples}}},\ \href@noop {} {Master's thesis},\ \bibinfo
   {school} {University of British Columbia} (\bibinfo {year}
  {2021})\BibitemShut {NoStop}%
\bibitem [{\citenamefont {Nandi}\ \emph {et~al.}(2018)\citenamefont {Nandi},
  \citenamefont {Scaffidi}, \citenamefont {Kushwaha}, \citenamefont {Khim},
  \citenamefont {Barber}, \citenamefont {Sunko}, \citenamefont {Mazzola},
  \citenamefont {King}, \citenamefont {Rosner}, \citenamefont {Moll},
  \citenamefont {K{\"{o}}nig}, \citenamefont {Moore}, \citenamefont
  {Hartnoll},\ and\ \citenamefont {Mackenzie}}]{Nandi2018}%
  \BibitemOpen
  \bibfield  {author} {\bibinfo {author} {\bibfnamefont {N.}~\bibnamefont
  {Nandi}}, \bibinfo {author} {\bibfnamefont {T.}~\bibnamefont {Scaffidi}},
  \bibinfo {author} {\bibfnamefont {P.}~\bibnamefont {Kushwaha}}, \bibinfo
  {author} {\bibfnamefont {S.}~\bibnamefont {Khim}}, \bibinfo {author}
  {\bibfnamefont {M.~E.}\ \bibnamefont {Barber}}, \bibinfo {author}
  {\bibfnamefont {V.}~\bibnamefont {Sunko}}, \bibinfo {author} {\bibfnamefont
  {F.}~\bibnamefont {Mazzola}}, \bibinfo {author} {\bibfnamefont {P.~D.~C.}\
  \bibnamefont {King}}, \bibinfo {author} {\bibfnamefont {H.}~\bibnamefont
  {Rosner}}, \bibinfo {author} {\bibfnamefont {P.~J.~W.}\ \bibnamefont {Moll}},
  \bibinfo {author} {\bibfnamefont {M.}~\bibnamefont {K{\"{o}}nig}}, \bibinfo
  {author} {\bibfnamefont {J.~E.}\ \bibnamefont {Moore}}, \bibinfo {author}
  {\bibfnamefont {S.}~\bibnamefont {Hartnoll}},\ and\ \bibinfo {author}
  {\bibfnamefont {A.~P.}\ \bibnamefont {Mackenzie}},\ }\bibfield  {title}
  {\bibinfo {title} {{Unconventional magneto-transport in ultrapure PdCoO$_2$
  and PtCoO$_2$}},\ }\href {https://doi.org/10.1038/s41535-018-0138-8}
  {\bibfield  {journal} {\bibinfo  {journal} {npj Quantum Materials}\ }\textbf
  {\bibinfo {volume} {3}},\ \bibinfo {pages} {66} (\bibinfo {year}
  {2018})}\BibitemShut {NoStop}%
\bibitem [{\citenamefont {Dingle}(1953)}]{Dingle1953}%
  \BibitemOpen
  \bibfield  {author} {\bibinfo {author} {\bibfnamefont {R.}~\bibnamefont
  {Dingle}},\ }\bibfield  {title} {\bibinfo {title} {{The anomalous skin effect
  and the reflectivity of metals I.}},\ }\href
  {https://doi.org/10.1016/S0031-8914(54)80042-5} {\bibfield  {journal}
  {\bibinfo  {journal} {Physica}\ }\textbf {\bibinfo {volume} {19}},\ \bibinfo
  {pages} {311} (\bibinfo {year} {1953})}\BibitemShut {NoStop}%
\bibitem [{\citenamefont {Takatsu}\ \emph {et~al.}(2007)\citenamefont
  {Takatsu}, \citenamefont {Yonezawa}, \citenamefont {Mouri}, \citenamefont
  {Nakatsuji}, \citenamefont {Tanaka},\ and\ \citenamefont
  {Maeno}}]{Takatsu2007}%
  \BibitemOpen
  \bibfield  {author} {\bibinfo {author} {\bibfnamefont {H.}~\bibnamefont
  {Takatsu}}, \bibinfo {author} {\bibfnamefont {S.}~\bibnamefont {Yonezawa}},
  \bibinfo {author} {\bibfnamefont {S.}~\bibnamefont {Mouri}}, \bibinfo
  {author} {\bibfnamefont {S.}~\bibnamefont {Nakatsuji}}, \bibinfo {author}
  {\bibfnamefont {K.}~\bibnamefont {Tanaka}},\ and\ \bibinfo {author}
  {\bibfnamefont {Y.}~\bibnamefont {Maeno}},\ }\bibfield  {title} {\bibinfo
  {title} {{Roles of high-frequency optical phonons in the physical properties
  of the conductive delafossite PdCoO$_2$}},\ }\href
  {https://doi.org/10.1143/JPSJ.76.104701} {\bibfield  {journal} {\bibinfo
  {journal} {Journal of the Physical Society of Japan}\ }\textbf {\bibinfo
  {volume} {76}},\ \bibinfo {pages} {104701} (\bibinfo {year}
  {2007})}\BibitemShut {NoStop}%
\bibitem [{\citenamefont {Levchenko}\ and\ \citenamefont
  {Schmalian}(2020)}]{Levchenko2020}%
  \BibitemOpen
  \bibfield  {author} {\bibinfo {author} {\bibfnamefont {A.}~\bibnamefont
  {Levchenko}}\ and\ \bibinfo {author} {\bibfnamefont {J.}~\bibnamefont
  {Schmalian}},\ }\bibfield  {title} {\bibinfo {title} {{Transport properties
  of strongly coupled electron–phonon liquids}},\ }\href
  {https://doi.org/10.1016/j.aop.2020.168218} {\bibfield  {journal} {\bibinfo
  {journal} {Annals of Physics}\ }\textbf {\bibinfo {volume} {419}},\ \bibinfo
  {pages} {168218} (\bibinfo {year} {2020})}\BibitemShut {NoStop}%
\end{thebibliography}%

\end{document}

\title{
    \texorpdfstring
        {
            Supplemental material:\\
            Non-local electrodynamics in ultra-pure PdCoO$_2$
        }
        {
            Supplemental material: Non-local electrodynamics in ultra-pure PdCoO2
        }
}
\date{\today}
\maketitle

\tableofcontents

\section{\texorpdfstring{S\lowercase{r}$_2$R\lowercase{u}O$_4$}{Sr2RuO4} preparation \& analysis} 

\subsection{Preparation}

The Sr$_2$RuO$_4$ sample was grown in an image furnace by the floating-zone method \cite{Mao2000,Bobowski2019}. The as-grown crystals are typically $5$ to $8\rm\ cm$ long rods and $2\times 3\rm\ mm^2$ in cross-section with the crystallographic $c$ axis oriented perpendicular to the long axis of the rod.  Samples were prepared by first cutting a $1$ to $2\rm\ mm$ section from the as-grown rod using a diamond saw. Platelet samples, approximately $50\rm\ \mu m$ thick, were then cleaved from the cut section such that the broad faces were aligned with the $ab$ planes. 

The highest-quality Sr$_2$RuO$_4$ single crystals have a sharp superconducting transition with a maximum transition temperature of $T_\mathrm{c} = 1.50\rm\ K$. AC susceptibility measurements are used to characterize the superconducting transitions of Sr$_2$RuO$_4$ crystals and gauge the sample quality \cite{Yonezawa2015,Bidinosti2000}. The sample used in this study had a $25\rm\ mK$ wide superconducting transition with $T_\mathrm{c} = 1.41\rm\ K$. The sharpness of the transition is a clear indication of the high-purity of the sample. The suboptimal $T_\mathrm{c}$ is likely due to a nonstoichiometric Ru content in the platelet sample. 

During the floating-zone growth, some RuO$_2$ is lost to evaporation such that excess RuO$_2$ is required when preparing the the initial feed rod. On the other hand, during the crystal growth, excessive RuO$_2$ leads to Ru segregation and the formation of a eutectic phase that causes internal strain and an enhanced $T_\mathrm{c}$ (the so-called 3-K phase) \cite{Maeno1998,Hicks2014}. It is, therefore, necessary to balance these two effects when preparing the initial feed rod. For the floating-zone growth conditions used at Kyoto University, 15\% excess RuO$_2$ in the feed rod has been found to be near optimal \cite{Mao2000,Bobowski2019}. The sample used in this study did not show any signs of the 3-K phase.

\subsection{Classical Skin Effect fit}

\begin{figure}[!htbp]
    \includegraphics{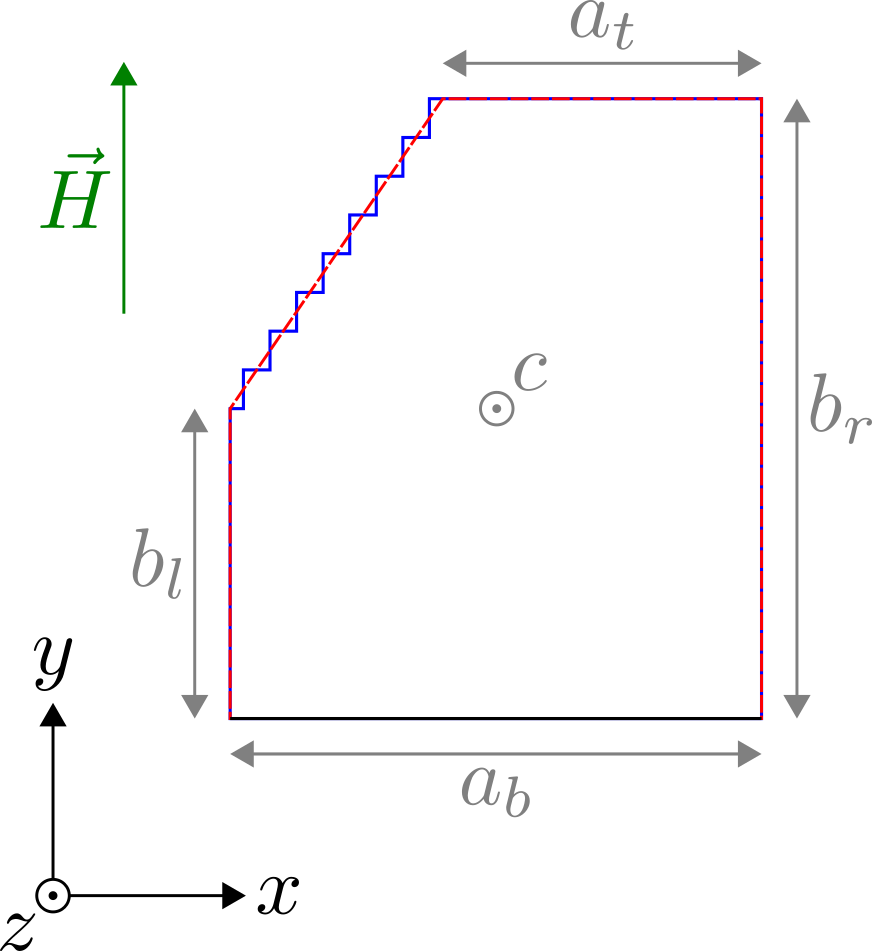}
    \caption{Geometry of Sr$_2$RuO$_4$ sample and alignment of microwave-frequency magnetic field. The idealized sample geometry is shown in red. The discretized sample geometry, as used in the extended fit of \cref{sec:corrections}, is shown in blue. The sample dimensions are given in \cref{tab:sr_dims}.}
    \label{fig:sr_dims}
\end{figure}

The measured data were fit to
\begin{equation}
    R=A\omega^{1/2}
\end{equation}
yielding a value of $A=\SI{7.5e-8}{\ohm\per\hertz\tothe{1/2}}$.
In our measurement, the microwave-frequency magnetic field was applied parallel to the $ab$ plane of a platelet sample, inducing current both in the $ab$ plane and along the $c$ axis of Sr$_2$RuO$_4$'s tetragonal crystal structure. When the skin depth is small compared to sample dimensions, the measured surface resistance can be expressed as
\begin{equation}
    R=f_{ab}R_{ab}+f_{c}R_{c}
\end{equation}
where $f_{i}=A_{i}/\sum_{j}A_{j}$ and $A_{i}$ is the area of faces in which currents run in the $i$ direction. Our sample dimensions, given in \cref{fig:sr_dims} and \cref{tab:sr_dims}, were such that current ran predominantly in the $ab$ plane, with $f_{ab}=0.94$ and $f_{c}=0.06$ \footnote{We took $A_{ab}=2(a_{b}b_{r}-(b_{r}-b_{l})(a_{b}-a_{t})/2)$ and $A_{c}=(b_{l}+\sqrt{(b_r-b_l)^2+(a_b-a_t)^2}+b_{r})c$.}. 
In the \gls{cse}, surface resistance for current in the $i$ direction is given by 
\begin{equation}
    R_{i}=\sqrt{\frac{\mu_{0}\omega\rho_{i}}{2}}
\end{equation}
where $\rho_{i}$ is the corresponding DC resistivity. 
Defining the resistivity anisotropy as $\alpha\equiv\rho_{c}/\rho_{ab}$ gives
\begin{equation}
    R=(f_{ab}+\sqrt{\alpha}f_{c})R_{ab}
\end{equation}
such that
\begin{equation}\label{eq:rho_ab}
    \rho_{ab}=\frac{2}{\mu_{0}}\left(\frac{A}{f_{ab}+\sqrt{\alpha}f_{c}}\right)^{2} .
\end{equation}
To determine $\rho_{ab}$ from our fit, we must assume a value for $\alpha$. 
Reported values of $\alpha$ in the literature range from 400 to 4000 \cite{Bergemann2003}, yielding a range from 38 to \SI{185}{\nano\ohm\centi\metre} for the $\rho_{ab}$ value of our sample. 
Recently reported residual resistivity values are \SI{100}{\nano\ohm\centi\metre} for $\rho_{ab}$ \cite{Barber2018} and 0.228 and \SI{0.278}{\milli\ohm\centi\metre} for $\rho_{c}$ \cite{Jerzembeck2022}, giving $\alpha=2280$  and 2780. Using these values of $\alpha$ in \cref{eq:rho_ab} gives $\rho_{ab}=$ 59 and \SI{50}{\nano\ohm\centi\metre}, leading to our reported value of approximately \SI{50}{\nano\ohm\centi\metre} in the main text. 

\begin{table}[!htbp]
    \centering
    \begin{tabular}{p{2cm}p{2cm}}
        \toprule
        Dimension & Value [\si{\micro\metre}] \\
        \midrule
        $a_{b}$ & 900 \\
        $a_{t}$ & 540 \\
        $b_{l}$ & 525 \\
        $b_{r}$ & 1050 \\
        $c$ & 46 \\
        \bottomrule
    \end{tabular}
    \caption{Dimensions of Sr$_2$RuO$_4$ sample. Labels refer to \cref{fig:sr_dims}.}
    \label{tab:sr_dims}
\end{table}

\subsection{Corrections to $\sqrt{\omega}$ behavior}\label{sec:corrections}

The Sr$_2$RuO$_4$ data exhibit small yet systematic deviations from $\sqrt{\omega}$ power-law behavior. Here, we show that these deviations are well-described by corrections coming from two effects: (1) the electromagnetic finite size effect---an extrinsic effect due to sample geometry---and (2) proximity to the relaxation regime. Both of these corrections fall within the scope of conventional, local electrodynamics, and do not change our interpretation of the data as arising from diffusive electron dynamics within the skin layer.

Here, we develop a model to account for these corrections in our Sr$_2$RuO$_4$ sample. To lay the groundwork for describing the full model, we start by reviewing the known solution for a simpler geometry. We consider an infinitely-long anisotropic metal with rectangular cross-section with dimensions $a$ and $c$. We consider a spatially-uniform, AC magnetic field with magnitude $H_{0}$ which is applied along the infinite dimension of the sample. We consider an effective surface impedance, defined so that the average power absorbed per unit area is $Z_{\text{eff}}H_{0}^{2}/2$. Assuming local electrodynamics, the effective surface impedance is given by \cite{Broun2000}\footnote{In \cref{sec:corrections} only, the formulae implicitly reflect a sign choice of $\exp(+i\omega t)$ time dependence so as to match Ref. \cite{Broun2000}. Everywhere else, the formulae implicitly reflect a sign choice of $\exp(-i\omega t)$ time dependence.}
\begin{equation}
    Z_{\text{eff}}^{\text{rect}}(a,c)
    =i\mu_{0}\omega\tilde{\delta}_{\text{eff}}
\end{equation}
with
\begin{equation}
    \tilde\delta_{\text{eff}}
    =\frac{ac}{a+c}\frac{4}{\pi^{2}}
    \sum_{\text{odd }n>0}\frac{1}{n^{2}}\left(
        \frac{\tanh\alpha_{n}}{\alpha_{n}}
        +\frac{\tanh\beta_{n}}{\beta_{n}}
    \right)
\end{equation}
where
\begin{equation}
    \alpha_{n}
    =\frac{c}{2\tilde{\delta}_{a}}
    \left[
        1+\left(\frac{n\pi\tilde{\delta}_{c}}{a}\right)^{2}
    \right]^{1/2}
\end{equation}
and
\begin{equation}
    \beta_{n}
    =\frac{a}{2\tilde{\delta}_{c}}
    \left[
        1+\left(\frac{n\pi\tilde{\delta}_{a}}{c}\right)^{2}
    \right]^{1/2} .
\end{equation}
The complex skin depth for current flowing along direction $i$ is given by 
\begin{equation}
    \tilde{\delta}_{i}
    =\sqrt{\frac{1}{i\mu_{0}\omega\sigma_{i}(\omega)}}
\end{equation}
where $\sigma_{i}(\omega)$ is the frequency-dependent conductivity in that direction.

For our Sr$_2$RuO$_4$ sample, the dimension along the $x$ axis is a function of the $z$ coordinate (see \cref{fig:sr_dims}). We may approximate the effective surface impedance of our sample by discretizing its geometry in terms of slices of rectangular cross-section \cite{Haug2019}. Assuming that the tangential magnetic field has a magnitude of $H_{0}$ at each slice, the effective surface impedance is then given by
\begin{equation}
    Z_{\text{eff}}
    =\sum_{i}Z_{\text{eff}}^{\text{rect}}(a_{i},c)\cdot\frac{b_{i}}{\sum_{i}b_{i}}
\end{equation}
where the factor $b_{i}/\sum_{i}b_{i}$ weights the contribution of each slice by its length along the $y$ axis.
To capture relaxation effects, we assume a Drude conductivity
\begin{equation}
    \sigma_{i}^{\text{Drude}}
    =\frac{1/\rho_{i}}{1+
    i\omega/\gamma_{\text{mr},i}}
\end{equation}
and, for simplicity, take $\gamma_{\text{mr},ab}=\gamma_{\text{mr},c}\equiv\gamma_{\text{mr}}$. 

\begin{figure}[!htbp]
    \centering
    \includegraphics{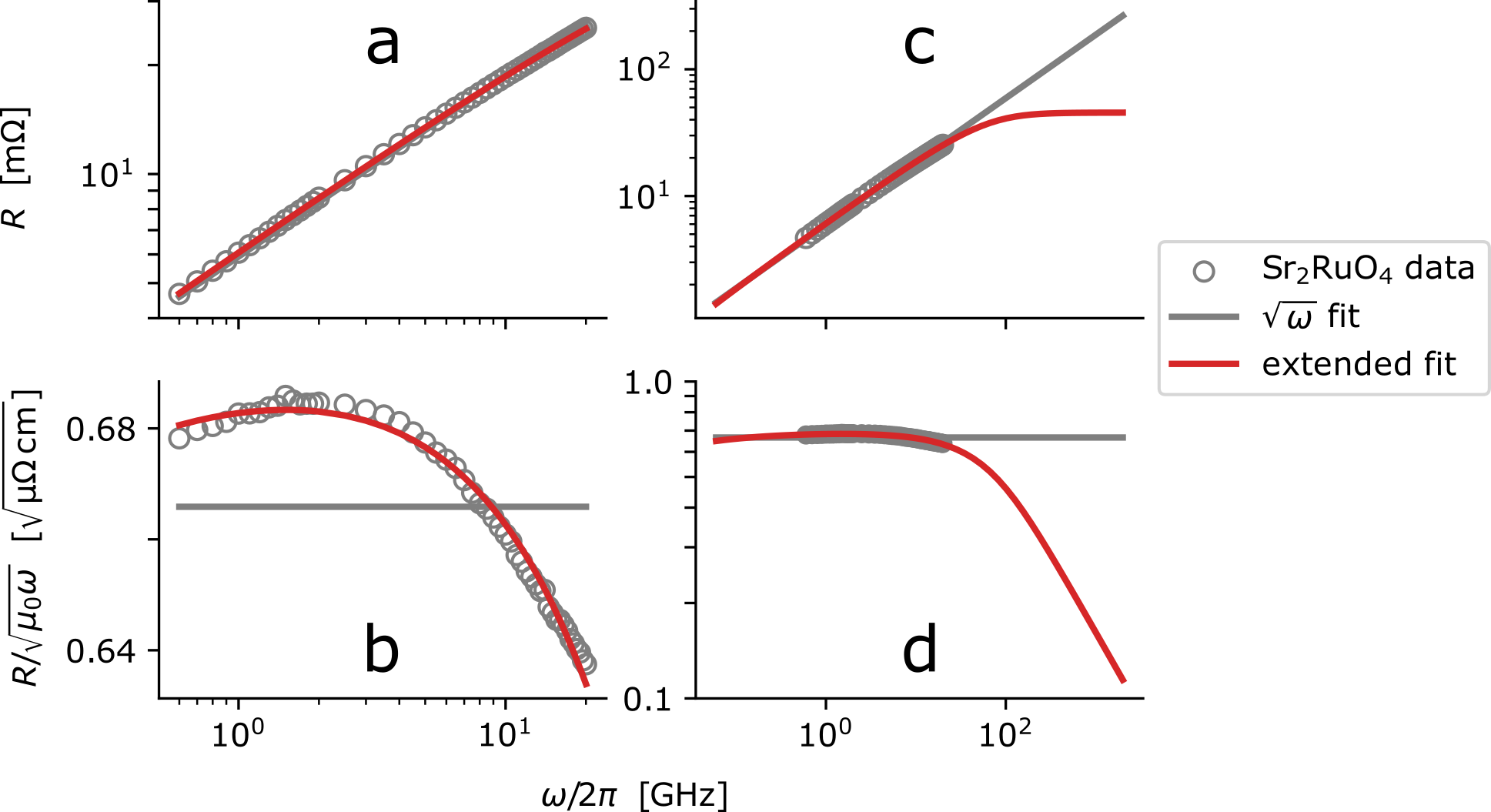}
    \caption{Fit of Sr$_2$RuO$_4$ to the extended surface resistance model, including corrections to $\sqrt{\omega}$ behavior. (a) The $\sqrt{\omega}$ and extended-model fits are difficult to distinguish on a log-log plot of surface resistance $R$ versus frequency $\omega$. (b) Upon dividing $R$ by $\sqrt{\omega}$, it can be seen that the extended fit is superior. (c), (d) Examining the extended fit over a wider frequency range shows that, despite the correction to $\sqrt{\omega}$ behavior stemming from relaxation effects, the data are below the frequency regime where these effects dominate the behavior of the surface resistance.}
    \label{fig:sr_full_fit}
\end{figure}

Using the above model, we performed a fit to the Sr$_2$RuO$_4$ surface resistance data with three free parameters: $\rho_{ab}$, $\rho_{c}$, and $\gamma_{\text{mr}}$. The number of slices was increased until the fit parameters converged; this was achieved by nine slices, as shown in \cref{fig:sr_dims}. The fit is shown in \cref{fig:sr_full_fit} and fit parameters are given in \cref{tab:sr_full_fit}. 
Both the $\sqrt{\omega}$ and extended fit yield similar values of $\rho_{ab}$.
While the difference between the two fits is difficult to distinguish on a log-log plot (\cref{fig:sr_full_fit}(a)), the superiority of the extended fit becomes visible upon dividing the data by $\sqrt{\omega}$ (\cref{fig:sr_full_fit}(b)). Even so, the data deviate from the $\sqrt{\omega}$ fit by less than $\pm5\%$. \Cref{fig:sr_full_fit}(c) \& (d) illustrate that, while there is a correction to $\sqrt{\omega}$ behavior owing to proximity to a crossover to the relaxation regime, the data are nonetheless below the frequency at which the relaxation regime is entered.

\begin{table}[!htbp]
    \centering
    \begin{tabular}{p{2cm}p{2cm}}
        \toprule
        Parameter & Value \\
        \midrule
        $\rho_{ab}$ & \SI{64.2}{\nano\ohm\centi\metre} \\
        $\rho_{c}$ & \SI{99.8}{\micro\ohm\centi\metre} \\
        $\gamma_{\text{mr}}$ & \SI{680}{\giga\hertz} \\
        \bottomrule
    \end{tabular}
    \caption{Best-fit parameters for the fit of the Sr$_2$RuO$_4$ data to the extended model.}
    \label{tab:sr_full_fit}
\end{table}

\subsection{Verification of expectation of Classical Skin Effect}

We wish to confirm that, according to conventional theory of the \gls{cse} and \gls{ase} \cite{Reuter1948,Sondheimer1954}, we should expect to observe the \gls{cse}. We will do so by determining the expected location of the crossover to the \gls{ase}. We note that we have already empirically determined the location of the expected crossover to the relaxation regime in \cref{sec:corrections}; from our fit we have that $\omega_{c\leftrightarrow r}/2\pi=\gamma_{\text{mr}}/2\pi=\SI{108}{\giga\hertz}$, which is indeed outside of our measurement range.  

Because Sr$_2$RuO$_4$ is tetragonal, the appropriate pre-existing model of the \gls{ase} is that for spheroidal Fermi surfaces within the relaxation time approximation, as considered by \citet{Sondheimer1954} in a follow-up to the original work on spherical Fermi surfaces by \citet{Reuter1948}. For an electronic dispersion given by 
\begin{equation}
    \mathcal{E}_{\bm{k}}
    =\frac{\hbar^{2}}{2m_{a}}(k_{x}^{2}+k_{y}^{2})
    +\frac{\hbar^{2}}{2m_{c}}k_{z}^{2} ,
\end{equation}
the classical surface impedance is
\begin{equation}\label{eq:Zc_sph}
    Z_{i}^{c}
    =\left[
        \frac{1}{\sigma_{0,ii}}
        \mu_{0}\omega
    \right]^{1/2}e^{-i\pi/4}
    =\left[
        \frac{1}{\epsilon_{0}\omega_{p,ii}^{2}\tau}
        \mu_{0}\omega
    \right]^{1/2}e^{-i\pi/4}
\end{equation}
and the anomalous surface impedance is
\begin{equation}\label{eq:Za_sph}
    Z_{i,j}^{a}
    =\beta\left[
        \frac{\sqrt{3}}{2\pi}
        \frac{l_{j}}{\sigma_{0,ii}}
        (\mu_{0}\omega^{2})
    \right]^{1/3}e^{-i\pi/3}
    =\beta\left[
        \frac{\sqrt{3}}{2\pi}
        \frac{v_{j}}{\epsilon_{0}\omega_{p,ii}^{2}}
        (\mu_{0}\omega^{2})
    \right]^{1/3}e^{-i\pi/3}
\end{equation}
where the following definitions have been used: 
\begin{equation}
    \sigma_{0,ii}
    =\epsilon_{0}\omega_{p,ii}^{2}\tau
\end{equation}
\begin{equation}
    \epsilon_{0}\omega_{p,ii}^{2}
    =\frac{ne^{2}}{m_{i}}
    =\frac{k_{a}^{2}k_{c}}{3\pi^{2}}
    \frac{e^{2}}{m_{i}}
\end{equation}
\begin{equation}
    k_{i}^{2}=\frac{2m_{i}\mathcal{E}_{F}}{\hbar^{2}}
\end{equation}
\begin{equation}
    v_{i}
    =\frac{\hbar k_{i}}{m_{i}}
    =\sqrt{\frac{2\mathcal{E}_{F}}{m_{i}}}
\end{equation}
\begin{equation}
    l_{i}=v_{i}\tau .
\end{equation}
We see that the resistivity anisotropy is related to the various quantities of interest via
\begin{equation}\label{eq:sr_anisotropy}
    \alpha
    \equiv\frac{\rho_{c}}{\rho_{a}}
    =\frac{\sigma_{0,aa}}{\sigma_{0,cc}}
    =\frac{m_{c}}{m_{a}}
    =\frac{\omega_{p,aa}^{2}}{\omega_{p,cc}^{2}}
    =\frac{v_{a}^{2}}{v_{c}^{2}}
    =\frac{l_{a}^{2}}{l_{c}^{2}} .
\end{equation}

While the location of the crossover from \gls{cse} to \gls{ase} can be estimated by the condition $\delta=l$, a more precise determination of the crossover frequency comes from equating the surface resistances in the two regimes. This leads to a crossover frequency given by
\begin{equation}
    \omega_{c\leftrightarrow a}
    =\frac{32\pi^{2}}{3\mu_{0}\sigma_{0,ii}l_{j}^{2}}
\end{equation}
where $i$ indicates the current direction and $j$ indicates the wavevector direction. Along the large faces of our platelet sample, current is in the $a$ direction and wavevector is along the $c$ direction; on the smaller side faces, current is along the $c$ direction and wavevector is along the $a$ direction. Because in the spheroidal model $\sigma_{0,aa}l_{c}^{2}=\sigma_{0,cc}l_{a}^{2}$, the same crossover frequency applies for all faces in our measurement.

In reality, Sr$_2$RuO$_4$ is known to have three Fermi surface sheets. In order to apply the spheroidal model to calculate $\omega_{c\leftrightarrow a}$, we must determine the quantity $\sigma_{0,cc}l_{a}^{2}$ (or $\sigma_{0,aa}l_{c}^{2}$) for an equivalent spheroidal Fermi surface. We use the material parameters from Ref.~\cite{Mackenzie1996}, as listed in \cref{tab:sr}. Following Ref.~\cite{Mackenzie1996}, the in-plane mean free path and conductivity in Sr$_2$RuO$_4$ can be related via
\begin{equation}\label{eq:sr_FS}
    l_{a}
    =C_{1}\sigma_{0,aa}
\end{equation}
with
\begin{equation}
    C_{1}\equiv\frac{hd}{e^{2}\sum_{n}k_{F}^{(n)}}
\end{equation}
where $d=\SI{6.4}{\angstrom}$ is the inter-layer spacing.
Using \cref{eq:sr_anisotropy,eq:sr_FS} we can write 
\begin{equation}
    \sigma_{0,cc}l_{a}^{2}
    =\frac{C_{1}^{2}}{\alpha}\sigma_{0,aa}^{3} .
\end{equation}
We can then obtain the crossover frequency using our experimentally-determined value of $\sigma_{0,aa}=1/\rho_{ab}=1/\SI{50}{\nano\ohm\centi\metre}$ and the corresponding value of $\alpha=2780$:
\begin{equation}
    \frac{\omega_{c\leftrightarrow a}}{2\pi} 
    =\frac{1}{2\pi}
    \frac{32\pi^{2}}{3\mu_{0}}
    \left(\frac{C^{2}}{\alpha}\sigma_{0,aa}^{3} \right)^{-1}
    \approx\SI{5}{\tera\hertz}.
\end{equation}
We see that we should not expect to observe the crossover from \gls{cse} to \gls{ase} according to conventional theory.

\begin{table}[!htbp]
    \centering
    \begin{tabular}{p{1.5cm}p{2cm}p{2cm}p{2cm}}
        \toprule
        Sheet 
        & $k_F$ (\si{\per\angstrom}) 
        & $m^*$ ($m_e$) 
        & $v_F$ (\SI[per-mode=symbol]{e6}{\metre\per\second}) \\
        \midrule
        $\alpha$ & 0.30 & 3.2 & 0.11 \\
        $\beta$ & 0.62 & 6.6 & 0.11 \\
        $\gamma$ & 0.75 & 12 & 0.072 \\
        \bottomrule
    \end{tabular}
    \caption{Fermi surface parameters for Sr$_2$RuO$_4$. The parameters $k_F$ and $m^*$ come from Ref.~\cite{Mackenzie1996}, while $v_F$ was calculated using $v_F=\hbar k_F/m^*$.}
    \label{tab:sr}
\end{table}

\section{S\lowercase{n} preparation \& analysis}

\subsection{Preparation}

The polycrystalline, platelet sample of Sn was prepared by cold-rolling 99.99\% pure Sn to the desired thickness and then cutting the lateral dimensions with a razor blade.

\subsection{Anomalous Skin Effect fit}

To fit the Sn data, we used the original theory of the \gls{ase} developed by \citet{Reuter1948}. It is based on the assumption of an isotropic, three-dimensional, free-electron dispersion. The asymptotic expression for the surface resistance in the \gls{ase} regime can be expressed as
\begin{equation}\label{eq:ase_vF}
    R
    =\left(
        \frac
            {3\sqrt{3}\pi\hbar^{3}\mu_{0}^{2}}
            {16e^{2}m_{e}^{2}}
            \frac{\omega^{2}}{v_{F}^{2}}
    \right)^{1/3}
\end{equation}
where we have assumed that electrons scatter diffusely from the sample's surface and have a band mass equal to the free electron mass $m_{e}$. We fit the Sn data to \cref{eq:ase_vF} with the Fermi velocity $v_{F}$ as the only free parameter.

\subsection{Verification of expectation of Anomalous Skin Effect}

Here, we wish to verify that we would expect to observe the \gls{ase} in Sn according to standard theory. As discussed in the main text, in order to observe the \gls{ase} over our entire frequency range, the classical skin depth must be less than the \gls{mr} mean free path over the entire frequency range of the experimental measurements: $\delta_{\text{cl}}<\lambda_{\text{mr}}$. A more precise but less physically insightful condition is that the asymptotic \gls{cse} surface resistance $R_{\text{cse}}$ should be greater than the asymptotic \gls{ase} surface resistance over the entire experimental frequency range: $R_{\text{cse}}>R_{\text{ase}}$. An upper bound on the DC resistivity comes from meeting this criterion at our lowest measured frequency. Using \cref{eq:Zc_sph,eq:ase_vF} to rewrite the inequality in terms of resistivity, we find 
\begin{equation}\label{eq:Sn_gamma_bound}
    \rho
    <\frac{2A^{2}\omega_{\text{min}}^{1/3}}{\mu_{0}}
    \approx \SI{20}{\nano\ohm\centi\metre} 
\end{equation}
where we have used the experimentally-determined value of $A$ to evaluate the expression.
We cannot deduce the DC resistivity of our sample directly from our measurement---a remarkable property of the \gls{ase} is that the asymptotic value of the surface resistance becomes independent of the DC resistivity \cite{Pippard1947}. 
Therefore, we turn to reported resistivity values for Sn to assess whether  the inequality expressed in \cref{eq:Sn_gamma_bound} can reasonably be met.
In Ref. \cite{Pippard1947}, the resistivities of two Sn samples from different commercial sources were reported to be 2 and \SI{7}{\nano\ohm\centi\metre} at \SI{4.2}{\kelvin}. 
This suggests that it is indeed reasonable for the inequality expressed in \cref{eq:Sn_gamma_bound} to be achieved at our measurement temperature of \SI{5.0}{\kelvin}.

\section{\texorpdfstring{P\lowercase{d}C\lowercase{o}O$_2$}{PdCoO2} preparation \& analysis}\label{sec:em}

\subsection{Preparation}

Single crystals of PdCoO$_2$ were grown using a mixture of PdCl$_2$ and CoO using the following methathetical reaction in an evacuated quartz ampule: PdCl\textsubscript{2} + 2CoO $\rightarrow$ 2PdCoO\textsubscript{2} + CoCl\textsubscript{2}. The ampule was heated to \SI{1000}{\celsius} for \SI{12}{\hour} and then to between \SI{700}{} and \SI{750}{\celsius} for \SI{5}{\day}. The product was washed and distilled with water and ethanol to remove CoCl\textsubscript{2}.

Platelet samples were cut into hexagons using a high-precision wire saw with a \SI{50}{\micro\meter} tungsten wire and \SI{50}{\nano\meter} Al\textsubscript{2}O\textsubscript{3} abrasive suspended in glycerin. The orientation of the crystals was determined via their growth edges, which are oriented perpendicular to the crystallographic axes. A goniometer mounted to the wire saw was used to rotate the samples in between cuts. Samples 1 and 2 were cut from the same original crystal. For Sample 3, a smaller thickness was desired, so it was cut from a separate crystal from the same batch. The sample dimensions listed in \cref{tab:dims} were determined via optical microscopy and were used for the data analysis in \cref{sec:em}.

\begin{table}[!htbp]
    \centering
    \begin{tabular}{p{3cm}p{3cm}p{3cm}p{3cm}p{3cm}}
        \toprule                            
        Sample
        & \begin{tabular}[b]{@{}l@{}}
            Hexagonal face\\area [\si{\milli\metre^2}]
        \end{tabular}
        & \begin{tabular}[b]{@{}l@{}}
            Hexagonal face\\perimeter [\si{\milli\metre}]
        \end{tabular}
        & Thickness [\si{\micro\meter}] 
        & Cut orientation \\ 
        \midrule                             
        1 & 0.187 & 1.72 & 88 & $a\parallel y$ \\
        2 & 0.169 & 1.62 & 79 & $a\parallel x$ \\
        3 & 0.191 & 1.76 & 49 & $a\parallel x$ \\
        \bottomrule       
    \end{tabular}
    \caption{Geometry of PdCoO$_2$ samples. The cut orientations refer to the directions defined in Figure 2 of the main text.}
    \label{tab:dims}
\end{table}

\subsection{Raw surface resistance}

To quantitatively interpret the measurements, it is necessary to account for the extrinsic geometric effects which lead to a difference between the externally applied, spatially-uniform magnetic field $\bm{H}_{0}$ and the magnetic field $\bm{H}(\bm{r})$ at the sample's surface. Consider a sample characterized by two dimensions: $c$, parallel to $\bm{H}_{0}$, and $a$, perpendicular to $\bm{H}_{0}$. In a conductor for which the skin depth $\delta\ll\{c,a\}$, eddy currents will act such that at the sample's surface, $\bm{H}(\bm{r})$ is nearly parallel to the surface. For a sample with $c\gg a$, $H(\bm{r})\approx H_{0}$ at the sample's surface. However, for $c<a$, as in the present work, $H(\bm{r})$ will have a non-trivial variation over the sample's surface. This effect is well-known in the context of magnetization measurements, where it is common practice to account for extrinsic geometric effects using demagnetization factors \cite{Prozorov2018}. In this work, we measured the power dissipated by the sample. In analogy with demagnetization factors, here we examine ``power factors'' to account for extrinsic geometric effects in our measurements of power absorption. A similar concept has been invoked previously in Refs. \cite{Fawzi1983,Landau1984}.

The time-averaged power dissipated in a sample is equal to the mean electromagnetic energy entering the sample per unit time:
\begin{equation}
    P=\int\bm{\bar{S}}\cdot d\bm{A}
    =\frac{1}{2}\text{Re}\left[\int dA\,E_{t}\times H_{t}^{*}\right]
    =\frac{1}{2}R_{\text{raw}}\int dA\,H_{\text{t}}^{2} 
\end{equation}
where $\bm{\bar{S}}$ is the time-averaged Poynting vector, $E_{\text{t}}$ and $H_{\text{t}}$ are the magnitudes of the tangential fields at the sample's surface, and $R_{\text{raw}}$ is the sample's effective surface resistance. When $\delta\ll\{c,a\}$, the variation of $H_{t}^{2}$ over the sample's surface is that of a perfect conductor of the same shape and is independent of absolute sample size \cite{Landau1984}. Therefore, we define the power factor for a given sample shape as
\begin{equation}
    \alpha\equiv\frac{1}{AH_{0}^{2}}\int H_{t}^{2}dA
\end{equation}
where $H_{t}$ is the tangential magnetic field strength experienced by a perfect conductor of the same shape. The surface resistances reported in Fig. 2(d) of the main text were found from measurements of power dissipation via
\begin{equation}\label{eq:reff}
    R_{\text{raw}}=\frac{2P}{\alpha A H_{0}^{2}} .
\end{equation}
$H_{0}$ is the magnitude of the applied magnetic field, as determined via an in-situ reference sample, and $A$ is the sample's surface area. 

To confirm the validity of taking $\delta\ll\{c,a\}$, we have estimated the skin depth across the range of measured frequencies using the classical expression and the published $ab$-plane residual resistivity of \SI{7.5}{\nano\ohm\centi\metre} \cite{Hicks2012}. We find that the classical skin depth is always $<\SI{0.1}{\micro\meter}$, and thus always more than two orders of magnitude less than the smallest sample dimension.

\begin{figure}[!htbp]
    \centering 
    \includegraphics{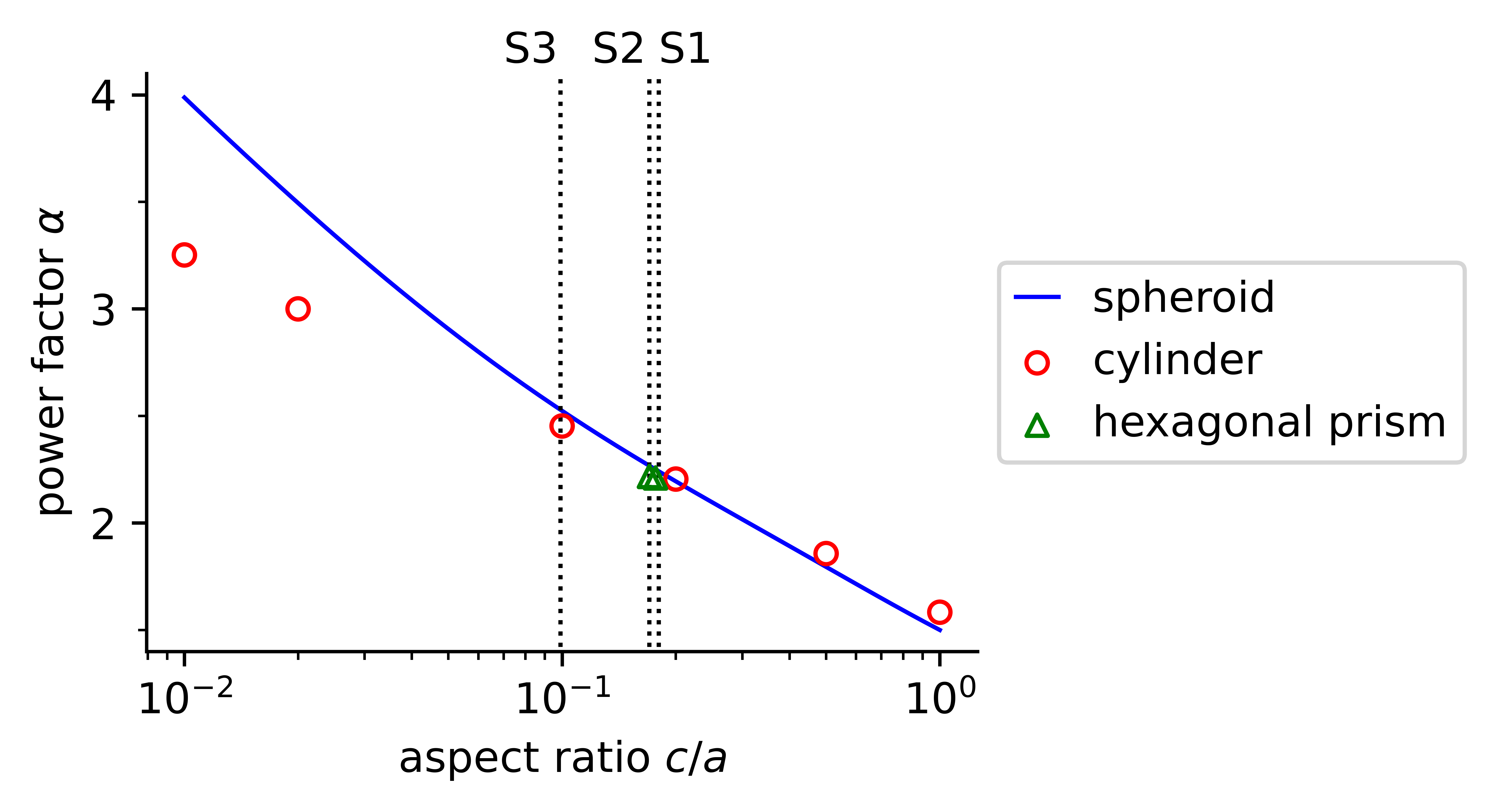}
    \caption{Effect of sample geometry on overall power absorption.}
    \label{fig:alpha}
\end{figure}

We studied the power factor $\alpha$ as a function of sample aspect ratio $c/a$ for several different shapes, as shown in \cref{fig:alpha}. For spheroids, it was possible to treat the problem analytically \cite{Baker2022}. For cylinders and hexagonal prisms, finite-element simulations were performed \cite{Branch2021}. Because the axial symmetry of cylinders allows for faster convergence of simulation results, this shape was studied over a wide range of aspect ratios. Since the simulation of hexagonal prisms was more resource-intensive, we only focused on directly relevant aspect ratios. Within the range of aspect ratios relevant to the samples studied here, all three shapes yielded similar results.

\subsection{Surface resistance components}\label{sec:components}

\begin{figure}[!htbp]
    \centering
    \includegraphics{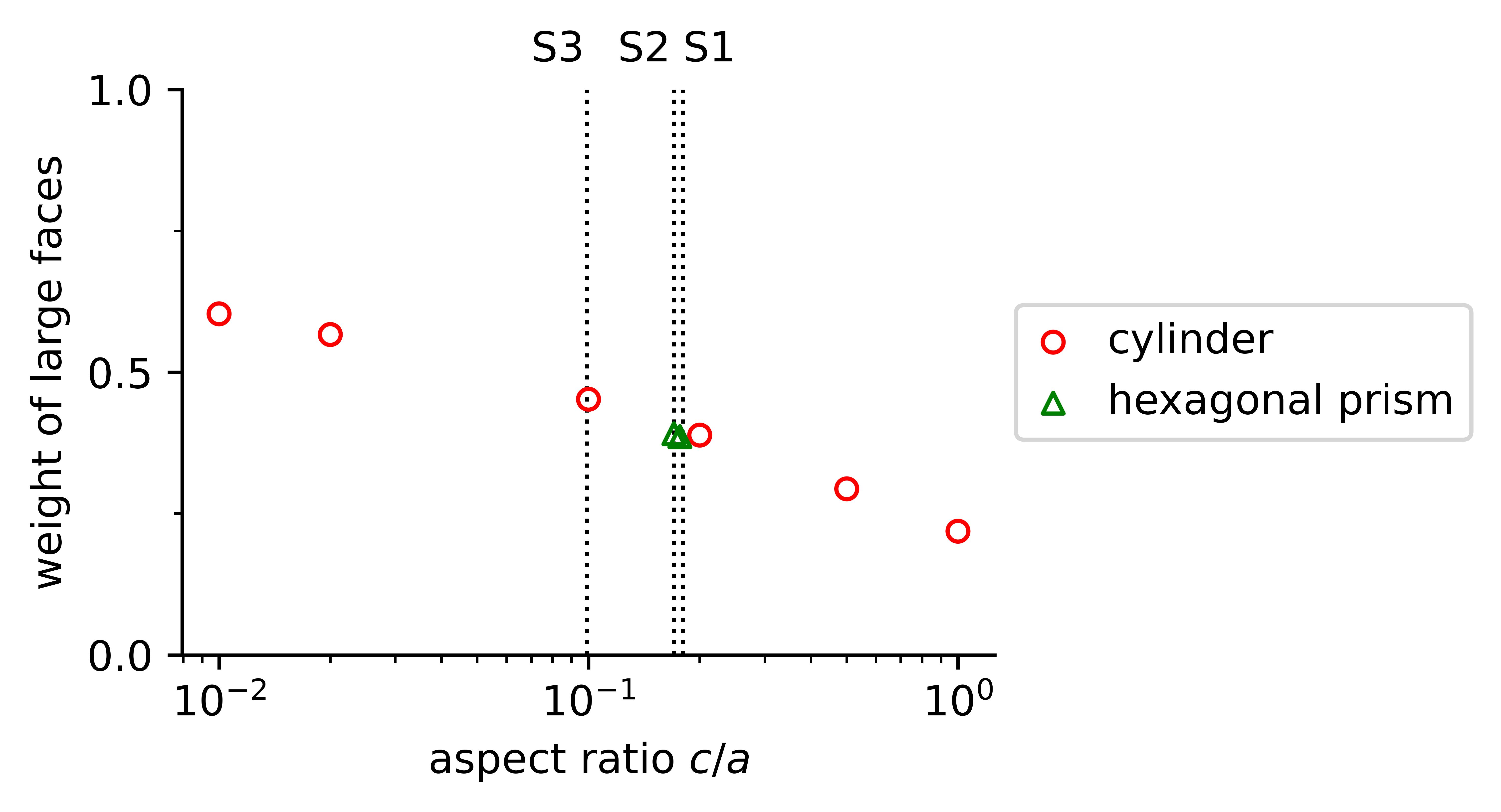}
    \caption{Effect of sample geometry on relative contribution of faces.}
    \label{fig:weights}
\end{figure}

Because $\delta\ll\{c,a\}$, $R_{\text{raw}}$ can be viewed as a sum of two independent components: $R_{\perp}$ comes from the faces perpendicular to $\bm{H}_{0}$ (the two large hexagonal faces) and $R_{\parallel}$ comes from the faces parallel to $\bm{H}_{0}$ (the six small rectangular faces). The relative contribution of the two components is set by a weight $w_{\perp}$:
\begin{equation}\label{eq:components}
    R_{\text{raw}}=w_{\perp}R_{\perp}+(1-w_{\perp})R_{\parallel}
\end{equation}
The weight $w_{\perp}$ can be found via the variation of the tangential magnetic field strength over the surface of a perfect conductor of the same shape:
\begin{equation}
    w_{\perp}=\left.\int_{A_{\perp}}H_{t}^{2}dA\middle/\int_{A_{\perp}+A_{\parallel}}H_{t}^{2}dA\right.
\end{equation}
where $A_{\perp}$ ($A_{\parallel}$) is the area of the perpendicular (parallel) faces. The weights determined by our finite-element simulations are shown in \cref{fig:weights}.

Applying \cref{eq:components} to the three samples that we measured, we get
\begin{equation}\label{eq:s1}
    R_{1}=w_{1}R_{a,c}+(1-w_{1})R_{30,0}
\end{equation}
\begin{equation}\label{eq:s2}
    R_{2}=w_{2}R_{a,c}+(1-w_{2})R_{0,30}
\end{equation}
\begin{equation}\label{eq:s3}
    R_{3}=w_{3}R_{a,c}+(1-w_{3})R_{0,30}
\end{equation}
where on the left side of the equations, $R_{i}$ refers to $R^{\text{eff}}$ for Sample $i$, and on the right side of the equations, the two subscripts refer to the directions $\bm{\hat{J}},\bm{\hat{q}}$.
Using \cref{eq:s2,eq:s3}, $R_{a,c}$ was found from the measurements of Samples 2 and 3:
\begin{equation}\label{eq:rac}
    R_{a,c}=\frac{w_{2}R_{3}-w_{3}R_{2}}{w_{2}(1-w_{3})-(1-w_{2})w_{3}} .
\end{equation}
$R_{a,c}$ was then used to determine $R_{30,0}$ and $R_{0,30}$ from \cref{eq:s1,eq:s2}:
\begin{equation}\label{eq:r300}
    R_{30,0}=\frac{R_{1}-w_{1}}{1-w_{1}}
\end{equation}
\begin{equation}\label{eq:r030}
    R_{0,30}=\frac{R_{2}-w_{2}R_{a,c}}{1-w_{2}} .
\end{equation}

The values of $\alpha$ and $w_{\perp}$ that were used in \cref{eq:reff,eq:rac,eq:r300,eq:r030} to arrive at the data presented in the main text are summarized in \cref{tab:cal_params}. For Samples 1 and 2, we used the values of $\alpha$ and $w_{\perp}$ from our simulations. For Sample 3, some adjustment relative to the simulation values was necessary in order to obtain physically plausible results for the components $R_{a,c}$, $R_{30,0}$, and $R_{0,30}$. We found that the magnitude of $R_{3}$ determined using the simulation value $\alpha_{3}^{\text{sim}}$ was larger than expected. Specifically, unless $R_{3}$ is of a comparable magnitude to $R_{2}$, then the decomposition into components via \cref{eq:rac,eq:r300,eq:r030} yields negative values, which is unphysical. A comparable magnitude of $R_{3}$ to $R_{2}$ can be accomplished by using$\alpha_{3}\approx 4$, whereas the simulation value is $\alpha_{3}^{\text{sim}}\approx2.5$. There are a couple potential explanations for this discrepancy. The first is that Sample 3, being cut from a different original sample, has a higher residual \gls{mr} scattering rate than Sample 2. In the \gls{cse}, $R\propto\sqrt{\gamma_{\text{mr}}}$, so a factor of $4/2.5\approx1.6$ difference in magnitude corresponds to a factor of $1.6^{2}\approx2.6$ difference in residual scattering rates. This is consistent with the variation in residual resistivity reported by \citet{Nandi2018}. The second potential explanation is that we failed to accurately account for geometric effects in our simulations, with $\alpha_{3}^{\text{sim}}$ being smaller than the true value of $\alpha_{3}$. This difference in magnitude would result if the power factor $\alpha$ increases more rapidly with decreasing aspect ratio $c/a$ than captured by our simulations. With $\alpha_{3}$ set to 4, next we turned to $w_{3}$. The simulation value of $\approx0.45$ also leads to negative decomposed values, whereas there is a wide range of $w_{3}$ values near $\approx0.6$ for which the decomposed values are positive and only weakly dependent on the specific choice of $w_{3}$. As with $\alpha$, the difference between empirical and simulated results may indicate that simulations are underestimating the magnitude of extrinsic geometric effects for small aspect ratios. Despite the necessity to adjust $\alpha_{3}$ and $w_{3}$ relative to simulation values to get physically reasonable results, the resulting values of $R_{a,c}$, $R_{30,0}$, and $R_{0,30}$ vary little over the range of $(\alpha_{3},w_{3})$ pairs leading to physically plausible results.

\begin{table}[!htbp]
    \centering
    \begin{tabular}{p{3cm}p{3cm}p{3cm}}
         \toprule
        Sample & $\alpha$ & $w_{\perp}$  \\
        \midrule
        1 & 2 & 0.4 \\
        2 & 2 & 0.4\\
        3 & 4 & 0.6 \\
        \bottomrule
    \end{tabular}
    \caption{Data calibration parameters.}
    \label{tab:cal_params}
\end{table}

\section{\texorpdfstring{P\lowercase{d}C\lowercase{o}O$_2$}{PdCoO2} surface resistance calculations}

Here we describe the details of the PdCoO$_2$  surface resistance calculations shown in the main text. First, we describe a generalized treatment of non-local electrodynamics that allows for arbitrary Fermi surface geometry and separate \gls{mr} and \gls{mc} scattering rates. Then, we describe the application of this treatment to PdCoO$_2$. 

\subsection{Distribution function}

We start with the Boltzmann equation for the evolution of the electronic distribution function $f_{\bm{k}}(\bm{r},t)$ in the presence of a spatially- and time-varying electric field $\bm{E}(\bm{r},t)$:
\begin{equation}\label{eq:boltz_rt}
    \left[
        \frac{\partial}{\partial t}
        +\dot{\bm{r}}\cdot\nabla_{\bm{r}}
        +\dot{\bm{k}}\cdot\nabla_{\bm{k}}
    \right]f_{\bm{k}}(\bm{r},t)=-\mathcal{C}_{\bm{k}}[f_{\bm{k}}(\bm{r},t)]
\end{equation}
with semi-classical equations of motion
\begin{equation}
    \dot{\bm{r}}
    =\bm{v}_{\bm{k}}
    =\frac{1}{\hbar}\nabla_{\bm{k}}\mathcal{E}_{\bm{k}}   
\end{equation}
and
\begin{equation}
    \dot{\bm{k}}
    =-\frac{e}{h}\bm{E}(\bm{r},t)
\end{equation}
and where $\mathcal{C}_{\bm{k}}[f_{\bm{k}}]$ is the collision integral.
We expand the distribution function around its equilibrium value $f_{0}$:
\begin{equation}
    f_{\bm{k}}
    =f_{0}+w_{\bm{k}}\psi_{\bm{k}}
\end{equation}
with 
\begin{equation}
    w_{\bm{k}}
    =k_{B}T\left(-\frac{\partial f_{0}}{\partial\mathcal{E}_{\bm{k}}}\right)
    =f_{0}(1-f_{0}) .
\end{equation}
We take the Fourier transform of \cref{eq:boltz_rt} and keep terms to linear order in $\psi_{\bm{k}}$ to obtain
\begin{equation}
    [-i\omega+i\bm{v}_{\bm{k}}\cdot\bm{q}]\psi_{\bm{k}}(\bm{q},\omega)
    +\frac{e}{k_{B}T}\bm{E}(\bm{q},\omega)\cdot\bm{v}_{\bm{k}}
    =-\hat{C}_{\bm{k}}\psi_{\bm{k}}(\bm{q},\omega)
\end{equation}
where $\hat{C}$ is the linearized collision operator
\begin{equation}
    \hat{C}\psi_{\bm{k}}
    =\frac{1}{w_{\bm{k}}}\int_{\bm{k}'}
    \frac{\delta\mathcal{C}_{\bm{k}}}{\delta\psi_{\bm{k}'}}\psi_{\bm{k}'} .
\end{equation}
Let $\psi_{\bm{k}}$ be an element of a function space with inner product
\begin{equation}
    \left\langle\phi\middle|\psi\right\rangle
    =\int_{\bm{k}}w_{\bm{k}}\phi_{\bm{k}}^{*}\psi_{\bm{k}} .
\end{equation}

We use as a basis the complete and orthonormal set of eigenfunctions $\chi_{\bm{k},m}$ of the collision operator:
\begin{equation}
    \hat{C}\chi_{\bm{k},m}=\gamma_{\bm{k},m}\chi_{\bm{k},m}
\end{equation}
with 
\begin{equation}\label{eq:complete}
    \sum_{m}\left|\chi_{\bm{k},m}\right\rangle\left\langle\chi_{\bm{k},m}\right|=1
\end{equation}
and
\begin{equation}\label{eq:orthonormal}
    \left\langle\chi_{\bm{k},m}\middle|\chi_{\bm{k},m'}\right\rangle
    =\delta_{m,m'} .
\end{equation}
We assume that the eigenfunctions $\chi_{\bm{k},m}$ include
\begin{equation}
    \chi_{\bm{k},0}
    =c_{0}
\end{equation}
and
\begin{equation}
    \chi_{\bm{k},i}
    =c_{i}c_{0}\hat{v}_{\bm{k},i}
\end{equation}
where $i\in\{x,y,z\}$ and that the eigenvalue spectrum is given by 
\begin{equation}
    \gamma_{\bm{k},m}
    =\begin{cases}
        0 & m=0\\
        \gamma_{\bm{k},i}^{\text{mr}} & m=i\\
        \gamma_{\bm{k}}^{\text{mc}} & \text{otherwise} .
    \end{cases}
\end{equation}
This describes a scenario in which collisions conserve charge, relax momentum in the $i$ direction at a rate $\gamma_{\bm{k},i}^{\text{mr}}$, and relax all other modes at a rate $\gamma_{\bm{k}}^{\text{mc}}$.
Using \cref{eq:complete}, we may write the collision operator as 
\begin{equation}
    \hat{C}
    =\gamma_{\bm{k}}^{\text{mc}}(1-\left|\chi_{\bm{k},0}\right\rangle
    \left\langle\chi_{\bm{k},0}\right|)
    -\sum_{i}\delta\gamma_{\bm{k},i}\left|\chi_{\bm{k},i}\right\rangle
    \left\langle\chi_{\bm{k},i}\right|
\end{equation}
where $\delta\gamma_{\bm{k},i}\equiv\gamma_{\bm{k}}^{\text{mc}}-\gamma_{\bm{k},i}^{\text{mr}}$. 

Throughout, we will use 
\begin{equation}
    \int_{\bm{k}}\cdots
    \equiv\frac{2}{(2\pi)^{d}}{\int}d\bm{k}\cdots
    =\frac{2}{(2\pi)^{d}\hbar}
    \int_{0}^{\infty}d\mathcal{E}
    \int_{\mathcal{S}(\mathcal{E})}\frac{dS}{v_{\bm{k}}}\cdots .
\end{equation}
Furthermore, we will assume that $T{\ll}T_{F}$ such that 
\begin{equation}
    \int_{\bm{k}}
    -\frac{{\partial}f_{0}}{\partial\mathcal{E}_{\bm{k}}}\cdots
    =\frac{2}{(2\pi)^{d}\hbar}\int_{S_{F}}\frac{dS}{v_{\bm{k}}}\cdots
\end{equation}
where $\mathcal{S}_{F}$ is the Fermi surface $\mathcal{S}(\mathcal{E}_{\bm{k}}=\mathcal{E}_{F})$. Finally, for simplicity, we will assume that on the Fermi surface the magnitude of the velocity is isotropic: $\bm{v}_{\bm{k}}=v_{F}\bm{\hat{v}}_{\bm{k}}$ for $\bm{k}$ on $\mathcal{S}_{F}$. 

The constants $c_{0}$ and $c_{i}$ are determined by \cref{eq:orthonormal}. We find
\begin{equation}
    \frac{1}{c_{0}^{2}}
    =\frac{2k_{B}T}{(2\pi)^{d}{\hbar}v_{F}}S_{F}
\end{equation}
and
\begin{equation}
    \frac{1}{c_{i}^{2}}
    =\int_{\mathcal{S}_{F}}\frac{dS}{S_{F}}\hat{v}_{\bm{k}i}^{2}
\end{equation}
where
\begin{equation}
    S_{F}\equiv\int_{\mathcal{S}_{F}}dS .
\end{equation}

We can now rewrite the collision integral  as
\begin{equation}
    \hat{C}\psi_{\bm{k}}
    =\gamma_{\bm{k}}^{\text{mc}}\psi_{\bm{k}}
    -\gamma_{\bm{k}}^{\text{mc}}n_{0}
    -\sum_{i}c_{i}^{2}\delta\gamma_{\bm{k},i}\hat{v}_{\bm{k}i}p_{i}
\end{equation}
where
\begin{equation}
    n_{0}\equiv\int_{\mathcal{S}_{F}}\frac{dS}{S_{F}}\psi_{\bm{k}}
\end{equation}
and
\begin{equation}
    p_{i}\equiv\int_{\mathcal{S}_{F}}\frac{dS}{S_{F}}\hat{v}_{\bm{k}i}\psi_{\bm{k}}
\end{equation}
The solution to the Boltzmann equation is then 
\begin{equation}
    \psi_{\bm{k}}(\bm{q},\omega)
    =\frac
        {
            -\frac{e}{k_{B}T}\bm{E}(\bm{q},\omega)\cdot\bm{v}_{\bm{k}}
            +\gamma_{\bm{k}}^{\text{mc}}n_{0}(\bm{q},\omega)
            +\sum_{i}c_{i}^{2}\delta\gamma_{\bm{k},i}\,\hat{v}_{\bm{k}i}\,p_{i}(\bm{q},\omega)
        }
        {\gamma_{\bm{k}}^{\text{mc}}-i\omega+i\bm{v}_{\bm{k}}\cdot\bm{q}} .
\end{equation}

Because we are ultimately interested in finding the transverse conductivity, we take $\bm{E}\perp\bm{q}$ with $\bm{E}\parallel\bm{\hat{\alpha}}$ and $\bm{q}\parallel\bm{\hat{\beta}}$. We define
\begin{equation}
    \braket{\mathcal{A}}
    \equiv\int_{\mathcal{S}_{F}}\frac{dS}{S_{F}}
    \frac
        {\mathcal{A}}
        {\gamma_{\bm{k}}^{\text{mc}}-i\omega+iv_{F}\hat{v}_{\bm{k}\beta}q} .
\end{equation}
We find that
\begin{equation}
    \begin{pmatrix}
        \gamma_{\bm{k}}^{\text{mc}}\braket{1}-1
        & c_{\beta}^{2}\delta\gamma_{\bm{k},\beta}\braket{\hat{v}_{\bm{k}\beta}} 
        & c_{\alpha}^{2}\delta\gamma_{\bm{k},\alpha}\braket{\hat{v}_{\bm{k}\alpha}}  
        & c_{\gamma}^{2}\delta\gamma_{\bm{k},\gamma}\braket{\hat{v}_{\bm{k}\gamma}}  \\
        \gamma_{\bm{k}}^{\text{mc}}\braket{\hat{v}_{\bm{k}\beta}} 
        & c_{\beta}^{2}\delta\gamma_{\bm{k},\beta}\braket{\hat{v}_{\bm{k}\beta}^{2}}-1 
        & c_{\alpha}^{2}\delta\gamma_{\bm{k},\alpha}\braket{\hat{v}_{\bm{k}\beta}\hat{v}_{\bm{k}\alpha}}  
        & c_{\gamma}^{2}\delta\gamma_{\bm{k},\gamma}\braket{\hat{v}_{\bm{k}\beta}\hat{v}_{\bm{k}\gamma}}  \\
        \gamma_{\bm{k}}^{\text{mc}}\braket{\hat{v}_{\bm{k}\alpha}} 
        & c_{\beta}^{2}\delta\gamma_{\bm{k},\beta}\braket{\hat{v}_{\bm{k}\alpha}\hat{v}_{\bm{k}\beta}} 
        & c_{\alpha}^{2}\delta\gamma_{\bm{k},\alpha}\braket{\hat{v}_{\bm{v}\alpha}^{2}}-1 
        & c_{\gamma}^{2}\delta\gamma_{\bm{k},\gamma}\braket{\hat{v}_{\bm{k}\alpha}\hat{v}_{\bm{k}\gamma}}  \\
        \gamma_{\bm{k}}^{\text{mc}}\braket{\hat{v}_{\bm{k}\gamma}} 
        & c_{\beta}^{2}\delta\gamma_{\bm{k},\beta}\braket{\hat{v}_{\bm{k}\gamma}\hat{v}_{\bm{k}\beta}} 
        & c_{\alpha}^{2}\delta\gamma_{\bm{k},\alpha}\braket{\hat{v}_{\bm{k}\gamma}\hat{v}_{\bm{k}\alpha}}  
        & c_{\gamma}^{2}\delta\gamma_{\bm{k},\gamma}\braket{\hat{v}_{\bm{k}\gamma}^{2}}-1  
    \end{pmatrix}
    \begin{pmatrix}
        n_{0} \\ p_{\beta} \\ p_{\alpha} \\ p_{\gamma}
    \end{pmatrix}
    =\frac{e}{k_{B}T}E_{y}v_{F}
    \begin{pmatrix}
        \braket{\hat{v}_{\bm{k}\alpha}} \\
        \braket{\hat{v}_{\bm{k}\beta}\hat{v}_{\bm{k}\alpha}} \\
        \braket{\hat{v}_{\bm{k}\alpha}^{2}} \\
        \braket{\hat{v}_{\bm{k}\gamma}\hat{v}_{\bm{k}\alpha}}
    \end{pmatrix}
\end{equation}
and assuming three mirror planes, we get
\begin{equation}
    \begin{pmatrix}
        \gamma_{\bm{k}}^{\text{mc}}\braket{1}-1
        & c_{\beta}^{2}\delta\gamma_{\bm{k},\beta}\braket{\hat{v}_{\bm{k}\beta}} 
        & 0 
        & 0  \\
        \gamma_{\bm{k}}^{\text{mc}}\braket{\hat{v}_{\bm{k}\beta}} 
        & c_{\beta}^{2}\delta\gamma_{\bm{k},\beta}\braket{\hat{v}_{\bm{k}\beta}^{2}}-1 
        & 0
        & 0  \\
        0
        & 0
        & c_{\alpha}^{2}\delta\gamma_{\bm{k},\alpha}\braket{\hat{v}_{\bm{v}\alpha}^{2}}-1 
        & 0 \\
        0
        & 0
        & 0
        & c_{\gamma}^{2}\delta\gamma_{\bm{k},\gamma}\braket{\hat{v}_{\bm{k}\gamma}^{2}}-1  
    \end{pmatrix}
    \begin{pmatrix}
        n_{0} \\ p_{\beta} \\ p_{\alpha} \\ p_{\gamma}
    \end{pmatrix}
    =\frac{e}{k_{B}T}E_{y}v_{F}
    \begin{pmatrix}
        0 \\
        0 \\
        \braket{\hat{v}_{\bm{k}\alpha}^{2}} \\
        0
    \end{pmatrix}
\end{equation}
To ensure that $\psi_{\bm{k}}=0$ when $\bm{E}=0$, we must also have $n_{0}=p_{\beta}=0$. For $p_{\alpha}$ we have
\begin{equation}
    p_{\alpha}
    =-\frac{eEv_{F}}{k_{B}T}\frac
        {\braket{\hat{v}_{\bm{k}\alpha}^{2}}}
        {1-c_{\alpha}^{2}\delta\gamma_{\bm{k},\alpha}\langle\hat{v}_{\bm{k}\alpha}^{2}\rangle}
\end{equation}
so that 
\begin{equation}
    \psi_{\bm{k}}
    =-\frac{eEv_{F}}{k_{B}T}
    \frac{1}{\gamma_{\bm{k}}^{\text{mc}}-i\omega+iqv_{F}\hat{v}_{\bm{k}\beta}}
    \frac{\hat{v}_{\bm{k}\alpha}}{1-c_{y}^{2}\delta\gamma_{\bm{k},\alpha}\langle\hat{v}_{\bm{k}\alpha}^{2}\rangle} .
\end{equation}

\subsection{Non-local transverse conductivity}

Electrical current is given by 
\begin{equation}
    \bm{J}
    =-e\int_{\bm{k}}\bm{v}_{\bm{k}}f_{\bm{k}} .
\end{equation}
Using
\begin{equation}
    J_{i}(\bm{q},\omega)
    =\sigma_{ij}(\bm{q},\omega)E_{j}(\bm{q},\omega)
\end{equation}
and assuming $\delta\gamma_{\bm{k},\alpha}$ to be $\bm{k}$-independent, we find the non-local transverse conductivity for $\bm{E}\parallel\bm{\hat{\alpha}}$ and $\bm{q}\parallel\bm{\hat{\beta}}$ as 
\begin{equation}
    \sigma(q,\omega)
    =\epsilon_{0}\Omega_{p}^{2}
    \frac
        {G_{0}(q,\omega)}
        {1-c_{\alpha}^{2}\delta\gamma_{\alpha}G_{0}(q,\omega)}
\end{equation}
where
\begin{equation}
    G_{0}(q,\omega)
    \equiv\langle\hat{v}_{\bm{k}\alpha}^{2}\rangle
    =\int_{\mathcal{S}_{F}}\frac{dS}{S_{F}}
    \frac
        {{\hat{v}_{\bm{k}\alpha}}^{2}}
        {\gamma_{\bm{k}}^{\text{mc}}-i\omega+iqv_{F}\hat{v}_{\bm{k}\beta}}  .
\end{equation}
and
\begin{equation}
    \Omega_{p}^{2}\equiv\sum_{i}\omega_{p,ii}^{2}
\end{equation}
where plasma frequency is given by
\begin{equation}
    \epsilon_{0}\omega_{p,ii}^{2}
    =\frac{e^{2}v_{F}^{2}}{{c_{0}}^{2}k_{B}T}
    \int_{\mathcal{S}_{F}}\frac{dS}{S_{F}}{\hat{v}_{\bm{k}i}}^{2} . 
\end{equation}

\subsection{Parameterization of Fermi surface}

For a Fermi surface $\mathcal{S}_{F}$ parametrized by the Fermi vector $\bm{k}_{F}(g,h)$ with $g\in\{g_{1},g_{2}\}$ and $h\in\{h_{1},h_{2}\}$, we define the vector $\bm{n}$ as 
\begin{equation}
    \bm{n}(g,h)
    =\frac{\partial\bm{k}_{F}(g,h)}{\partial g}
    \times\frac{\partial\bm{k}_{F}(g,h)}{\partial h} .
\end{equation}
Then the Fermi surface integral is given by 
\begin{equation}
    \int_{\mathcal{S}_{F}}dS\cdots
    =\int_{g_{1}}^{g_{2}}dg\int_{h_{1}}^{h_{2}}dh\,n(g,h)\cdots
\end{equation}
with $n\equiv|\bm{n}|$. The unit vector normal to the Fermi surface (parallel to the Fermi velocity) is given by 
\begin{equation}
    \bm{\hat{n}}(g,h)
    =\frac{\bm{n}(g,h)}{n(g,h)}.
\end{equation}

\subsection{Surface resistance}

For specular scattering of electrons at the sample's surface, surface impedance is given by \cite{Reuter1948}
\begin{equation}
    Z
    =i\mu_{0}\omega\frac{2}{\pi}\int_{0}^{\infty}dq\left[
        i\mu_{0}\omega\sigma(q,\omega)
        +\frac{\omega^{2}}{c^{2}}
        -q^{2}
    \right]^{-1}
\end{equation}
while for diffuse scattering, it is given by \cite{Dingle1953}
\begin{equation}
    Z
    =i\mu_{0}\omega\pi\left(
        \int_{0}^{\infty}dq\,\ln\left[
            \frac{i\mu_{0}\omega\sigma(q,\omega)}{q^{2}}
            +\frac{\omega^{2}}{c^{2}q^{2}}
            -1
        \right]
    \right)^{-1}. 
\end{equation}
Finally, surface resistance is given by 
\begin{equation}
    R=\text{Re}(Z) .
\end{equation}

\subsection{Calculations for PdCoO\texorpdfstring{\textsubscript{2}}{2}}

We used the Fermi surface parameterization from \citet{Hicks2012}:
\begin{equation}
    \bm{k}_{F}(\phi, \phi_{0}, k_{z})
    =\rho(\phi-\phi_{0}, k_{z})[\cos\phi\,\bm{\hat{\imath}}+\sin\phi\,\bm{\hat{\jmath}}]+k_{z}\bm{\hat{k}}
\end{equation}
where
\begin{equation}
    \rho(\phi-\phi_{0},k_{z})
    =\sum_{\mu,\nu}k_{\mu\nu}\cos[\mu(\phi-\phi_{0})]
    \begin{cases}
        \sin[\nu dk_{z}] & k_{31} \\
        \cos[\nu dk_{z}] & \text{otherwise}
    \end{cases}    
\end{equation}
with $d=c/3$ where $c=\SI{17.743}{\angstrom}$ and with the Fermi surface harmonics listed in \cref{tab:harmonics}. The angle $\phi_{0}$ sets the in-plane rotation of the Fermi surface relative to the coordinate system. To take advantage of the simplifications arising from three mirror planes, we set $k_{31}=0$. We assumed diffuse surface scattering when calculating the surface resistance. We used the reported experimental parameters as given in \cref{tab:parameters}. We took $\gamma_{\bm{k}x}^{\text{mr}}=\gamma_{\bm{k}x}^{\text{mr}}=\gamma^{\text{mr}}$ using the value from \cref{tab:parameters}, and took $\gamma_{\bm{k}}^{\text{mc}}=\gamma^{\text{mc}}$ with $\gamma^{\text{mc}}$ as a free parameter.

\begin{table}[!htbp]
    \centering
    \begin{tabular}{p{1.5cm}p{1.5cm}p{2cm}}
        \toprule
        $\mu$ & $\nu$ & $k_{\mu,\nu}$ \\
        \midrule
        0 & 0 & 0.9538 \\
        6 & 0 & 0.040 \\
        12 & 0 & 0.007 \\
        0 & 1 & 0.0107 \\
        0 & 2 & -0.009 \\
        3 & 1 & 0.0010 \\
        \bottomrule
    \end{tabular}
    \caption{Harmonics for parameterization of Fermi surface of PdCoO\textsubscript{2} from \citet{Hicks2012}.}
    \label{tab:harmonics}
\end{table}

\begin{table}[!htbp]
    \centering
    \begin{tabular}{p{3cm}p{3cm}}
        \toprule 
        Parameter & Value \\
        \midrule
        $v_{F}$ 
        & \SI[per-mode=symbol]{7.5e5}{\metre\per\second} \\
        $\omega_{p,ab}$ 
        & \SI{7.2e15}{\hertz} \\
        $\rho_{ab}(T=\SI{2}{\kelvin})$ 
        & \SI{7.5}{\nano\ohm\centi\metre} \\
        % $\rho_{c}(T=\SI{2}{\kelvin})$ 
        % & \SI{7.5}{\nano\ohm\centi\metre} \\
        $\gamma^{\text{mr}}(T=\SI{2}{\kelvin})$
        & \SI{34}{GHz} \\
        \bottomrule
    \end{tabular}
    \caption{Parameters for PdCoO\textsubscript{2}. All parameters are directly from \citet{Hicks2012}, except for $\gamma^{\text{mr}}$ which was found as $\gamma^{\text{mr}}=\epsilon_{0}\omega_{p,ab}^{2}\rho_{ab}$.}
    \label{tab:parameters}
\end{table}

\section{Estimate of momentum-conserving electron-phonon scattering in P\lowercase{d}C\lowercase{o}O\texorpdfstring{$_2$}{2}}

\begin{figure}[!htbp]
  \centering
  \includegraphics{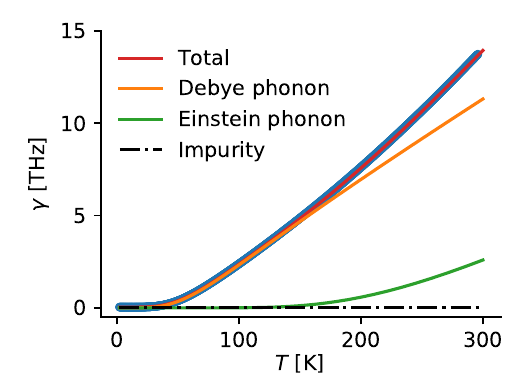}
  \caption{Fit of momentum-relaxing scattering rate in PdCoO$_2$ showing the contribution of each scattering mechanism. The data (in blue) were obtained from bulk resistivity measurements.}
  \label{fig:scat_rate}
\end{figure}

\begin{figure}[!htbp]
  \centering
  \includegraphics{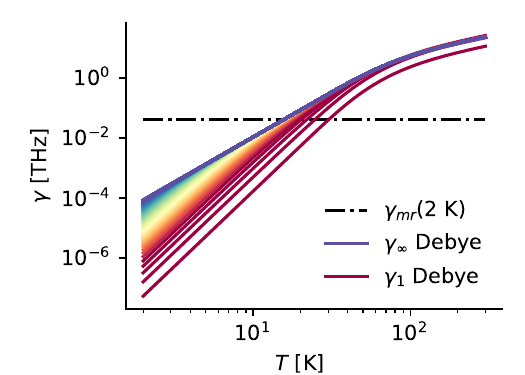}
  \caption{Spectrum of scattering rates for electron-Debye phonon scattering compared with the \SI{2}{\kelvin} momentum-relaxing scattering rate.}
  \label{fig:mc_scattering}
\end{figure}

Here we estimate the rate of \gls{mc} electron-phonon scattering in PdCoO\textsubscript{2}. First, we obtained an experimental MR scattering rate as $\gamma_{1}^{\text{exp}}=\rho_{xx}/\epsilon_{0}\omega_{p,xx}^{2}$ using $\rho_{xx}$ for a \SI{155}{\micro\metre} channel from \citet{Nandi2018} and using $\omega_{p,xx}$ = \SI{7.2e15}{Hz} from \citet{Hicks2012}. Next, we performed fits to $\gamma_{1}^{\text{exp}}$. The results are shown in \cref{fig:scat_rate} and the fit parameters given in \cref{tab:fit}. 
To find the MR electron-impurity scattering rate $\gamma_{1}^{\text{imp}}$, we fit $\gamma_{1}^{\text{exp}}$ to a constant over the range $\SI{2}{\kelvin}<T<\SI{10}{\kelvin}$. 
Next we considered electron-phonon scattering. It has previously been noted that at high temperature $\gamma_{1}^{\text{exp}}\propto T^{\alpha}$ with $\alpha>1$, in contrast with expectation that $\alpha=1$ within the Bloch-Gr{\"u}neisen treatment of electron-acoustic phonon scattering. This discrepancy has been attributed to electron-optical phonon scattering \cite{Takatsu2007,Hicks2012}. Therefore, we fit $\gamma_{1}^{\text{exp}}-\gamma_{1}^{\text{imp}}$ to a sum of Einstein and Debye contributions using 
\begin{equation}
  \hbar\gamma_{1}^{\text{Ein}}
  =\frac{\pi}{2}\lambda_{E}k_{B}T_{E}
  \frac{T_{E}/T}{\sinh^{2}(T_{E}/2T)}
\end{equation}
and
\begin{equation}\label{eq:Debye_modes}
  \hbar\gamma_{l}^{\text{Deb}}
  =4\pi\lambda_{D}k_{B}T
  \left(\frac{T}{T_{D}}\right)^{2}
  \int_{0}^{T_{D}/T}dx\, 
  \frac{x^{3}}{\cosh(x)-1}
  \left[
    1-P_{l}\left(1-(T/T_{D})^{2}x^{2}\right)
  \right]
\end{equation}
with $l=1$ (cf. eq. A.45 from \citet{Levchenko2020}). The fit returned Debye and Einstein temperatures $T_{D}$ and $T_{E}$ and transport electron-phonon couplings $\lambda_{D}$ and $\lambda_{E}$. Following \citet{Hicks2012}, the fit was performed over the range $\SI{60}{\kelvin}<T<\SI{300}{\kelvin}$ .
Having determined $T_{D}$ and $\lambda_{D}$, we then used \cref{eq:Debye_modes} to determine $\gamma_{l}^{\text{Debye}}$ for all $l$, as shown in \cref{fig:mc_scattering}. We see that at \SI{2}{K}, all $\gamma_{l}$ for electron-Debye phonon scattering are less than the experimentally determined MR scattering rate. Therefore, electron-phonon scattering is unlikely to be responsible for the MC scattering inferred from our measurements.

\begin{table}[!htbp]
  \centering
  \begin{tabular}{p{1.5cm}p{1.5cm}}
      \toprule
      $\gamma_{1}^{\text{imp}}$ & \SI{39}{\giga\hertz} \\
      $\lambda_D$ & 0.049 \\
      $T_D$ & \SI{331}{\kelvin} \\
      $\lambda_E$ & 0.030 \\
      $T_E$ & \SI{1120}{\kelvin} \\
      \bottomrule
  \end{tabular}
  \caption{Parameters for fit of momentum-relaxing scattering rate in PdCoO$_2$.}
  \label{tab:fit}
\end{table}

\newpage
\bibliography{supplement.bib}